Andrea Scharnhorst[1], Werner Ebeling[2]

# Evolutionary Search Agents in Complex Landscapes – a New Model for the Role of Competence and Meta-competence (EVOLINO and other simulation tools)

[1] Virtual Knowledge Studio, Royal Netherlands Academy of Arts and Sciences, Joan Muyskenweg 25, 1090 HC Amsterdam, Netherlands
[2] Humboldt University Berlin, Institute for Physics, Newtonstr. 15, 12489 Berlin, Germany

# 1 Abstract

The acquisition of competence is a key element in the ability to assert oneself in the complex and rapidly changing modern worlds of work. This is true of almost all areas of professional activity. In a society characterized by the acceleration of progress and globalization, the ability to adapt rapidly to new conditions is one of the main requirements of employees. Thus, flexibility is gaining in significance. This report, which involves the generalization of a previously developed concept, examines the evolution of competence, i.e. the role of competences in an evolutionary problem-solving process, and the role of flexibility as a meta-competence from the perspective of the general concept of *G*eometrically-*O*riented *E*volution *THE*ory (G_O_E_THE).

The evolution of competence is described as the characteristic of a collective search for better local solutions by interacting groups of individuals in a multi-dimensional problem space in which a value landscape is defined. In modelling terms, we refer to these individuals and groups as search agents (i.e. agents of the competence search). A search agent operates in an abstract value landscape and simulates important aspects of the actual process of the search for competence. The value landscape acts as a fitness landscape. It represents an evaluation of certain competences, i.e. the evaluation of problem solutions achieved using certain competences.

As in previous research, the dynamic of the search process is characterized using evolutionary search strategies; strategies of the types developed by Darwin and Boltzmann are used in particular. The geometrically-oriented representations of competence development in characteristics spaces or problem spaces are analyzed and previously developed concepts are systematically generalized. The concept of the “evolutionary search agent” is introduced and different interactive simulations developed. Evolutionary search agents are abstract models of real individuals and groups who seek group or maximum-competence solutions and have a reservoir of opportunities available to them in their search. The crucial difference between these search agents and the agents already widely used in research is the fact that the properties of the new agents are largely based on the methods of evolutionary search strategies.

The concepts previously developed as part of the general approach G_O_E_THE are systematically consolidated here and developed further from the perspective of competence development. As a new approach to the modelling of meta-competence, extending the earlier models we will assume that the population densities can depend not only on the characteristics (locations) of the individuals, but also on the velocities at which the characteristics change. Furthermore, we assume that velocity-dependent forces exist that influence the search process. Using this approach, it is possible to model specifically the role of meta-competence as flexibility in the adaptation of certain competencies.

Thanks to this new extension of the model, we obtain a far richer dynamics and have the possibility of using certain results from studies on Boltzmann strategies and on the theory of active Brownian motion. Thus, the central focus of this report is the implementation of this new concept in the context of a formal model and the analysis of approaches arising from this concept for the understanding of the evolution of competency and the role of flexibility.

Partly interactive simulations were developed for the different modelling approaches.

## 2 Introduction – Modelling of Competence as a Self-Organization Process

Learning has been modelled in different ways in studies on the theory of complexity and self-organization. (Kühn, Menzel et al. 2003) In this study, we describe learning in two different ways:

In the first model, we present individual development in the sense of the use and development of different competences, depending on the relevant learning situation. Building on an earlier model (Scharnhorst 1999), we imagine that individuals in different learning situations draw on their competences in different ways. A competence profile emerges which changes over time in the course of a learning process. In this type of modelling, abilities or properties relevant for competences form the axes of a characteristics space (analogous to the phenotypic characteristics space in biological evolution). These properties can be measured (KODE®) and have been labeled as "competencies" in a specific approach (Erpenbeck, Rosenstiel 2003). In this approach the notion "competence" is used in slightly different way to the understanding of somebody being characterized as competent. In the later case the label of competence is already connected to an act of valuation of the actions of a person or a group. In the Erpenbeck approach competencies are seen as prepositions or dispositions for certain forms of action. Therefore, we model them as characteristics of a person.

We observe agents (isolated individuals and groups of individuals) who develop in such a competence space, i.e. have a dynamics and interact with each other. The agents search for maxima in an evaluated competence landscape. In other words they search for competence profiles, which correspond to a high overall competence or to a certain norm and value formed in the group. Not the absolute increase of each of the competencies reaches the goal but a differentiated combination or a certain relation of strengths of the different competencies might improve the behaviour and the success of an individual or group. Finally, we describe using an evolutionary search algorithm the optimization of the use of certain characteristics.

In a second model, we model the problem-solving process of individuals and groups of individuals in a problem space. Here the different aspects of the problem are represented on the axes of the space and the difficulty in solving the problems is represented as the height of the landscape. In this model, competences are the system parameters that characterize the dynamics of the search process. An agent (or a group of agents) with good problem solving abilities finds the maximum very quickly. In this modelling process, the space is given by the class of the problems to be resolved. In the abstract sense, the searcher is an agent that simulates typical traits of the search for solutions making the best possible use of its competencies. Individuals (agents) seek solutions to problems individually or collectively and in the process apply different competences in the sense of different dynamics of the elementary processes of a complex search process.

The two types of modelling are based on a generalized modelling approach, a geometrically-oriented evolution theory (G_O_E_THE, (Scharnhorst 2001)). In this modelling approach, evolution is described as a search process in a highly complex valuation landscape. Values and norms also play an important role in self-organized learning processes. By referring back to models of search processes in valuation landscapes in this project we create an operationalization of value formation processes in the mathematical language of self-organization and evolutionary models.

Evolutionary models contain models of self-organization. We understand self-organization as the emergence of structure and order. Thus, evolution incorporates the process of the instabilization of an existing pattern or structure and of transition to a new structure. The modelling and simulation of such transition processes is the central focus of this study. In the picture of the evolutionary search in a landscape, this corresponds to the transition from one peak to another. The problem with such a transition lies in achieving an overall improvement while allowing instances of deterioration locally (you have to be in a position to go through the valley).

For the modelling of competencies, we will refer firstly to dominant competence profiles which are established in a group as the norm. These represent the existing structures. The finding of a new competence profile if the group is in a learning situation is the task to be fulfilled in a process of trial and error. The new competence profile is the new peak to be climbed. This is the first model view. Simulations investigate the processes that promote such a transition, the temporal course of the transition (i.e. gradual or erratic) and the role played by group interaction.

In the second model view, in which competences are understood by agents in the sense of dynamic characteristics and boundary conditions of a problem-solving search, the focus is on the resolution of problems. A new peak stands for a new solution to the problem. In this model, the competences are linked with dynamic elementary mechanisms of an evolutionary search, such as selection, mutation and imitation. Simulations are used to explain the interaction of evolutionary factors that facilitates the search for the solution or why the group is unable to find a solution.

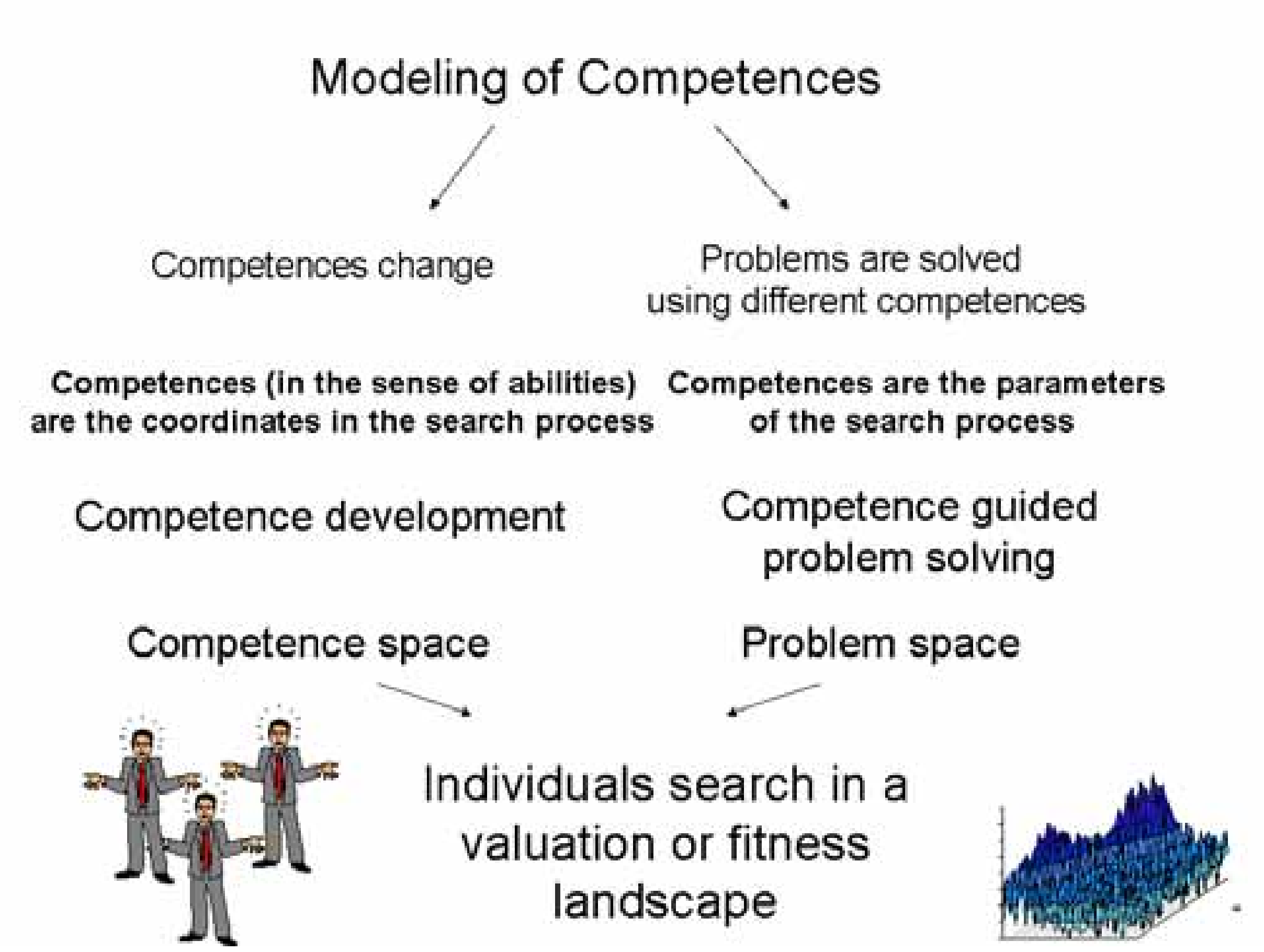

**Figure 1: A model approach and two different models for competence – an example of a generalization and two different re-specifications**

This dual interpretation of a general model approach is also reflected in the structure of this study. The chapters correspond to the different phases of abstract model building. From a formal point of view, both models are treated by the same mathematical tools however with a different interpretation. We start by introducing the picture of a geometrically-oriented theory of evolution and consider populations, i.e. groups of individuals in different landscape spaces. We examine the dynamic of the development of populations as units of evolution. The starting point is provided by a generalized population concept.[3]

[3] In the original population-dynamics approach, the "population" concept stands for a group of organisms that act as a unit in terms of evolution and ecology (Roughgarden, J. (1979). Theory of population genetics and evolutionary ecology. New York, Macmillan.

. In self-organization theories, this concept has been extended to groups of elements of a random nature. Thus, in addition to particles, molecules and organisms, individuals, organizations and institutions can function as elements. Their belonging to a group is based on specified individual characteristicss or characteristics, e.g. the reaction capacity of molecules in the case of chemical species, the political preferences of individuals in the case of voter groups and the use of certain technologies by companies in the case of technological populations.

In the other part of this study we develop a mathematical model for the dynamic of populations of search agents in the evolution phase space. This is followed by a more detailed analysis of this dynamic. The model's solution behaviour is examined systematically in simulations. We start by examining simple cases, combining the formal analysis of the evolution dynamics with interpretations based on competence development that are as concrete as possible. What is involved here – formally also – is a new field in which many questions remain open. Thus, it should be mentioned that the model formation, which was based on the requirements specified in the call for proposals this project replied to, ultimately led to models that have not yet been considered in this way in physics. This supports the idea that social phenomena also pose new challenges for model formation which can also result in a contribution to the available mathematical instruments in the context of a mutual interdisciplinary exchange of information. Thus, a mathematical-physical study that was carried out as a result of this project has been included in the Annex. In the final section of the study, the models developed by this study are analyzed critically and the prospects for possible additional applications are presented.

In the course of this report, we repeatedly refer to the two aforementioned interpretations of the evolution model: the modelling of the dynamics of competences (competence space) and the modelling of the effect of competencies on the dynamics of a problem-solving process (problem space).

Simulations played a major role in the conceptual development of the models. Five different types of simulation were developed in the course of the project. The programming of the interactive simulations was carried out by Thomas Hüsing who also made significant contributions to the design of the content of the simulations.

The simulations include:

1. Metaphorical simulation: visualization of the investigation of new areas of experience and an associated growth in competence (interactive)
2. SimKom: simulation of a group-dynamic interaction of individuals who use different competences. The group norm (order parameter) is constantly redefined as part of this process and the group development is measured on the basis of the distance travelled by the order parameter. (interactive)

---

The groups or populations are relatively autonomous interacting sub-systems which are subject to competition and selection processes in the system.

3. SimKom_Berg: simulation of a group-dynamic interaction of individuals who use different competences and simultaneously change the group norm (internal norm) and move in a simple landscape (external norm). The aim here is to find optimal competence profiles for learning situations. (interactive)
4. EvoKom: simulation of an evolutionary search process in a problem space in which the competencies stand for different elementary process of the search and a successful composition of competences for a learning process is tested. (interactive)
5. Brownian agents: simulation of a group of individuals as a Brownian motion of evolutionary search agents in the competence or problem space with a certain mean competence profile in a creative learning situation. Systematic investigations of the transition rates.

The simulations also ultimately contributed to the development of an understanding of how the role of competences must be formalized in a learning process and how it can be operationalized to be made accessible to mathematical modelling. The following questions arise in this context: Which use of terminology must be reinterpreted? Which effects do we expect from a simulation? How must an interactive simulation look to convey to the user both an understanding of our approach and a sense of joy in the game? Simulations mainly assumed the role of a “trading zone” (Galison 1997) in this project for the purpose of also stimulating the communication between experts from different disciplinary backgrounds in the project.

The different simulations are introduced in different parts in this report. The web-based simulations (1-4) can be played on the EVOLINO web site (go to www.virtualknowledgestudio.nl, members, Andrea Scharnhorst, Current activities and projects). A CD with all results can be requested from the authors.

## 3 The Formalization of Competences, Measurability and Temporal Development

The acquisition of competence is a key element in the ability to assert oneself in the complex and rapidly changing modern worlds of work. This applies to the entire area of professional activity. It is equally true for engineers and teachers, for clerical and manual workers. In a society characterized by the acceleration of progress and globalization, the ability to adapt rapidly to new conditions is one of the main requirements of employees. People who are incapable of adapting quickly are constantly at risk of losing their jobs if the company, faculty or job is restructured. Thus, flexibility is gaining in significance.

Competences have been discussed widely in the literature for many years. (Rychen and Salganik 2001; Rychen and Salganik 2003; Sydow, Duschek et al. 2003; Reglin and Hölbing 2004) Changes in the worlds of work provided the context for these debates. The development towards a knowledge economy and information society (Webster 1995) is generating different requirements in particular, for example greater flexibility and open learning situations.

The topic of “competence” is a key focal point in management theory (in the context of “core competences”) (Hamel and Prahalad 1994; Bond 2000), in evolutionary economics (the competence of a company or region) (Knudsen and Foss 1996), in the computer sciences (in connection with problem-solving algorithms) (Wielinga, Akkermans et al. 1998) and in learning theory or the psychology of learning (e.g. the link between competence and academic learning). (Bernal and Inesta 2001)

Some authors have hitherto rejected the concept of competence due its vague definition. They argue that the previous concept of qualifications or skills is continues to suffice. However, competences are increasingly referred to instead of skills and qualifications in various fields. This shift is not coincidental and it is not a matter of fashion. In the area of advanced vocational training, this change reflects the growing dynamic and increasing complexity of processes in the world of work. Competences are associated with flexibility, the ability to change and the need to engage in problem-solving processes under uncertainty. (Anonymous 2000)

The fundamental change that has taken place in the worlds of work in recent decades cannot be denied and will continue in the future. The political world tries to keep abreast of this change by establishing various promotion and research programmes, for example the German Federal Ministry of Education and Research’s “*Lernkultur Komptenzentwicklung*” (“Learning Culture and Competence Development”) programme http://www.bmbf.de/591_827.html. It is essential that theoretical research reflect on this altered reality if this research is to fulfil tasks in the context of political consultancy. The concept of competence, which remains vague and in need of further clarification, appears to correspond to this need more than other terms associated with concepts which reflect the working worlds of the past.

A special approach to competence was proposed in Germany in recent years. This approach links the development of competences with the abilities of individuals to learn in a self-organized way. (Erpenbeck and Heyse 1999) Our project is primarily based on this approach. Self-organized learning incorporates processes of problem-solving under uncertainty, in which personal and social competences stand on an equal footing with professional competence.

With an orientation based on processes of learning under uncertainty, it is also possible to seek relevant concepts, methods and models in other disciplines that deal with similar problems. Hitherto, the concept of self-organized learning was primarily associated with theories of self-organization in both the social and natural sciences. Empirically, studies carried out in company environments were to the forefront here. Academic research has hitherto been seldom regarded as an area for the study of self-organized learning although research *per se* is learning under uncertainty.

While conceptual, historical and empirical dimensions of the concept of competence have been explored in other projects in the basic research area of the "Learning Culture and Competence Development" programme (http://www.abwf.de/main/programm/frame_html ), this study concentrates on the question of the capacity of competences to be modelled and thus advances on issues of educational research in a very specific way. Approaches from the theory of self-organization and evolution, in particular in association with the work carried out by Erpenbeck, have been introduced into this area. (Erpenbeck and Weinberg 1992; Erpenbeck 1996; Erpenbeck and Heyse 1999) For this study on the question of the capacity of competence development to be modelled we decided to build on the concept developed by John Erpenbeck and others. According to this concept, competences are defined as "dispositions for self-organized action" (Erpenbeck and Rosenstiel 2003) (p. XI). A crucial difference between competences and qualifications lies in the fact that competences are not fact centred or result centred, but subject centered and action oriented. (Erpenbeck and Rosenstiel 2003) Competences become visible in actions, they describe preconditions for action.

A distinction is made between four main basic competences which behave in different ways in terms of the individual and his objective and social environment.

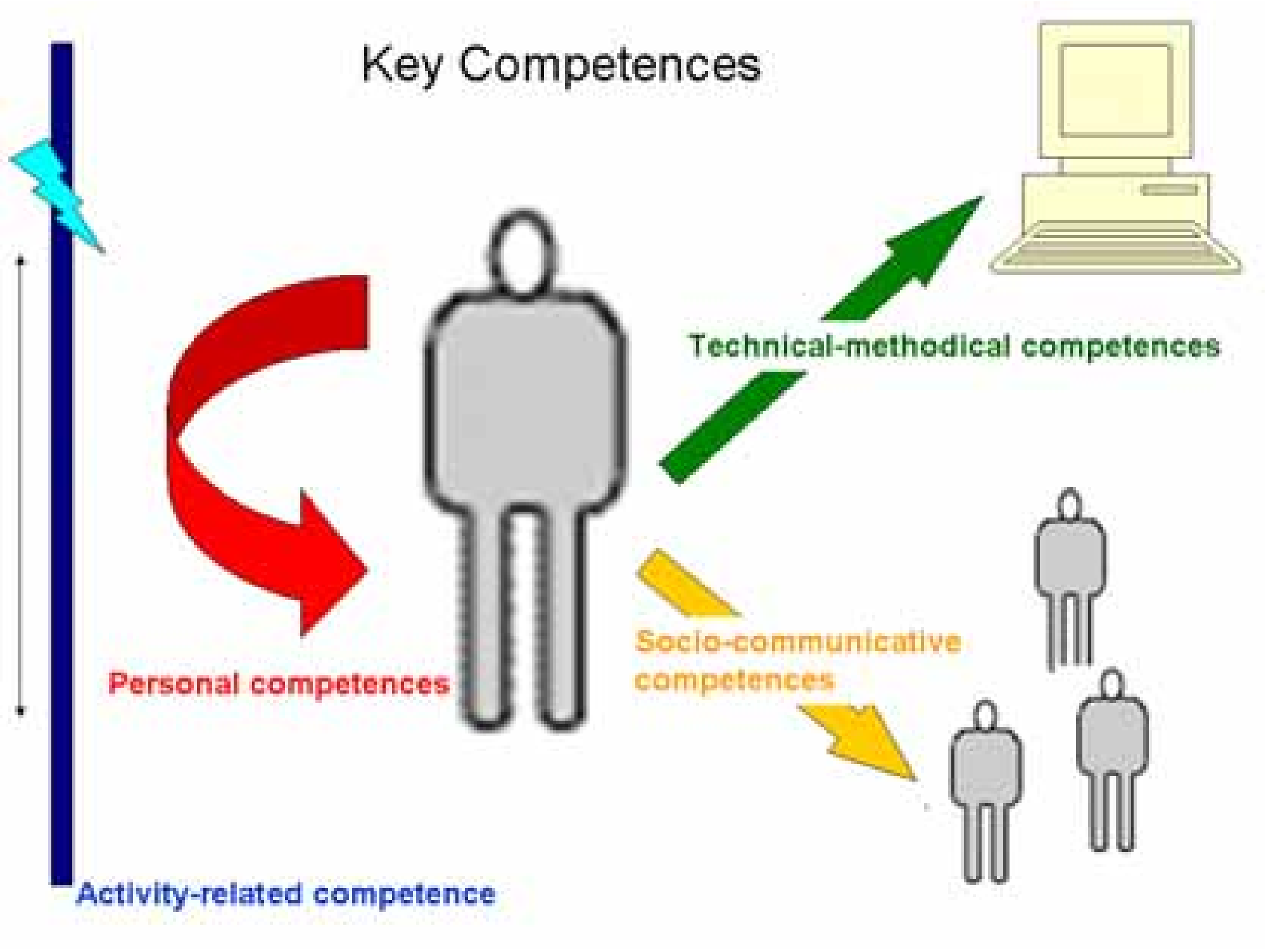

**Figure 2: The four basic types of competence (key competences)**

Within this classification, the four competences are defined as follows (Erpenbeck and Rosenstiel 2003)(p. XVI)):

“Personal competences (P): The dispositions of an individual to act in a reflexive self-organized way, i.e. to assess himself, to develop productive attitudes, values, motifs and self-images, to develop his own talents, motivation and performance suggestions and to develop and learn creatively, both within and outside the work context.”

“Technical-methodical competences [*or knowledge-related competences –AS/WE*] (K): The disposition of an individual to act in an intellectually and physically self-organized way when resolving factual-objective problems, i.e. to resolve problems creatively using technical and instrumental knowledge, skills and abilities and to classify and evaluate knowledge based on its sense; this includes dispositions for the structuring of activities, tasks and solutions in a methodically self-organized way and to further develop the methods creatively himself.”

“Socio-communicative competences (S): The dispositions to act in a communicatively and cooperatively self-organized way, i.e. to agree and disagree creatively with others, to behave in a group-oriented and relationship-oriented way and to develop new plans, tasks and objectives.”

While these three competences represent individual characteristics (or, more accurately, preconditions for action), the fourth activity-related competence describes in a general way how competences are mobilized for actual actions. Thus, this competence can be understood in the sense of an activity level (as with a volume adjuster) which, nonetheless, has its own independent character.

“Activity-related competence (A): The dispositions of a person to act in an active and generally self-organized way and to direct this action at the implementation of intentions, schemes and plans – either for himself or for other and with others, in a team, in a company in an organization. Thus, these dispositions include the ability to integrate the individual’s own emotions, motivations, skills and experiences and all other – personal, technical-methodical and socio-communicative – competences into the individual’s own volitional drives and to successfully carry out actions.”

Erpenbeck’s taxonomy represents an attempt to describe basic competencies. As is the case with all attempts to classify human behaviour, the boundaries between the different behavioural categories are flexible. In general, different aspects of different competences will always be found in a specific learning situation or action situation. However, the classification proposed by Erpenbeck fulfils an essential precondition for the operationalization of competences in the sense of model building. Based on the systematic introduced here, it is possible to define variables that are quantifiable and, as we shall see, can be formalized in different ways. If we initially work on this taxonomy, questions may arise in the course of the model building (including the creation of simulations) with respect to the regrouping of the competences (cf., for example, the metaphorical simulation, Chapter 5.3). The broad scope for the interpretation of the basic competences also accommodates different transmission options in the area of the mathematical modelling and evolutionary-theory interpretation of the basic competences. Beyond the two aforementioned basic modellings of competences, it is possible to associate the latter with evolution mechanisms in different ways. We shall return to his point later.

The measurability of competences has been demonstrated in various studies based, for example, on the KODE® questionnaire. The measurements have shown that competences are dynamic quantities. Important findings in this regard, to which we shall return in the discussion of model building, include the fact that:

- Competences can be defined for individuals and for groups.

- In groups, dominant behaviour patterns can be identified as dominant competence profiles, the outcome of negotiation processes between the group members.
- In groups, it is possible to observe individual variability beyond the group norm in the use of competences by group members.
- The use of competences is context-dependent, i.e. different competence ranges will be used according to the needs of different situations (different learning situations) (cf. Figure 3).
- Thus, competences can be evaluated, for example, on the basis of the group's success in a learning situation.
- Competence development is path-dependent, i.e. the current competence profile is the result of an irreversible development process.

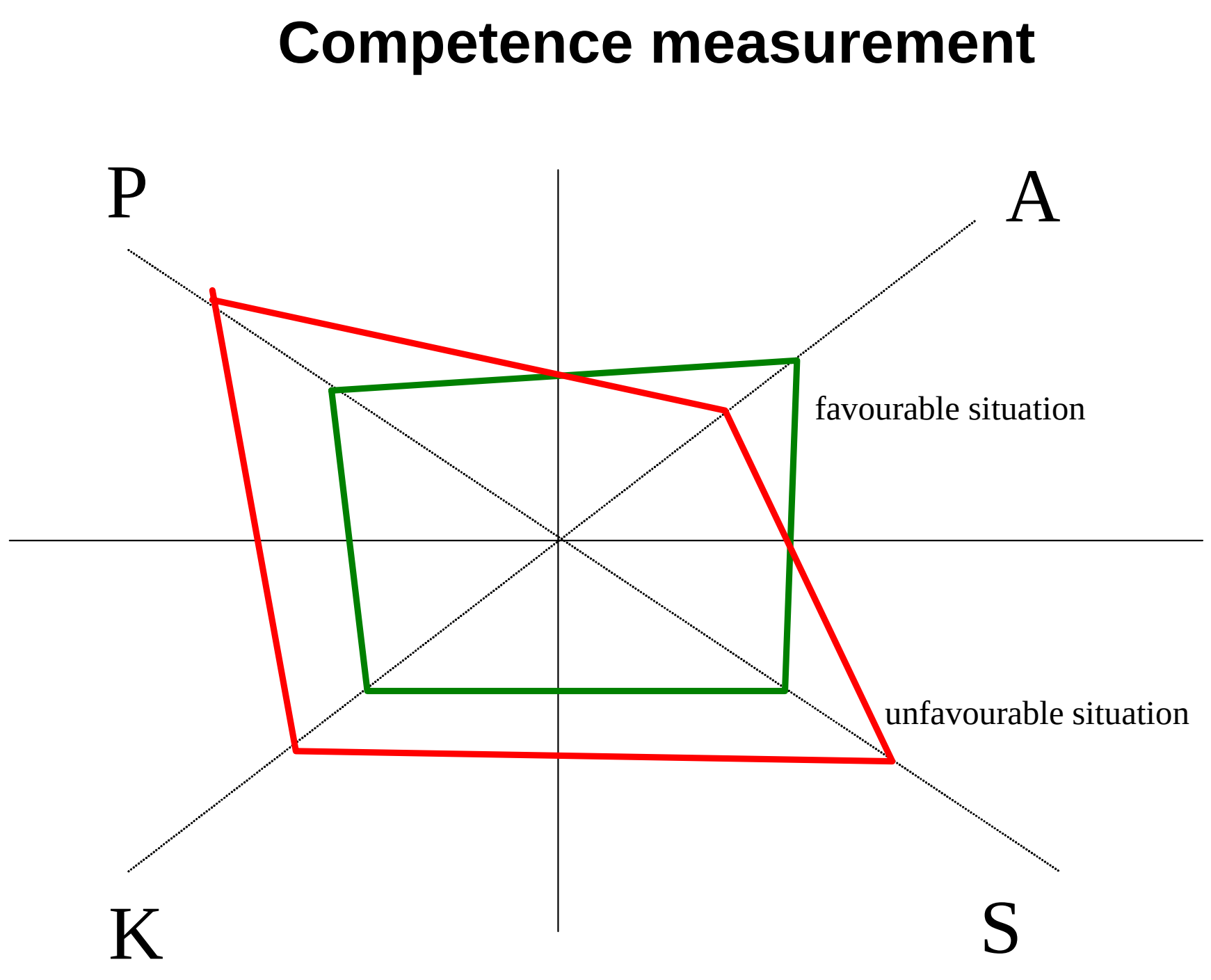

**Figure 3: The use of different competences in a group in different learning situations (with and without stress). Diagram adapted from Erpenbeck and Heyse 1999 pp. 84/85.**

Competences that are identified using questionnaires like the KODE® questionnaire can be scaled and, as shown in Figure three, and plotted in a spider map. This converts them into quantifiable entities which can be used in a mathematical model as coordinates, variables or parameters. Empirical studies, like the example shown here, significantly influenced our choice of evolutionary model which is based on a spatial representation of search processes. We shall present this model in the next chapter and discuss how competence profiles become competence spaces and landscapes.

# 4 G_O_E_THE: Introduction to Geometrically-Oriented Evolution Theories and Models

Mathematical models that build on the evolution concept have a permanent place in biology. They abstract in different ways from the concrete properties of the developing population. While population genetics describes the structure of the genotypic properties of the population and their alteration using frequency distributions, most of the ecological models relate to the size and distribution of interactive populations.

Given that evolutionary thinking has become established in almost all areas of the sciences, the importance of mathematical models that build on the concept of evolution is also growing. A specific discipline has emerged, for example, in economics, i.e. “evolutionary economics” which has its own journals, anthologies and institutions. (Jimenez-Montano and Ebeling 1980; Nelson and Winter 1982; Saviotti and Metcalfe 1991; Witt 1993; Saviotti 1996)

The methods based on the theory of self-organization or synergetics, which have developed extensively over the past 30 years and have been further advanced in recent years under the heading of complexity research, play an important role here. (Haken 1973; Haken 1975; Ebeling 1976; Haken 1977; Nicolis and Prigogine 1977; Ebeling and Feistel 1982; Haken 1983; Nicolis and Prigogine 1987; Maturana and Varela 1988; Ebeling, Engel et al. 1990; Ebeling, Engel et al. 1990; Ebeling, Engel et al. 1990; Ebeling and Feistel 1990; Haken and Haken-Krell 1992; Kauffman 1993; Ebeling and Feistel 1994; Haken 1995; Mainzer 1997) Self-organization and evolutionary models are also increasingly common in sociology. (Weidlich and Haag 1983; Prigogine and Sanglier 1985; Weidlich 1991; Troitzsch 1996; Troitzsch 1997; Weidlich 2000) It is also possible to observe that evolutionary models are increasingly prominent in the philosophy of science. (Bruckner, Ebeling et al. 1990; Gilbert 1997; Ahrweiler and Gilbert 1998; Leydesdorff 2001; Scharnhorst 2003)

A specific thrust in evolution theory involves the modelling of the evolution of populations through the change in density distributions in abstract high-dimensional spaces. The metaphor of “evolution as a high-climbing process in a fitness landscape” is based on a concept coined by Wright who considers evolution as “a mechanism by which the species may continually find its way from lower to higher peaks in such a field.". (Wright 1932, p. 358)

Wright’s general approach, which was further developed by Conrad and others, provides a firm foundation for the mathematical models to be developed here. (Conrad 1978; Ebeling and Feistel 1982; Feistel and Ebeling 1982; Conrad 1983; Feistel and Ebeling 1989)

For a long time, abstract high-dimensional spaces were only found in mathematics and theoretical physics. However, in recent times, such spaces are playing an increasingly important role in association with landscape models, in the visualization and analysis of dynamic processes in different scientific disciplines. The representation and analysis of dynamic systems in status spaces with landscapes has already made a significant contribution to the understanding of non-linear evolution phenomena. (Conrad 1978; Ebeling and Feistel 1990) Such ideas have also been recently playing an important role in knowledge transfer to other areas (cf., for example, Krugman 1996).

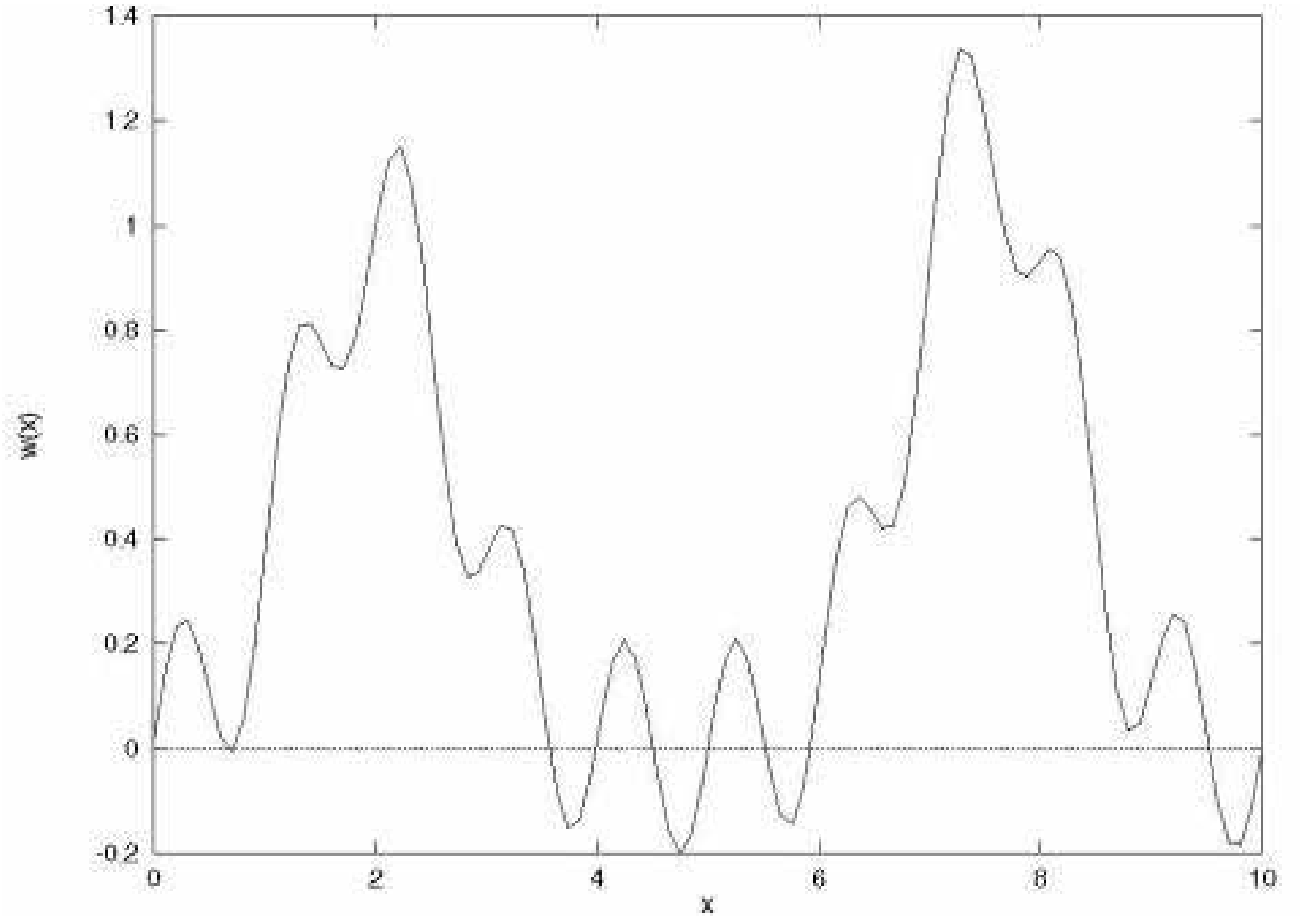

**Figure 4: Illustration of a one-dimensional section through a search landscape over the phase space of an evolving system**

By way of illustration, Figure 4 shows an example of a landscape (in one-dimensional section) with numerous maxima and minima. Metaphorically speaking, evolution would consist in the fact that the system approaches the main maximum (on the right) through the many lateral maxima. The dynamics of this search process must be envisaged as a stochastic process which, although it moves “upwards” on average, can move “downwards” in between (necessarily).

The following concepts are central to geometrically-oriented evolution theories:

- SPACE: in the sense of a phenotypic space or, more generally, a characteristics space, the characteristics describing the searching agents. The introduction of a space enables the definition of concepts such as proximity and distance and – something of particular significance for evolution theory – the consideration of evaluation and fitness landscapes. Furthermore, important concepts like evolution rate can be introduced into this picture.
- SEARCH AGENTS (GROUPS, POPULATIONS): search agents occupy specific locations in the characteristics space; they change (develop) in that they search for better characteristics. This characteristics change leads to the movement of the agents in the characteristics space. Given that our agents are subject to a constant search dynamics, they are not evenly distributed in the characteristics space. Structures become visible in the accumulation or clustering of agents in particular places in the characteristics space.
- OCCUPATION LANDSCAPE: groups or populations are formed by the clustering of searchers in the landscape as generated by the search dynamic. Overall, the frequency distribution of the agents in the space forms an occupation landscape.
- VALUATION LANDSCAPE: in the sense of a fitness landscape or value landscape, this may be predetermined externally or internally by the system dynamic itself. The valuation landscape forms a second landscape over the space or, in other words, the occupation landscape develops in parallel to the valuation landscape. It is also possible to talk about an exploration and occupation of the valuation landscape.
- SYSTEM DYNAMICS: the system dynamics characterizes the dynamics of the search in a valuation landscape. At its most simple, it involves a local gradient dynamics, i.e. locally only improvements are accepted. However, far more complex cases exist, in which the population distribution influences the form of the landscape in a non-linear way and hence also the search in the landscape.

- EVOLUTION MECHANISMS: mechanisms, according to which the expression of certain characteristics (or in other words their occupation) in one or more populations changes. Competition, selection and mutation are described as basic or elementary mechanisms. In terms of social systems, imitation is also involved as an important new mechanism of evolution. The evolution mechanisms can also be understood in the sense of the evolution strategies of the searching agents. These look for maxima in the valuation landscape. Thus, evolution can be understood as an optimization process.

### 4.1 Spaces and Occupation

In general, as an abstract mathematical space of state variables, the state space or phase space is extremely important in mathematics and physics. The state space contains all possible configurations of characteristics and their population; it corresponds in the generalized sense to the characteristics space.

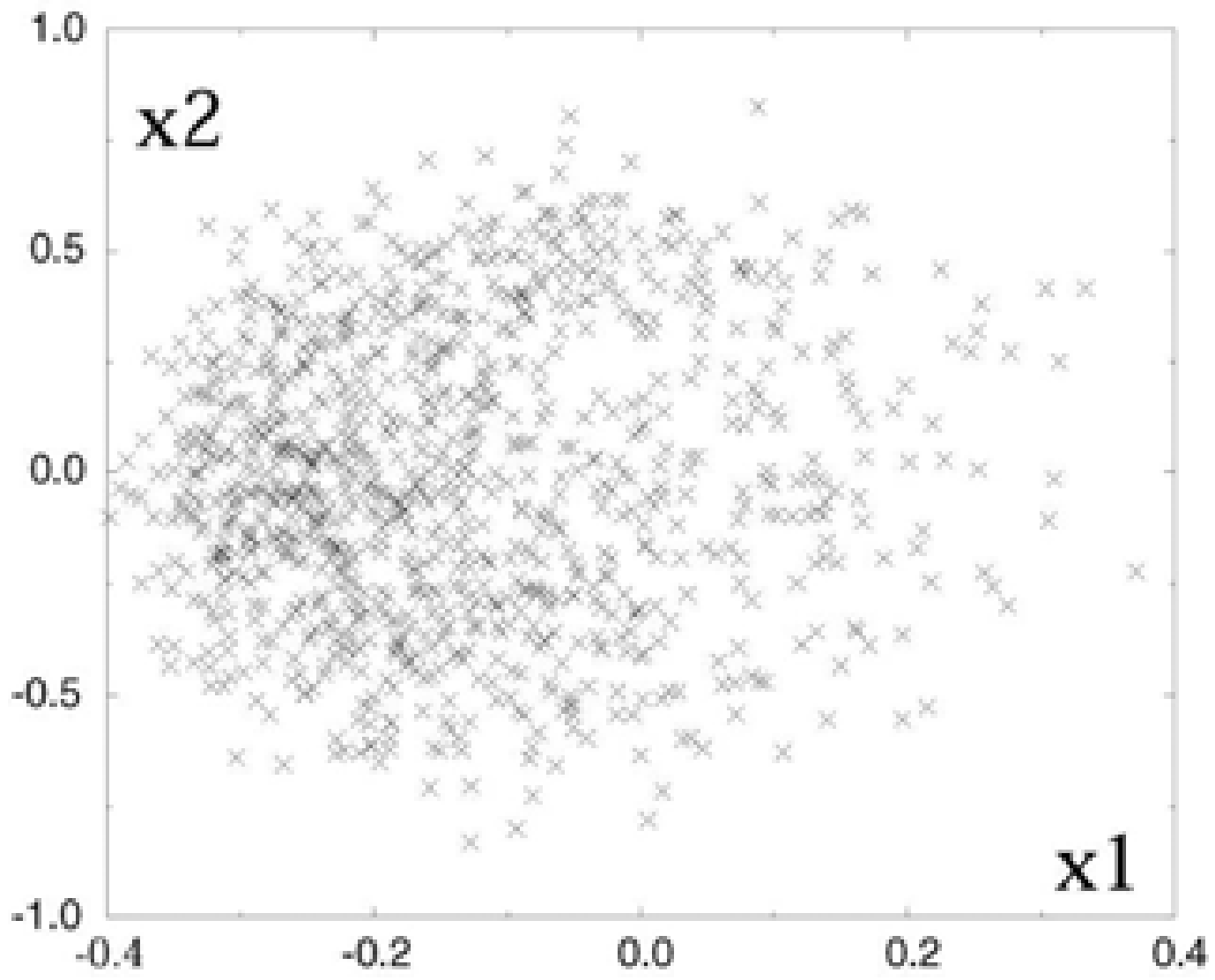

**Figure 5: A point cloud in the phase space which represents the state of the population at a specific point in time. Each point corresponds to the location of an individual (agent) in this search space. Accumulations of points represent groups which can be defined by a common or similar characteristic.**

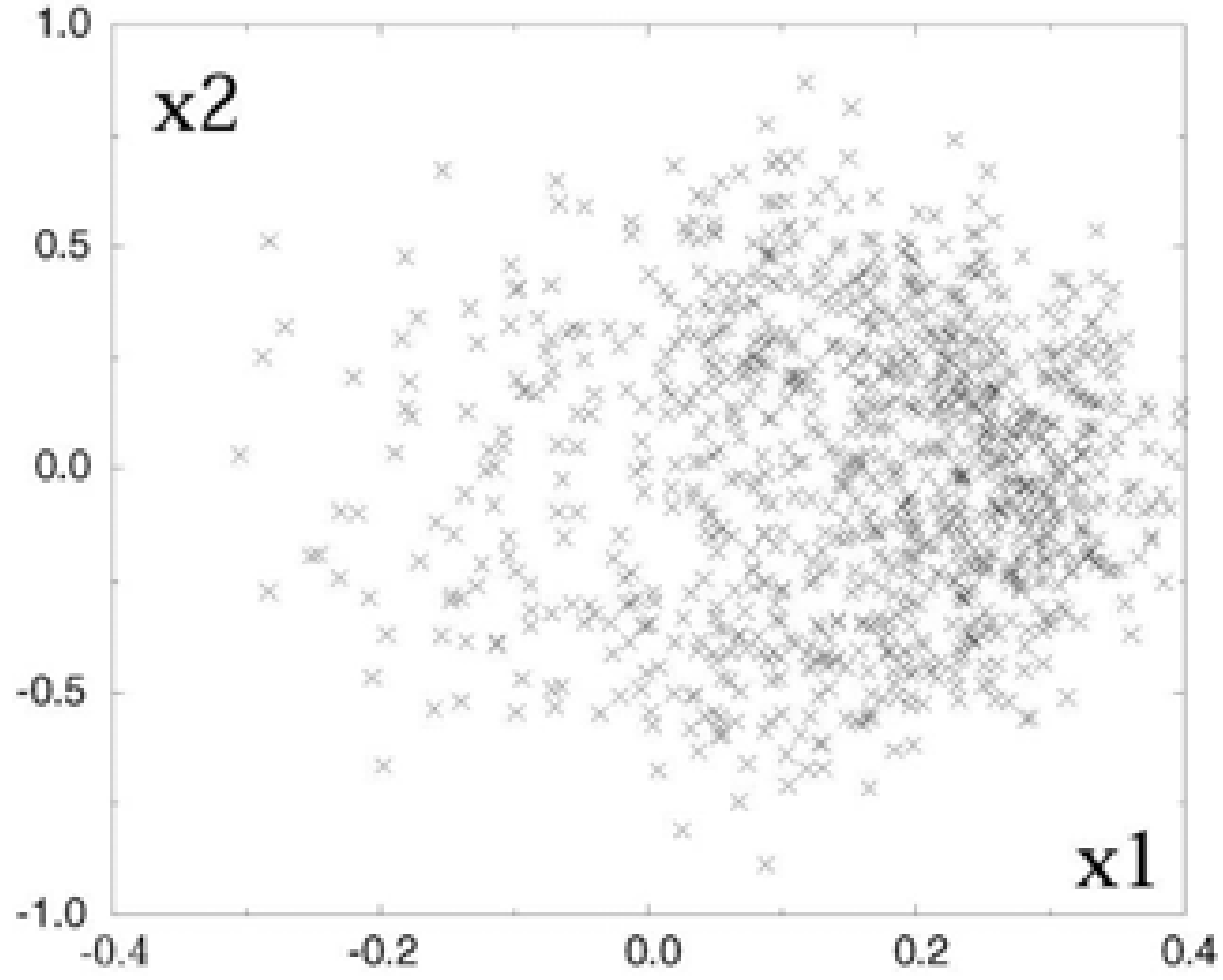

**Figure 6: A point cloud in the phase space which represents the population at a later point in time. The changed situation corresponds to an altered configuration of characteristics. The group has moved, changed and developed in terms of its characteristics.**

Depending on the concrete system being observed, it is possible from time to time to define functions over the state space, whose extremal properties provide information about the stationarity and stability of states and processes.

In classical physics, the phase space (coordinate, impulse) is the space in which potential energy and the Hamilton function for mechanical systems are defined as scalar functions over the phase space. This function's minima characterize special configurations that represent stable situations in the damped system. In thermodynamics, macroscopically-defined variables or state variables – such as pressure, volume and temperature – form the space in which thermodynamic potentials are defined as special state variables. The analysis of the extremal properties of these functions enables the identification of state changes (trajectories in phase space) and of the stability of equilibrium states. If thermodynamic equilibrium processes in closed systems are characterized by a maximization of the entropy, linear non-equilibrium processes can be characterized by a minimization of the entropy production (Ebeling and Feistel 1982; Ebeling and Feistel 1990; Ebeling and Sokolov 2005).

Another prominent example is provided by the so-called gradient systems examined in the context of catastrophe theory (Thom 1972). Here, the trajectories of the system dynamics follow the gradient line of a potential function and a topological analysis of the shape of the potential leads to stationary states and their stability. However, for most complex systems only local criteria, if any, exist for the stationarity of processes and the stability of stationary states. (Ebeling and Feistel 1982; Ebeling and Feistel 1990).

### 4.2 The landscape picture – valuation and strategies

In this study, we refer to landscapes in an *evolution theory* context. The transition from general dynamic systems to evolutionary processes takes place here through competition and selection. Selection always involves an evaluation and, hence, the existence of a standard of comparison. As a function of the state variables, such a selection criteria – even if it is only locally defined – can be visualized as a landscape over the state space.

In the *landscape picture*, metaphorically speaking, selection leads to a climbing towards the nearest peak. The transition from a selection process to an evolution process occurs through successions of such local evaluative elementary processes. The question that arises here, in particular, is how peaks once reached can be left, i.e. how stationary states become unstable again. Mutations provided the key to this process. Innovations are special mutations, i.e. those that lead to states that have not arisen previously and are of great consequence for the whole system dynamics. The testing of new states constitutes an essential element for the gaining of advantages from competition. The occupation of a local peak describes the emergence of a temporarily stable order state through self-organization. Thus, evolution can be understood as a sequence of self-organization steps.

The idea of a “fitness landscape” was first developed by Wright (1932) in the context of biological evolution. Wright describes the existence of different maxima of adaptivity in a high-dimensional space of possible gene combinations. According to Wright, the problem for evolution consists in how the species find the path from lower to higher mountain peaks and overcome the valleys in between. This generally concerns the problem as to how local optima can be found in a complex landscape and, once found, left again.

The idea of a scalar fitness function is, without doubt, too simple to depict the complex processes of biological evolution. However, landscape models make it possible to examine fundamental characteristics of the evolution process.

Studies can currently be found on “fitness landscapes” in the context of molecular evolution (Schuster and Stadler 1994), complexity research (Kauffman 1995) and evolutionary algorithms. (Asselmeyer, Ebeling et al. 1996; Rosé 1998) The concept of adaptive and fitness landscapes is being discussed increasingly with regard to the description and understanding of social evolution. (Allen 1995; Kauffman 1995; Westhoff, Yarbrough et al. 1996; Ebeling, Karmeshu et al. 1998)

The valuation landscape with local attractors functions as a kind of evolution principle which describes the qualitative features of the possible selection and evolution processes in the system. Hence, evolution becomes describable as the local optimization of certain parameters or functions.

Global evolution criteria only exist in exceptional dynamics cases. (Ebeling and Feistel 1982; Feistel and Ebeling 1982; Ebeling and Feistel 1990) The assumption of the existence of an valuation function does not mean that it is already known to the actors prior to the search process.

Analytical descriptions of valuation, target or fitness functions are generally only assignable for very simple problems. The solution of complex problems does not follow an algorithm, but must instead be understood as a (blind) groping search process. Nonetheless, we assume that once found, i.e. occupied, the practicability of problem definitions (or proposed solutions) can be defined, be it in comparison to the assumption of another problem definition. Without evaluation and comparison, a learning process is inconceivable.

The connection between evolution and optimization can already be visualized in simple evolution models. In the example of the Fisher dynamics (Ebeling and Feistel 1990), the reproduction rates decide the course of selection. In the classical Fisher model without mutations and under the boundary conditions of a constant population size, the species with the highest self-reproduction rate (fitness) wins, while all other competitors are selected out. If better new species are constantly introduced into the system, it can be shown that the average fitness of a population group will increase. A complete evolution principle applies for this case. (Ebeling and Feistel 1982)

The model description in this example contains all of the elements of evolutionary search (reproduction, competition, selection, mutation) without necessarily settling them in a search space. Populations function as units of the evolution. These are viewed in discrete evolution models and in other traditional population-dynamics models, for example in Lotka-Volterra systems (Peschel and Mende 1986) and in replicator systems (Hofbauer and Sigmund 1984), as classifiable and enumerable, i.e. typologically described in a certain sense.

The emergence of the new in the system is, therefore, associated with the emergence of a new type with new characteristics. In order to associate this kind of *discrete description* with a spatial concept and to present valuation or selection criteria of different types of populations as a landscape, it is first necessary to define a neighbourhood relation between species.

Another possibility for the description of evolution dynamics in landscapes is reached when we work from the assumption of an – abstract – characteristics space (similar to a phenotypic characteristics space) from the outset. In this space, individual characteristics structures are represented through places and populations through groups of populated locations. In accordance with the *landscape picture,* the dynamics of the occupation of the characteristics space follows an evaluation function which is then understood as fitness in a general sense. Evolution is described as the process of *mountain climbing* or *hill-climbing* in this fictitious landscape.

The approach of a random but correlated landscape the problem of the definition of a fitness function and the knowledge of its concrete form into account. As a result, it is possible to establish links with the physics of disordered systems[4] (Anderson 1983) and the description of the dynamics in fitness landscapes. The latter are examined both in the context of macro-molecular evolution (Fontana, Stadler et al. 1993) and the context of complex optimization problems (Kauffman 1993; Schwefel 1995; Rosé 1998). The concept of adaptive and fitness landscapes is increasingly also discussed in relation to the description and understanding of socio-technological evolution. Apart from its clarity, the advantage of this approach lies in the fact that processes of populations and their amalgamation or differentiation arise endogenously from the system dynamic and do not require any taxonomical intervention. Individual variability can be described explicitly. In this study, we will highlight the heuristic value of such models in terms of evolution and innovation processes as compared with discrete descriptions. In doing this, we concentrate on approaches for the modelling of competing populations in adaptive landscapes. (Ebeling, Engel et al. 1984; Feistel and Ebeling 1989) These are based on a continuous characteristics space and are referred to below as *continuous models*. Discrete and continuous models are equivalent in a certain way.[5] They enable different perspectives on the view of one and the same development process. The two model approaches, however, differ in their conceptual approaches and formal informational value. This plays an important role in concrete model descriptions and, in particular, in the use of social phenomena.

[4] With the link to the theory of disordered systems in solid state physics the circle closes, in a certain sense, between landscapes conceived on the basis of evolution theory and the “traditional” landscape concepts in physics mentioned at the beginning.

### *4.3 Populations and Qualitative System Behaviour – Number of Order Parameters and Structures*

As shown above, population is the central concept in the evolutionary context. In generalizations of the population concept, for example in synergetics and organization sociology (Hannan and Freeman 1989; Haken 1995) and in studies on the diffusion of technological innovations (Mahajan and Peterson 1985), models of population dynamics are mostly used (Lotka-Volterra model).

In terms of the theories and models of technological change, this led to the criticism of the excessive description of **inter**-technological interactions and the neglect of **intra**-technological processes. (Saviotti and Metcalfe 1991; Saviotti 1996) This criticism is made in the context of basic considerations of the relevance of the biological theory of evolution for economic, sociological and technological processes in particular. (Metcalfe 1989; Metcalfe and Gibbons 1989; Andersen 1996) The significance of *population thinking*, which views the typical characteristics as abstractions of a widely varying behaviour in reality and shifts the central focus to variability, is given particular emphasis. In contrast, *typological thinking* is associated with the focussing on a few characteristics in the differentiation of the populations. Geometrically-oriented evolution theories highlight this distinction in a certain way. They make it possible to follow both the behaviour of individual agents and the behaviour of groups of agents through the concept of frequency or density distribution. As a result, the possibilities for the qualitative discussion of system dynamics are extended.

In population models, the dimensionality of the system, i.e. the number of interacting groups or populations, is crucial to the possibilities open to the system. (Nicolis 1995) This is applies in particular to the so-called stationary stable states, i.e. the attractors approached by the

[5] Comparable, for example, to the equivalence between different formulations of quantum theory by

system. In the case of one group (one-dimensional systems), a status is specified as stationary if the group has reached a certain size and the group size does not change anymore. The system runs into an attractor (final state). The behavioural variety in the system already increases with the existence of a second group. From now on, attractors can be considered for the two groups. These compete with each other. There is a possibility that one of the groups will win, that both will co-exist or that the groups will alternate in their dominance (oscillation). Chaotic attractors may even arise in the case of three interacting groups. In this case, the system behaviour is no longer predictable, although the system dynamic is clearly determined. The possibilities for high-dimensional systems reproduce accordingly. (Feudel and Grebogi 1997)

Thom's catastrophe theory (Thom 1972) contained basic statements on a special class of systems in terms of the conditions (change of system parameters) under which and way in which order states occur. Qualitative statements can also be made for continuous models in the evolutionary search in the landscape picture. In this case, the dimensionality of the characteristics space constitutes the decisive criterion. The system's order parameters form the coordinates of the characteristics space.

Very generally, the local shape of the valuation or fitness function $w(q_1, q_2, ...q_n; a_1, ...a_m)$ can be described using an approximation by a polynomial $p(q_1,q_2,\ldots q_n;a_1,\ldots a_m)$. Here, $q_1,...q_n$ are the order parameters and $a_1,...,a_m$ the other (constant or slowly variable) system parameters which are also referred to as bifurcation parameters. In the simplest case, $p(q) = E(q) - \langle E \rangle$, i.e. the local deviation from the mean develops in a Taylor series after $q_1,...q_n$. If $p > 0$ applies for the fitness function, the corresponding population grows locally, i.e. the group becomes bigger and gains new members. If $p < 0$ applies to the fitness function, agents leave this area. In other words: two areas with different signs for p drift apart. In biology, this is described as "speciation", i.e. the formation of new species. In a social interpretation of the landscape picture, in which groups search for different norms and values, this process corresponds to the formation of a group with relatively uniform norm which break away qualitatively from other groups. This can extend as far as the formation of interest groups, clubs, associations, organizations etc. In our context, the only important issue is the qualitative distinguishability and independence of the new emerging group.

---

Heisenberg's matrix formulation, on the one hand, and Schrödinger's wave mechanics, on the other.

In the case of a one-dimensional characteristics space (one order parameter), the behaviour can be presented in graphic form. The nodes (groups) prefer to move in the direction of the increase in the valuation function. This leads, at the same time, to a local deformation of the landscape.

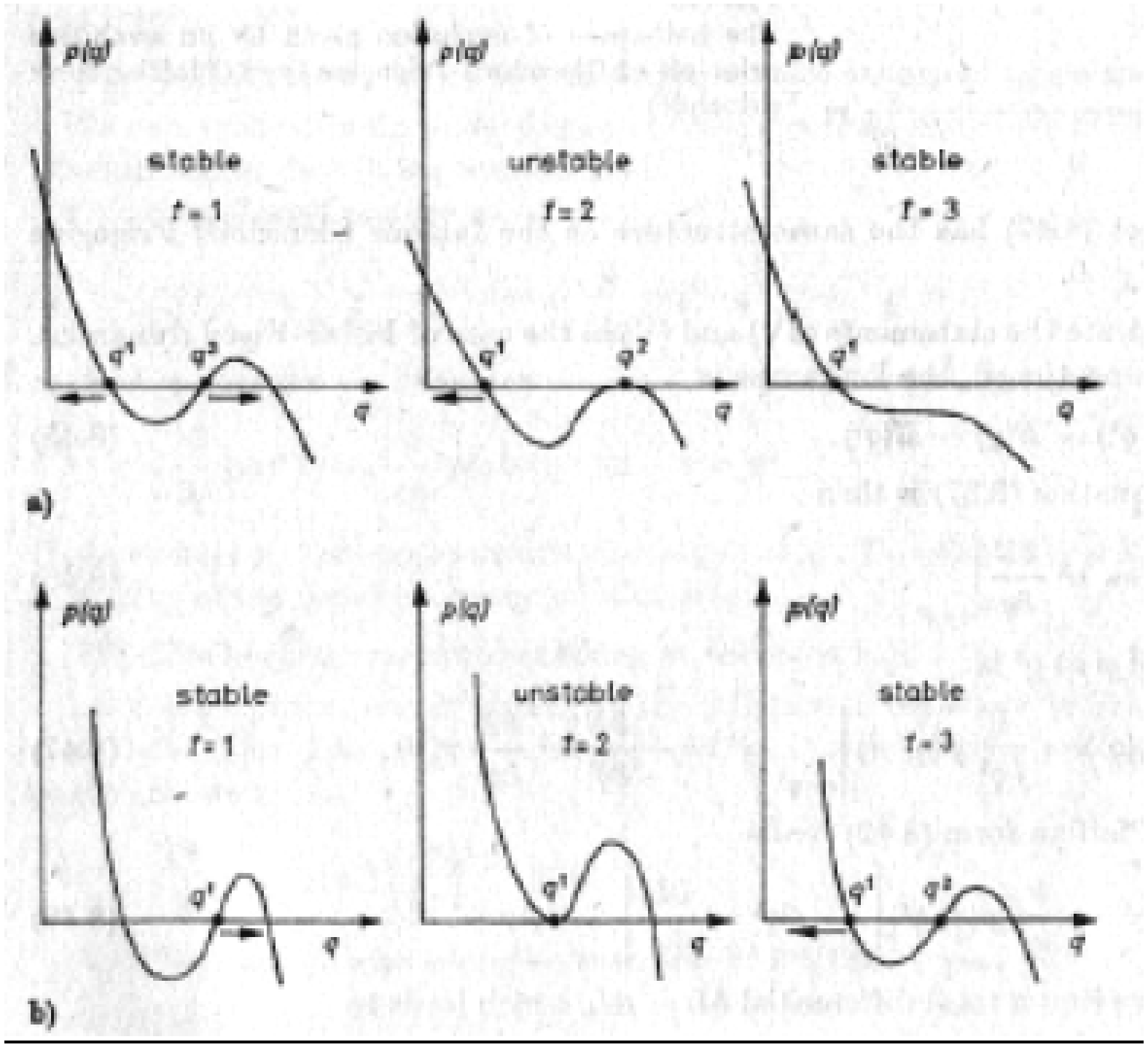

**Figure 7: Example of the disappearance (case a) or formation of an independent group (case b).**

Figure 7 shows two qualitatively different cases:

a) Due to a change in the landscape, one group disappears (the right hill $p > 0$).

b) A new group forms (the right hill $p > 0$).

These and other qualitative changes of a higher order can be classified in accordance with the catastrophe theory depending on the system parameters $a_1,...,a_m$.

Let us now consider the case whereby there are two different order parameters. In this case, completely new possibilities arise for the formation of separate groups. A particularly important variant was presented in Figure 8. A uniform area with positive fitness $p > 0$ is locally constricted by approaching a “saddle”. This may be clearly visualized as a mountain range with a saddle which disintegrates into two mountain ranges with rising water. Suddenly there is no path leading from one hill to the other. In the metaphorical sense, this means that two groups have suddenly formed out of a uniform group and they continue to differentiate and as a result move away from each other. A possible example of a social process would involve, for example, the introduction of a new distribution of tasks between two groups. From the modelling perspective, at least two order parameters are required for this for geometrical reasons. In other words, group differentiation that accompanies the distribution of tasks requires at least two relevant order parameters.

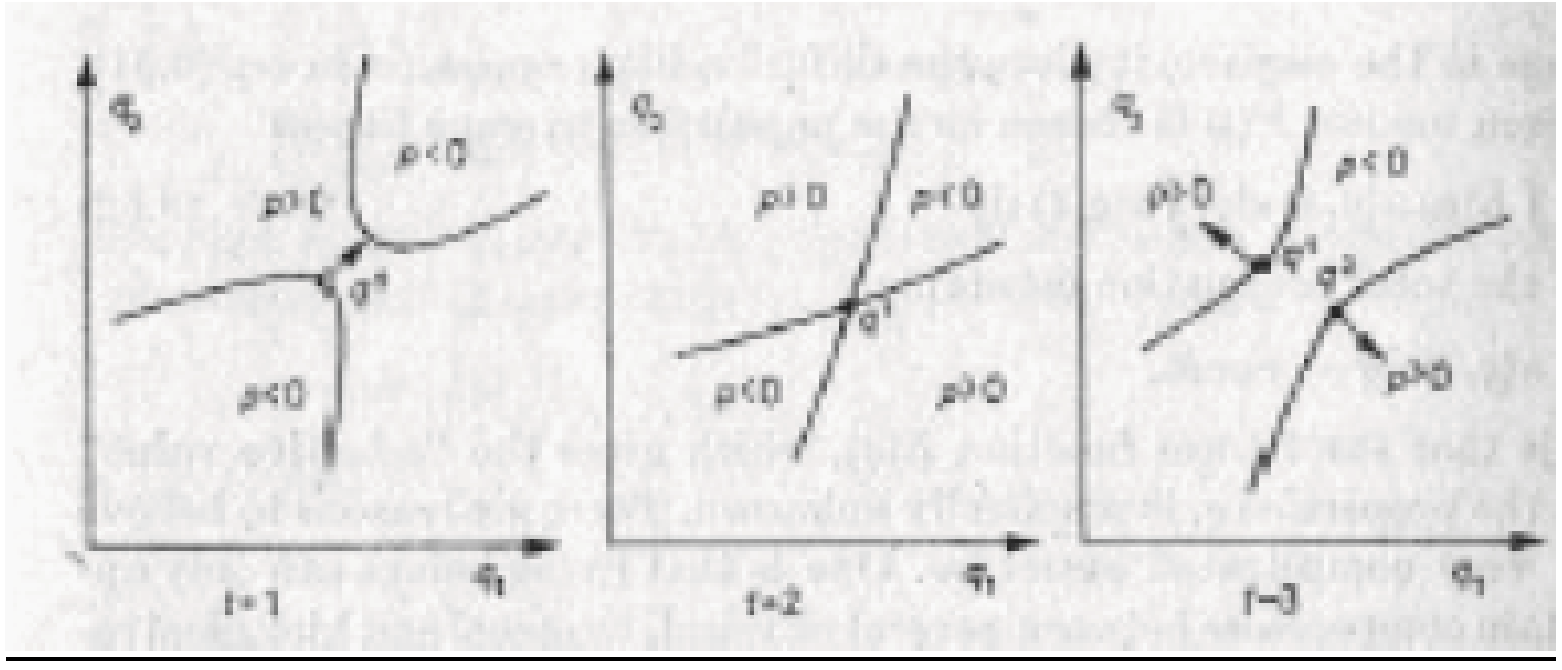

**Figure 8: Example of the simultaneous formation of social groups. One original uniform group splits into two separate groups through a simple shift of its social characteristics. Such processes can be observed, for example, in the context of the distribution of tasks between two groups.**

Thus, we can reach the basic conclusion that competence development is largely dependent on the number of order parameters involved in the process. This has already become part of everyday experience, for example people speak of getting around a problem. This requires at least two order parameters in the sense of G_O_E_THE.

# 5 The Modelling of Competences in Competence and Problem Spaces – Competences as Evolutionary Search Object and Instruments – Competence Development through Growth in Experience in a Metaphorical Simulation

In this section, we develop two different models based on the above-developed landscape picture of an evolutionary search. The four basic competences introduced above provide the starting point for a formalization and mathematical description.

In the first model, we assume that an individual changes his competence profile through the development of his "phenotypic" competence properties. For example, the ability to use computers, the acquisition of a foreign language, the completion of management and communication training can lead to a different combination of competences being drawn on in certain situations.

In the second model, we imagine that competence development accompanies the problem resolution process. Problem solving is understood as a process similar to hiking in a hilly landscape or, more accurately, mountain climbing.

In both models, the individual is, metaphorically speaking, a “mountain climber”. This mountain climber or a group of mountain climbers constantly tries to climb to the top of the next peak. Without stopping to enjoy the view for long, he aspires to new and greater heights. Thus, just as skiers compete in winter sports and the fastest usually wins the laurel wreath, the speed at which the peaks are reached will also be given specific consideration in a special variant of our model (presented in Chapter 7). What is of particular interest to us here is the process of the acquisition of competence in the social context.

The two interpretations share the idea of an evolutionary search in a complex landscape. The link between competence and space implies a visualization of the system dynamics. Different visualizations of competence development were created in the course of the project. Some resulted from simulations, with the help of which the solution behaviour of the different models was numerically investigated (*Evolino*, *Brownian Agents*), others represent visualizations of concepts such as the concept of space and motion in space (*metaphorical simulation*), others again were the result of an intensive discussion of the role of emergent structures arising through self-organization and their representation in a computer game (*SynKom*). An introduction to so-called *metaphorical simulation* is provided in the final section of this chapter (5.3). However, we shall first present the link between landscape models and competence development. In doing this, we use the two previously introduced model views: 1. Competence development as a search in the competence space; and 2. learning as a competence-controlled evolutionary search in the problem space.

### 5.1 Competences as Order Parameters – the Competence Development of Individuals and Groups – Competences as Objects of the Evolutionary Search

In an earlier model, one of the authors of this study (Scharnhorst 1999) discussed the fact that it is possible to observe competences or competence profiles in groups which represent different values and norms. The norm of referring to technical-methodical competences in a learning situation can be represented in a group by a specific number of people. It can compete with the norm of acting in a socio-communicative way in a learning situation or with a mixture of social competence and activity-related competence. In that earlier model we imagined that different competence profiles or competence types can initially coexist in a group. However, they then compete with each other and one competence type ultimately becomes dominant for the group. We described this dominant competence type as the system's order parameter (order status). It represents the norm to which the other group members must adapt.

In mathematical terms, a population dynamics approach was used in the earlier model. This corresponds (cf. Chapter 4.2) to a so-called discrete approach. In a group of individuals, different sub-groups develop which stand for different norms and values. Individuals can decide to adopt a norm and hence shift from one competence type to another. If the model also incorporates "unoccupied" or "empty" subgroups in the sense of possible competence types, the emergence of a new innovative competence type can also be modelled in the sense of a transition of individuals to this new subgroup.

Enumerable groups or subgroups are used in this discrete approach. Depending on their characterization, these could also be accommodated in one space (Figure 9).

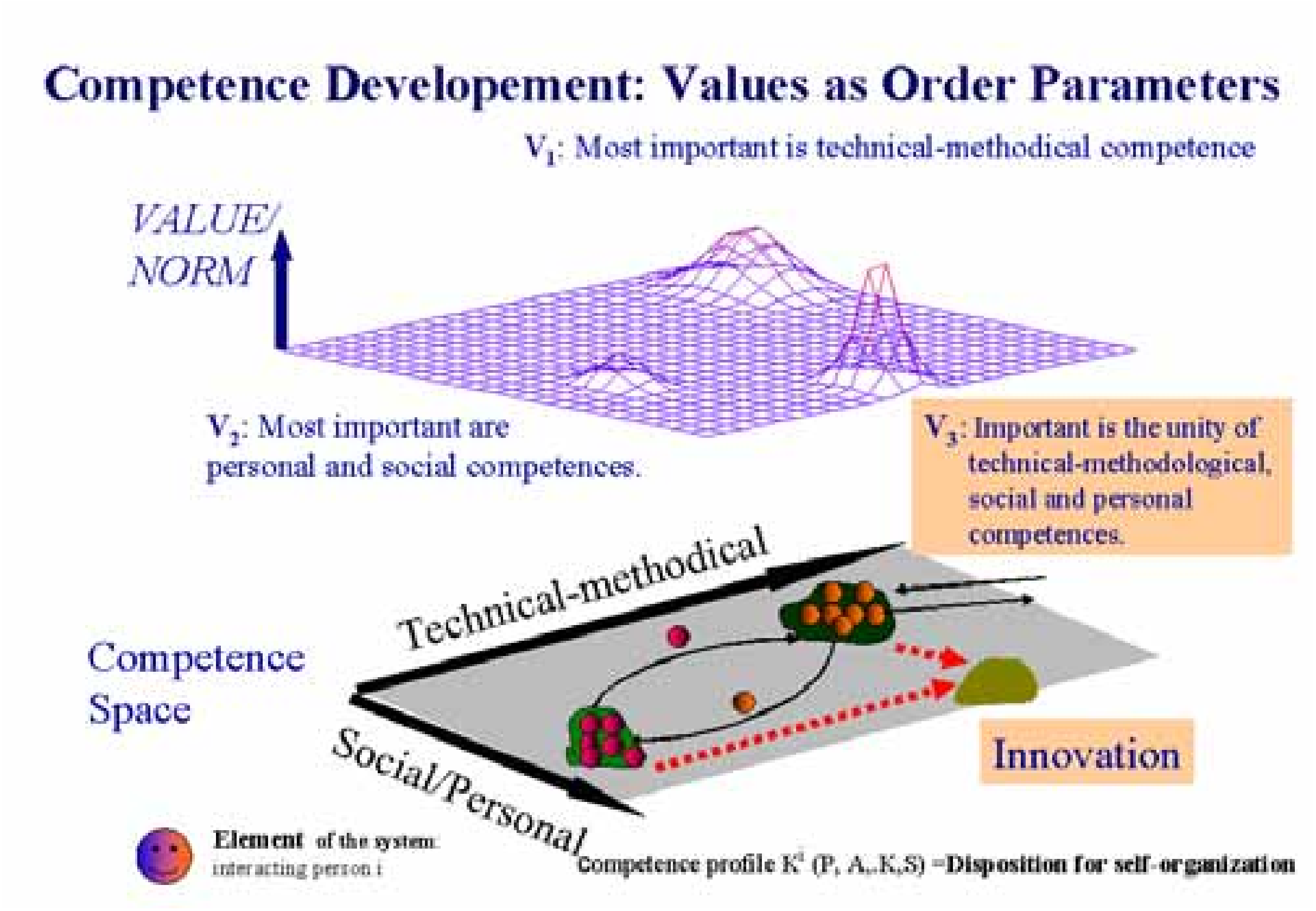

**Figure 9: Model for the interaction of individuals in a system with three possible competence types (Based on Erpenbeck, Scharnhorst: "Zur Modellierung von Kompetenzen" ("On the Modelling of Competences"), Lecture at the 12th Autumn Academy, Jena October 2004, cf. http://www.virtualknowledgestudio.nl/en/vks_members/homepage_andrea_scharnhorst/projects_presentations_reports/models_of_competencies/N%3A%5CSEC%5CNERDI%5CWebsite%5Cmodels+of+competencies.pdf )**

The step from this representation to a continuous model is just a small one. In our new extended model, each individual is described by a number of characteristics, i.e. competence characteristics. In this case, the phase space (or characteristics space) represents the group's range of competence, the axes of the space correspond to different competences which may be more or less pronounced. A point in this space represents a certain profile of competences. In the simplest case, what is involved is a four-dimensional space of the basic competences. However, it is also possible to introduce more finely-tuned classifications and to increase the number of characteristics correspondingly. An improvement of all different competences (moving on the axes to higher values) does not automatically correspond to an improvement of the competence profile (moving towards a higher value in the valuation landscape over the competence space). Otherwise, the goal of the search would be quite simple. What can be observed in groups is that an optimal competence profile for a certain task requires a specific combination of strengths in different competences. It is the relative relation between competences (or more specific between abilities to use competences) what matters.

The individuals in a group will generally have different competence profiles, i.e. the group members are located in different parts of the competence space. At the same time, it may be assumed that the competence profiles of the group members are not distributed randomly in the space of possible ranges of competences. Norms and values within a group ensure that group norms prevail at the individual level, also in relation to the use of certain competences. In our landscape picture, that means that certain group members will group around a certain competence profile. This group profile or the average value of the group corresponds to the competence type in the discrete model. It represents the order parameter. The population of the competence space by individuals and agents creates an occupation landscape.

The group tries to agree on a value (or norm) in a learning process and to form an “optimal” competence spectrum.

In accordance with the evolution landscape picture we introduced in the last chapter, we also assume here that an evaluation can be given to the different locations in the competence space. Depending on the relevant situation with respect to requirements for learning, one or other combination of basic competences will prove more favourable and useful. This evaluation is defined, in part, objectively outside the group, for example when the group has to meet external requirements; however, it can also take place internally through self-evaluation. In terms of the model, however, all that is important here is that such an evaluation function exists in a complex form.

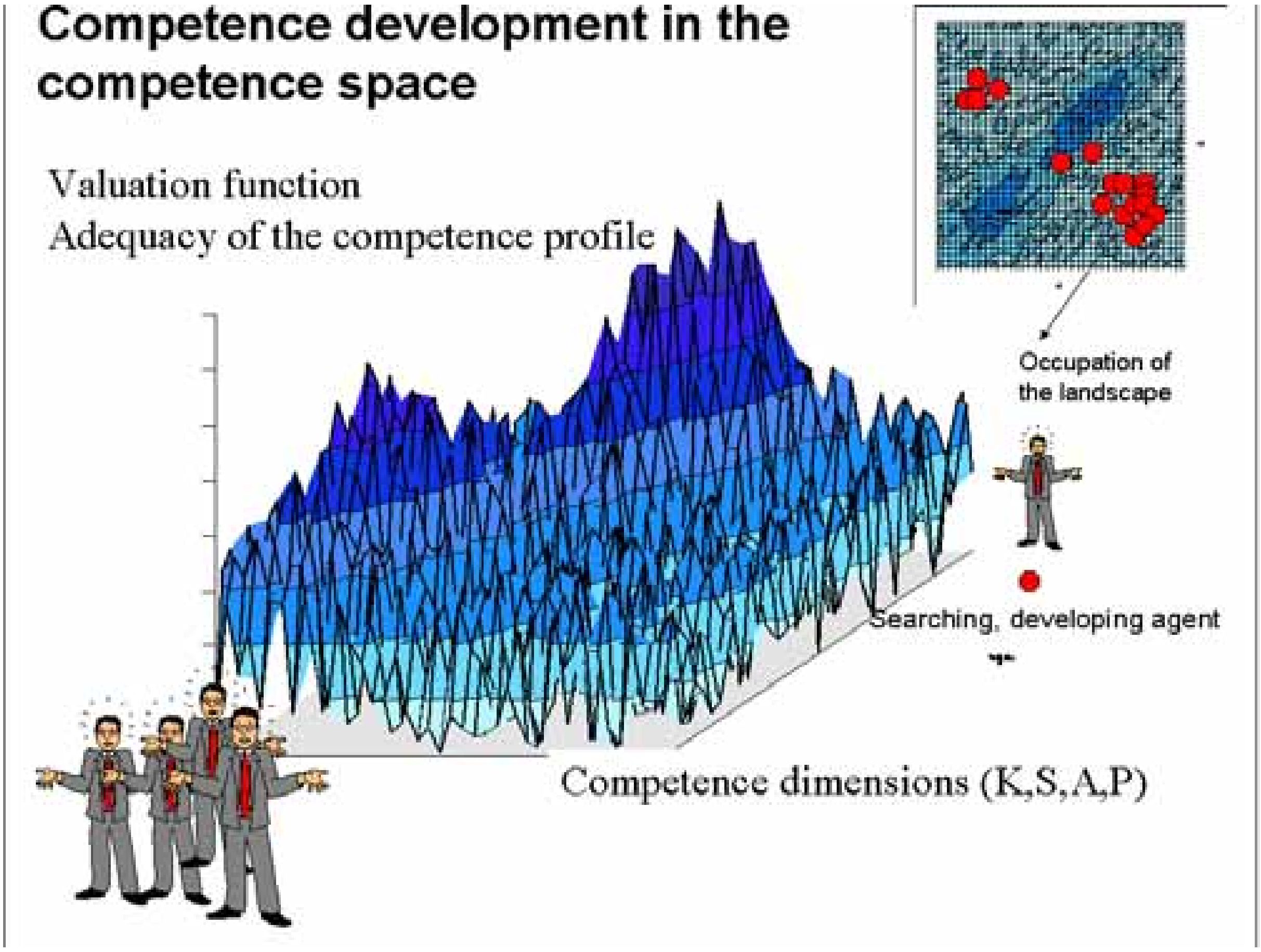

**Figure 10: Searching agents in a competence landscape**

In this model, the learning process targets a change in the applied competences, the individual learns to use his competences differently, both individually and as a member of the group. The elementary processes of the evolutionary search incorporate:

- the comparison of the competences between different group members in relation to their valuation (fitness) and individual decisions to adopt better or better-adapted competence profiles (selection);
- the (random) testing of different competence variants (mutation);
- and, where applicable, the comparison of the population of certain competence profiles and individual decisions to adopt the more frequently occurring profiles (imitation).

With this formalization, the competence development itself becomes the objective of a search process. The peaks of the valuation landscape correspond to the stationary states in the system that are stable and towards which our search agents are heading as objectives. They attract the searching individuals. The aim of the search process is to occupy as high a maximum as possible, i.e. to establish a competence profile as the norm in the group that best corresponds to a given learning situation.

In the course of time, the trajectory of a searching individual signals an individual competence development or management strategy. Elementary processes like selection and imitation stand for group-oriented strategies, while the mutation corresponds to a purely individually-based strategy.

The mathematical formulation of this model is presented in the next chapter. We will first turn our attention to an alternative formalization of competences in this kind of an evolutionary landscape picture.

### *5.2 Competences as evolutionary mechanisms in a problem-solving process – competences as instruments of the evolutionary search*

According to the literature on the role of competences, the latter have a role to play in particular when a learning situation is open. (Erpenbeck and Weinberg 1992) The theory of self-organized learning does not work on the assumption of a linear chain model of knowledge transfer from the teacher to the student, but focuses on the active role of the learning individual. Self-organized learning occurs when problems have to be solved under uncertainty. The uncertainty here can relate to both the objective of the learning process (learning with open objectives to be defined by the learner) and the process of learning (how do I achieve a certain aim?). In the case that the objective or goal of the learning is not defined and expectations are merely made of the learning process, the group members are forced to fall back on their entire range of competences. As a typical example of such a situation, Erpenbeck refers to an advertising company which is given the task of developing an advertising presence for a client. The precise form this presence should take is initially completely open; the only criterion defined for the search process being it should be convincing and attractive.

Such situations do not only arise in business contexts, they are generally characteristic of creative processes. Competence has a specific role to play in such situations. A range of competences are also required, i.e. social and personal competences are required along with technical competences. This specific aspect of self-organized learning has been intensively debated both in the context of problem-solving processes in companies and learning processes in the area of company training. As opposed to this, little attention has been paid hitherto to the role of competences in other creative work situations, e.g. in scientific research.

On the other hand, the examination of problem-solving processes is one focus in the history and philosophy of science and science studies. The representation of research fronts and knowledge domains as networks and maps, on which the location of research fronts, the increasing differentiation of research areas due to specialization and the increasing networking between the actors (scientists) is depicted, play an important role here. Today, the visualization of so-called *knowledge domains* (Chen 2003) constitutes an independent research area. Its findings are also adopted for strategic decisions about research investments (VxInsight, (Börner, Chen et al. 2002)) in the sense of data mining. What the visualization efforts have in common is that they make problem areas visible. The development of new technologies is also associated with learning processes. Technology development is exemplary of learning processes in complex systems. Another process is the introduction of new instruments into scientific practice, such as, for example, the development of scientific databases. In all of these cases, the development processes can be understood in the sense of problem solving. Such problem spaces can be interpreted in the sense of phenotypic characteristics spaces. (Bruckner, Ebeling et al. 1990; Scharnhorst 2001) The problem-solving process is thus interpreted as an evolutionary search process in the sense of our landscape picture.

This literature influences the formation of a second model approach whereby competences are not the object of the evolutionary search, but are translated into mechanisms thereof.

To our knowledge, the role of competences in the scientific problem-solving process has not previously been discussed in the literature. On the other hand, extensive literature exists on the conditions of the research process which uses other concepts and terms. In science studies, there has been intensive debate to the effect that a common basis, in the sense of an overlap of competences, a shared language or a shared understanding, must first be created to enable successful interdisciplinary work. The concepts of "*trading zones*"[6] (Galison 1997) and of "*validation boundaries*" (Fujigaki and Leydesdorff 2000; Fujigaki 2001) represent theoretical approaches which attempt to make the emergence of innovations on the boundaries of scientific disciplines and between science and the public understandable.

[6] Trading zones represent locations, in the metaphorical sense, in which different concepts and theories are negotiated. The term "validation boundaries" concerns the different area of validity of theories and concepts.

For our model, we assume firstly that the characteristics space is a problem space. Problems can be described using keywords and it is possible to imagine that these keywords form a high-dimensional characteristics space. The frequency with which the keywords arise in a problem description represents a kind of weighting or scaling of a keyword. A vector of word frequencies provides a possible way of characterizing a problem and representing it graphically in a location. (Chen 2003) The literary model of science is often used to represent problems in the literature. This means that what is depicted in such spaces are the products of scientific communication, such as scientific articles or patents. (Leydesdorff 1995) Thus, one point in such a characteristics space corresponds to a certain scientific article. For our model, it is essential that the actors or agents are themselves represented. Thus, we assume that a particular location in the problem space, i.e. a certain formulation of a problem, corresponds to the view of one or more individuals of the problem. It is possible that individuals can be allocated to problems or, more precisely, problem representations through the survey or analysis of documents about the authorship of oral and written statements. In such an individualized problem space, a change of location stands for a changed problem view on the part of the actors.

We also assume that an evaluation function can be defined through the problem space. This evaluation function expresses whether a certain problem can be resolved and the benefit associated with this solution. With this presentation, we assume that some problem solutions are better than others. The solution of a problem is thus equated with a certain formulation of the problem. We assume hereby that the change in the individual problem perception follows a problem-solving process. In other words, the search for optimum solutions is formulated as the search for the best possible problem definitions.

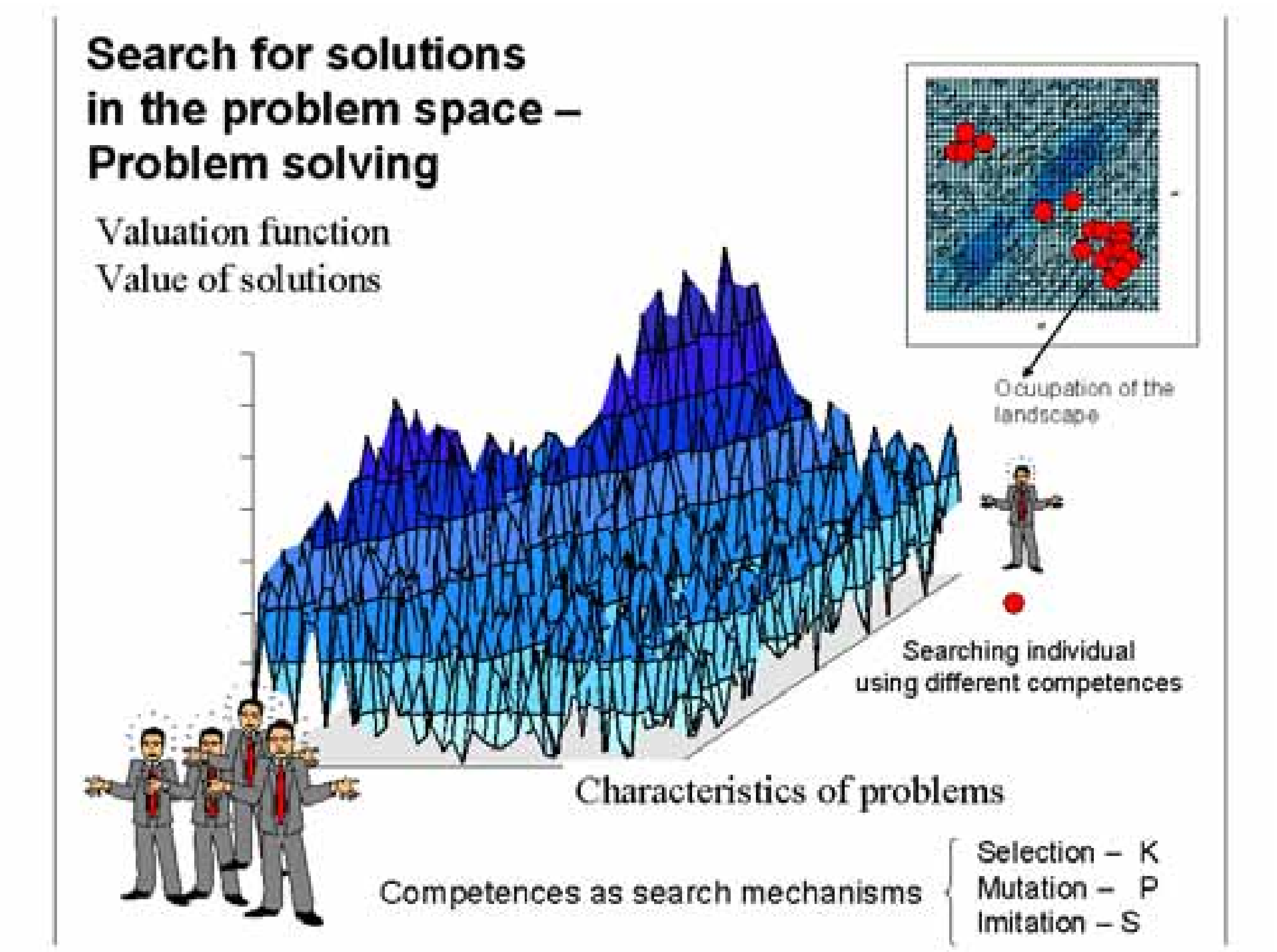

**Figure 11: Competences as mechanisms of an evolutionary problem-solving search**

The evolution of competence is described as an accompanying process of the collective search by interacting populations for local better solutions in a high-dimensional phenotypic phase space (problem space). Thus, competences are parameters that accompany a problem-solving process in the form of boundary conditions. The elementary processes of the evolutionary search, which have already been introduced (i.e. selection, mutation and imitation), are allocated to the application of specific competences.

The elementary processes of the evolutionary search contain:

- The comparison of the problem definitions of different group members in terms of their solvability (fitness) and individual decisions to adopt better solutions (selection). We assume that in such a comparison, recourse is mainly made to technical-methodical competences.
- The (random) testing of different problem definitions and solution variants (mutation). We assume that recourse is mainly made to personal competences here.

- And, where applicable, the comparison of the population of certain problem areas and individual decisions, the problem views that are predominantly represented in a group, i.e. adopting proposed solutions (imitation). We assume that recourse is mainly made to socio-communicative competences here.

The proposed attribution of basic competencies to evolution mechanisms (use of personal competence = mutation; use of technical-methodical competence = selection; and use of socio-communicative competence = imitation) can, without doubt, be explored further. Just as different competencies are ultimately combined in the actions of people, competences are not completely separable in the context of an evolution theory modelling. The selective comparison of different problem views can imply, for example, that there is communication about these different positions. Given that, in contrast with imitation, with this comparison the problem definition must be evaluated, we assume that technical-methodical competences are central to it. The situation with regard to an imitative comparison is different. Here, the frequency with which the two particular problem views are represented in the group are compared with each other. This presupposes above all competences for which the group is apprehended in its diversity, i.e. socio-communicative competences.

## 5.3 *Operationalization of Competence and Learning, Visualization in an Interactive Game – Metaphorical Simulation*

Both model approaches link competences with movements in an abstract space. In the course of the project we used simulations to solve certain mathematical systems of equations. Examples of these are provided in Chapter 7. These simulations generate visualizations, as shown, for example in Figure 5 and Figure 6. The execution of these simulations generally presupposes that the user has skills to run an executable program based on FORTRAN or another programming language.

One of the aims of our work on the project consisted in the creation of interactive simulations that can also be implemented by users with no previous knowledge of programming languages. The quest for the clearer visualization, interactive design and conceptual representation of concepts like space and movement in the context of competences led to a series of simulations which are linked with mathematical models to varying extents. These include a completely new type of simulation, which was initially proposed by Thomas Hüsing and further developed into its final form in the context of a discussion process. Because it primarily represents the concept and does not numerically solve equations from the mathematical model, we refer to this simulation as a *metaphorical simulation*. An alternative name for it is “competence development through increase in experience”. The approach used in this simulation is described below.

The purpose of this simulation is to represent certain concepts of individual competence development and their significance for learning processes in groups. In doing this, the metaphorical simulation takes up ideas of an evolution dynamic in landscapes and visualizes certain concepts without directly solving the mathematical equations numerically. Insights from competence research are visually implemented based on a series of simple rules combined with random elements. Interactive elements make it possible to test the consequences of certain conceptional ideas for the individual and group dynamic in the context of a game. Figure 12 shows the simulation’s starting picture.

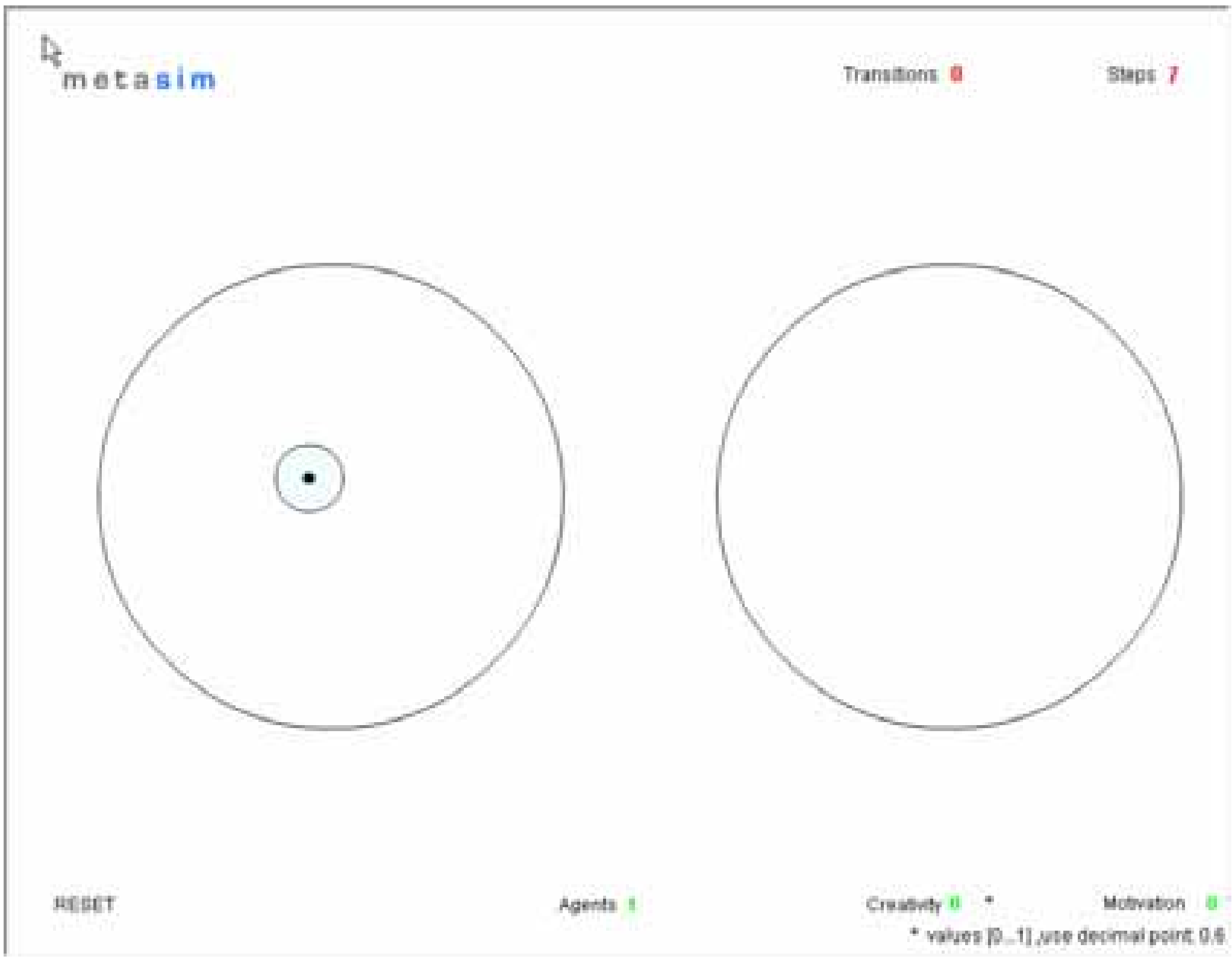

**Figure 12: Start picture of the “competence development” metaphorical simulation**

### 5.3.1 Short description of metaphorical simulation

Two circles can be seen on the screen. They each symbolize a certain area of experience. The actors or agents always start in the left experience area. Each individual is represented by a small circle with a black nucleus. Three parameters can be entered:

- the number of individuals
- creativity (the frequency with which they generate ideas (a number between zero and one))
- the motivational strength (a number between zero and one).

A "reset" button restarts the game with the agent's initial status and the two parameters at zero.

If an agent leaves the old experience area and goes over to the new one, this transition is counted in a window on the top left of the screen and the individual is marked with a small black line. Both experience areas can be shifted across the play ground. Thus, transitions can also be triggered manually.

Certain actions are accompanied by an acoustic signal. These tones are associated with the input of the number of agents, the emergence of an idea and a transition.

### 5.3.2 Concept of metaphorical simulation (with Thomas Hüsing)

The metaphorical simulation is based on the assumption of a group of people acting. The number of group members can be entered. It is also possible to experiment with a single person. The group of people move in a space and different areas of this space can symbolize different experiences, different problem solutions, different norms and values. Each individual starts with a certain set of personal, social, technical-methodical and activity-related competences and is represented in the space by a small circle with a central point.

People are not static, they reflect and act constantly. This is visualized as the movement of the points in the space. We assume here that actions are carried out on a goal-oriented basis. We assume, moreover, that these goals are defined internally and individually. We will later describe how such a goal can relate to an externally defined objective.

The people initially move in a certain experience area. This area represents the set of possible target coordinates, i.e. goals that can be considered and formulated by all of the individuals (within the corresponding reference framework). Thus, it corresponds to the specified experience space (socially accepted world view, reality island, common sense, meaning).

If one wanted to observe individual differences, each person would have to be given its own experience space (circle) in reference to common sense. In this simulation we consider persons as particles in a swarm (group) in a space, with are differentiated from each other in an only limited way. The process of emerging ideas is modelled as random search (mutation). This randomness represents one kind of individuality. We will discuss later that in the process of making a transition the person also acquires additional skills making him different from the rest of the group. The next diagram, Figure 13, shows the different possible goals the people may have. The grey circle stands for "common sense"; goals that represent undiscovered, unthought-of, visionary and "crazy" experiences lie outside the grey circle; the red, green and blue circles symbolize individual sets of possible goals and decisions.

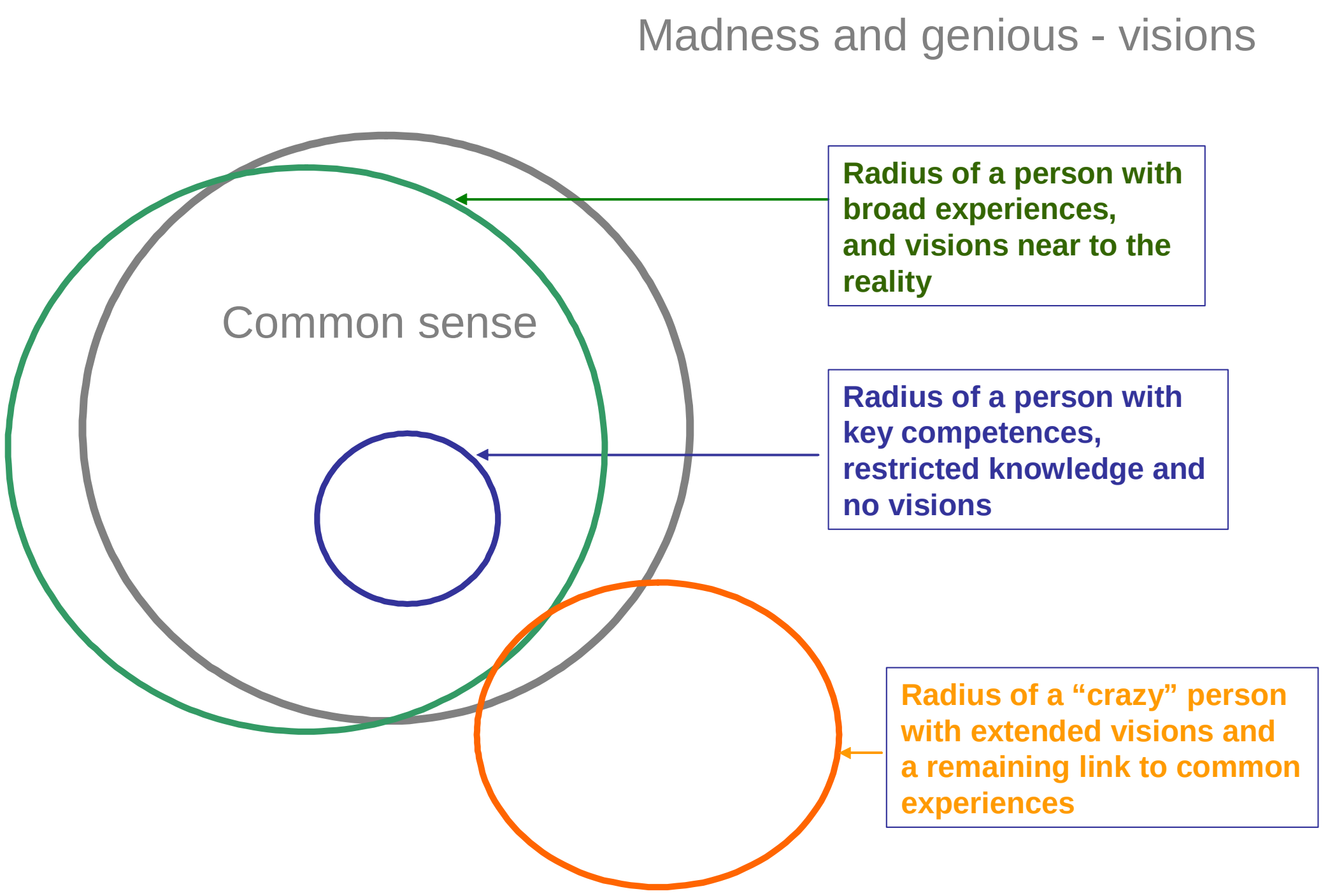

**Figure 13: Group experience area (group order parameter = common sense) and possible individual mobility spaces which represent different experiential scopes (Illustration proposed by Thomas Hüsing).**

We firstly assume that each person selects target coordinates which are controlled over a certain period of time "t". The time "t" is the activity time and is defined internally by the program. The selection of these coordinates should be carried out on the basis of a trial and error process and is technically implemented as a random process. (Figure 14)

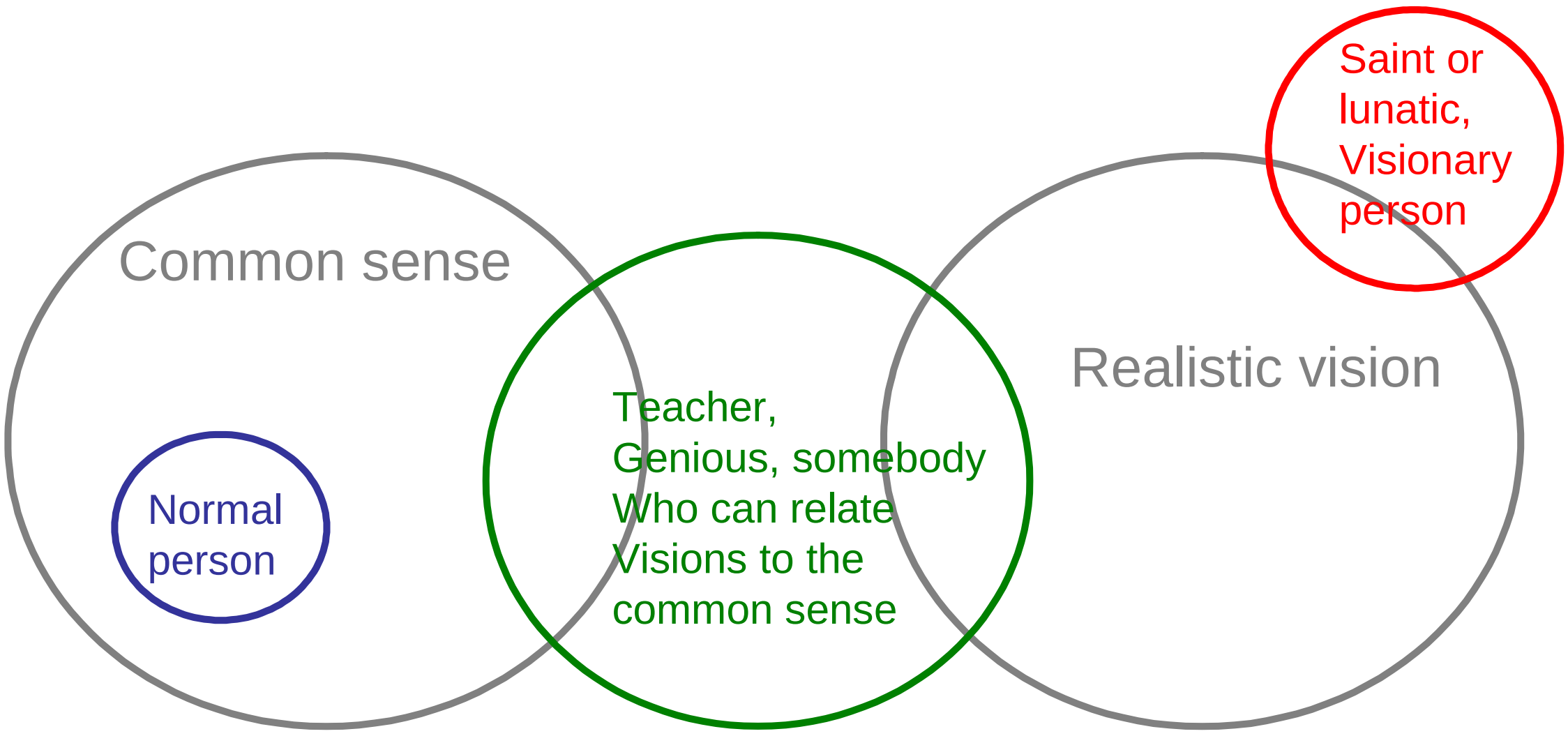

**Figure 14: Selection of goals in relation to the current treasury of experience (common sense) and a possible extension of the latter (meaningful visions = new area of experience)**

These target coordinates are related to a defined area within the space (common sense). We assume that this area has a centre that is surrounded by a catchment area (attractor basin). This centre and its catchment area stand for the people's current experiences, problem solutions or values. A second experience area (meaningful, realistic visions) can also be seen beside this in the space. This area is not populated at the beginning.

*Basic movement*

The people start their exploration processes in a certain area of the experience space. The new target coordinates lie by default in the catchment area of previous experiences. If the motion impulse is exhausted (t < a preset parameter), the distance to the two centres is checked. We assume here that each area acts as a gravitation centre that attracts the people into it. For this reason, the selection of goals is directed at the centre from which the people start at the beginning of the simulation. This is demonstrated in the simulation by a grey circle. This circle does not, however, represent any clear boundary. Instead, the people can also move beyond it, but they are still in the field of attraction of their previous experience area.

The movement to the self-selected goals is damped. In order to achieve an “organic” effect, friction is applied to the movement. Using this velocity-dependent representation we symbolize the fact that the setting of goals (definition of the target coordinates) cannot automatically be equated with their achievement. Instead, effort is required to achieve the goal. The closer the people come to the objective, the slower they move. The old goals are also constantly replace by new ones. After a certain time (initially set as a defined parameter[7]), the people re-orient themselves and aim towards a different objective. Goals are only selected in the values area of the current experience. However, with the help of a second process (see below), the aims can be shifted to an area outside the previous experiences.

In the picture of evolution theory, this motion corresponds to the spontaneous emergence of a mutation, through which the position of the person in the experience space is changed. The people generally carry out movements that lead them to the neighbourhood of their previous positions. This corresponds to the fact that mutations generally involve small changes.[8] In certain cases, the target coordinates selected may lie outside the previous area of experience. In this case, we may also refer to rare large mutations.

So far, the target coordinates consistently lie within the initial circle of experiences, problem solutions or values.

---

[7] This parameter (persistence or reaction time) incorporates the possibility of a further control parameter. Different levels of persistence can be selected for people. If the time is reduced, people quickly shift their attention between different goals. Such a person reacts more quickly and sets himself new objectives more quickly. As a certain time is required to achieve an objective, a shorter reaction time simultaneously leads to a reduction in persistence. This means that the objective may not be reached at all. A slower reaction time or greater persistence leads to greater proximity to the objective. As opposed to this, excessive persistence results in the person achieving the goal with increasing precision, but being seldom open to new goals. He is trapped, so to speak, in the approach to the goal. Obviously, the aim here is to achieve an optimum balance between flexibility (reaction capacity) and persistence.

[8] At present, the next environment is not being explored, but the action radius is extended according to the “motivation” parameter. The idea of a gradual extension of the action radius appears to make sense and be more subtle. In order to make this process into a real exploration process by the group in the next environment, one could, for example, model a dependence between presence in the external thought space (circle radius – y) and mutation – if the element has been in the external space x times, its individual value area of possible targets expands (circle radius + y), the individual feels its way layer by layer (onion model)

*Ideas and visions*

In our simulation two characteristics must coincide for a person to be able leave the previous area of experience.

We call the first of these characteristics strength of will, goal-orientedness, endurance, visionary strength or **motivation.**[9] We assume here that this characteristic extends the search radius of the previous experience circle. Technically, the motivation value specifies the extent to which the action radius penetrates into the area of the new experience area. With a motivation value of 0.5, the new goals will generally lie on the boundary between the two spheres of influence, at 0.8 they lie on a radius around the old centre which covers 80% of the distance between the two centres. Through this extension of the projection of the goals we make it possible for people to extend their exploration radius and achieve more remote goals, despite all resistance (visualized by the damped motion).

To ensure that the goal definition also lies within this extended search space, a second characteristic, **creativity**, is required which is interpreted as the occurrence of an impulse, an inspiration or idea.[10]

Firstly, the frequency of ideas is simultaneously regulated for all individuals – through the “creativity” parameter. Thus, creativity describes more a “learning effort”. It can also be said that “creativity” is the effort that must be made to realize an idea. The frequency of ideas can assume a value between 0 and 1. If it is equal to 0, goals outside of the current centre are never considered; if it is equal to 1 goals outside the current centre and within the extended search space are considered with every new motion impulse (examples are 0, 1, 0.1 or 0.5). This means that if new goals will be focused on in the “extended search area” at a frequency of 0.5, statistically, a new mode of action (in the sense of the testing of the extended search area) is adopted for every second attempt.

In terms of evolution theory, ideas are potentially long-trajectory mutations, how long-trajectory they are depends, in turn, on the strength of motivation. In the simulation, such an event (emergence of an idea) is indicated acoustically by a “crackling” signal. The search radius is not increased by the “strength” parameter until an idea is born; a target coordinate is then selected randomly within this extended search radius. Behind this lies the interpretation that the realization of a vision requires an idea **and** the ability to assert oneself. Creativity and motivation can vary independently of each other. If the motivation is zero, ideas can also arise and the search is limited to the original area.

[9] Other possible associations: strength, will, endurance, ability to assert oneself, budget, fuel, life-force, energy resources.

Through the selection of these two characteristics (processes) controlled by the parameters “creativity” and “motivation”, we want to demonstrate the connection between setting a goal, pursuing a goal and learning. Learning is understood here as reaching a new experience area. Setting a goal (having an idea or being creative) and pursuing a goal (strength, motivation) are components of the basic competences; goal-pursuit can be understood as an activity-related competence and goal-setting as personal competence. Without will and goal-orientedness, the achievement of adopted goals does not lead beyond the previous experience area. Without creativity, the maximum effort is of no avail; the person will not leave his experience area. Although the search radius may be theoretically larger in this case, the ideas, on the basis of which the search is actually undertaken, do not emerge.

If creativity is combined with motivation, the person may succeed in leaving the sphere of influence area of previous experience and reaching the sphere of influence a new experience and, therefore, learn. In the simulation, the person then moves over to the area of the other circle. The gain in experiences is visualized by means of a mark in the circle that symbolizes the person. Acoustically, this kind of transition is represented by a “piling”. All of the transitions between worlds of experience are counted on the top edge of the screen. In the simulation, the person can also leave the new circle again and return to the old one. After this, he can make for the new circle on the right again and this is interpreted as the gain of another new experience.

Such a transition can also be interpreted as a display of metacompetence, i.e. self-organized learning. The metacompetence results from the interaction of different personal competences (metacompetence = spontaneity (ideas) + resources (will, strength)). The person who has completed the learning process also changes in relation to his parameters; his creativity increases.

[10] Other possible associations: brainwave, inspiration, spontaneity, spontaneous mutation, potential, impulse.

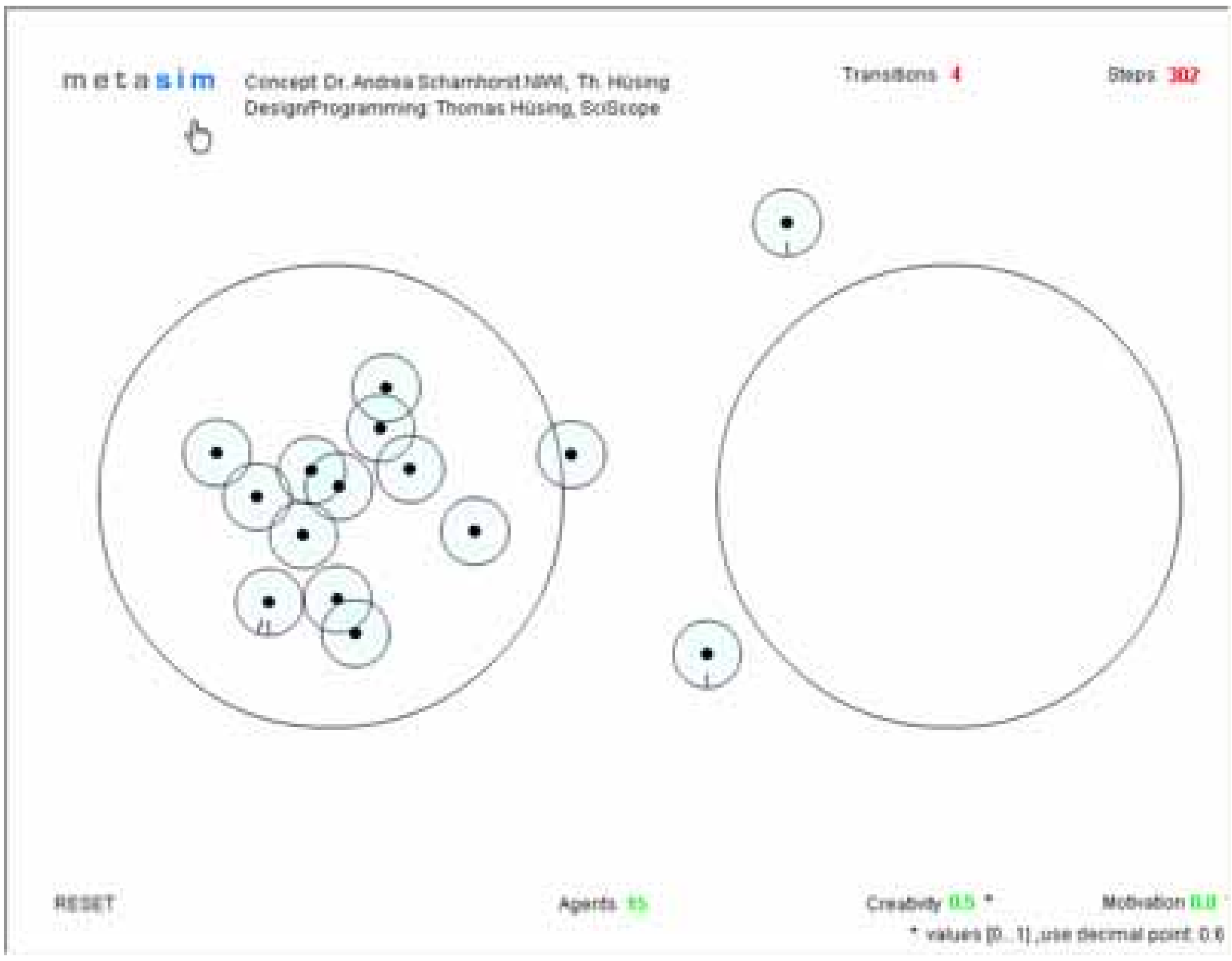

**Figure 15: Intermediate state of play with 15 agents, of whom two have just completed a transition and one has completed two transitions.**

The simulation enables one other interaction. The two centres can be moved with the mouse and the persons located within them follow the movement of their experience centre. To do this, the user clicks on the left or right circle and keeping the mouse button pressed drags the circle to the new position. Through the movement of a new experience centre, it is possible to experiment with the link between the distance between two different experience centres, the frequency of ideas of individuals and their strength of will. If the distance between the two centres increases, the motivation strength must also increase.

The movement of a centre can also be understood in the sense of the saying “if the prophet does not go to the mountain, the mountain must come to the prophet”. In other words, even if creativity and motivation are zero, particles can still be “caught” by allowing the new experience area to overlap with the old one.

Thus, the link between heteronomy and autonomy can be expressed in the goal-setting. The second experience area represents an “externally” defined goal that the people should attain through individual goals combined with basic competences. This external goal can be located at varying distances from the previous experience area and, subject to this distance, specific personality characteristics are required to gain a new experience. It is also imaginable, however, that the new experience area will move closer to the previous one through the processes of negotiation between the participants as the new learning goals will be related to what has already been learned. Hitherto, it was only possible to present this advance as an external process. However, it is possible to imagine that this distance itself diminishes in the course of the game through group processes (irrespective of how these are visualized).

#### 5.3.3 The limits and possible extension of metaphorical simulation

All of the people in the simulation are equipped with the same development possibilities. The parameters that are entered are group parameters, i.e. they define the extent to which certain characteristics are realized on average in the group. The random components in the selection of the individuals lead to an individualization of individual profiles, but these are still subject to the group parameters which act as boundary conditions for each individual. The people are not "individualized" until the transition to a new area. Another approach would involve giving each individual his own set of parameters from the outset. This problem will arise again in connection with the other simulations. We simulate group processes and not individual processes. The apparent advantages of an individual modelling leads to a strong increase in the dimensionality of the parameter spaces. The more parameters involved, the more difficult it is to obtain general statements for system processes. In the case of metaphorical simulation, the selection of the parameters serves in the clarification of a certain concept. We show, for example, the role that chance plays in exploration processes. The game is too simple, however, to allow for surprising effects. That changes with the other simulations which incorporate a broader spectrum of possible behaviour. The emergent idea here is to describe unpredictable characteristics (e.g. a transition and the way it takes place), for which only the boundary conditions are defined, through collective behaviour.

Another limitation of metaphorical simulation lies in the fact that individuals act independently of each other, their association is only directly defined by the fact that they all share one and the same experience area. The brainstorming (goal-setting) and striving towards the goals takes place individually. It would be expected of a group of people that they interact with each other. Thus, it could be imagined that the group members would attract each other and a social competence would be represented as a result. If such an attraction also leads to the adjustment of the individual goals, the people then search in similar directions. This kind of coherence can be helpful if the “right” direction leading to the new experience area is taken. However, it can also be obstructive if the direction initially taken is not very promising. The question that arises here is how many “mavericks” who resist the peer pressure does a group need to keep all search directions open. The group members, who have reached the new experience area, should have a signal effect on the others. Their “weight” for the group can differ here (perhaps personal competences). This interaction will be the focus in the development of the other simulations (*Evolino, SynKom, EvoKom, Brownian Agents*) which are not metaphorical, but based on concrete mathematical models.

Different competences have been described hitherto in the form of creativity and motivation. It remains open how these competences can affect the group. If the “mavericks” take other group members with them, they change the centre of the group. This is taken into account in *SynKom*, another model-based simulation (cf. Chapter 6.3.2), in that the group centre is formed inherently and also changes through the actions of the group members.

Metaphorical simulation primarily involves a visualization of certain ideas and concepts, such as space and mobility. It has little to do with the mathematical description of the evolutionary search.

In the next chapter we return to the general model framework of an evolutionary search in a valuation landscape over a problem space introduced in Chapter 4 and develop the mathematical models that describe this search and simulations based on these models.

# 6 The Mathematical Modelling of the Evolutionary Search of Groups of Individuals in Characteristics Spaces

## *6.1 Description of Groups or Populations in the Characteristics Space*

The classical population-dynamics approach processes the evolutionary units as distinguishable from each other and hence countable and classifiable. Each unit (species) is assigned a number, i.e. an integer $i = 1,2,3...$ . The population or, generally in the case of social systems, the groups thus form a countable set. Each group “i” characterizes a variable, i.e. a quantifiable, time-dependent value, a real number $x_i(t)$ representing the size of the group.

In the case of evolutionary ecology, these are the densities or number of individuals of competing species (e.g. in predator-prey systems), in the theory of molecular evolution, they are chemical concentrations of different macro-molecular species and in the case of innovation diffusion, the number (or fraction) of users of a technology. In the case of competence modelling, they are groups of individuals, e.g. working groups in a company, or groups with similar functions, such as the group of work council members from different companies. The competence within a group is more or less uniform, the spatial location in the same place is not crucial, but the relative equal competence of the group members.

Synergetics calls these quantitative variables "order parameters" which represent the populations at the macro level. The temporal dependence of the order parameters is represented by a mathematical function, generally defined by standard differential equations. As a result, in a certain sense, there is abstraction from the characteristics structure of the individuals in the population and their change and the competition between different populations becomes the central concern. In the context of mathematical description, this approach leads to systems of non-linear standard differential equations.

As explained in the previous chapters, in this study we use a characteristics-oriented continuous description through real numbers, landscapes and continuous dynamic models (partial differential equations) as an alternative to the discrete description using indices i = 1,2,…, in accordance with former approaches (Ebeling, Engel et al. 1990; Ebeling, Karmeshu et al. 1998).

As in our earlier approach (Ebeling, Engel et al. 1990; Ebeling, Karmeshu et al. 1998), we firstly use a description of the population through characteristics which are characterized by a set of *d* real-value variables $\vec{q} = \{q_1, q_2, ..., q_d\}$. In this way, an abstract characteristics space $\boldsymbol{Q}$ is defined. The main characteristics of the model are summarized in brief below:

The individual characteristics $q_i$ constitute the coordinates of the characteristics space $\boldsymbol{Q}$ which has the dimension *d*. As a rule, there are many characteristics, i.e. *d* is a large number. In typical cases *d* might be of the order of 100 or 1000. The coordinates $q_1,...,q_d$ characterize the expression of the different characteristics. Thus, a point in the space $\boldsymbol{Q}$ characterizes the actual status of an individual through its characteristic configuration. The alteration of implemented characteristics leads to a movement of the corresponding points analogous to the motion of particles through temporally varying coordinates in the location space. The movement of these points is not described individually in the continuous model, but through a density function $x(\vec{q},t)$. The density function is a real-value non-negative function over the space $\boldsymbol{Q}$ which has a complicated structure with many maxima. Most places in the $\boldsymbol{Q}$ space are unoccupied, i.e. the density function there is zero and local maxima of the density function are only available in a few "favourable" places. The density function $x(\vec{q},t)$ takes the place of the discrete functions $x_i(t)$.

Populations are formed from groups of elements/individuals with similar characteristics which are spatially adjacent to each other. Geometrically, a population corresponds to a local maximum of the density function. The spatial arrangement of the population corresponds to the proximity or distance of their characteristic structure. The dynamics of populations, in the sense of a characteristics distribution, and its variability correspond to changes on the characteristics level which are reflected through the temporal-dependence of the function $x(\vec{q},t)$. It can also be said that in such a characteristics space, characteristics combinations are characterized by implemented or populated locations, populations are characterized as local accumulations of such populated places and dynamics is characterized by the temporal shifts of the local maxima.

The system dynamics defines the rules governing the temporal change of the occupation. This was determined using a partial differential equation for the function $x(\vec{q},t)$. In order to quantify the evolution dynamics, it was assumed that the characteristics of different locations differ with respect to additional criteria. Hence, we introduced an evaluation function over the characteristics space of the competences whose topological structure was linked with the system dynamics. This kind of evaluation – in the simplest case given by a scalar function $w(\vec{q},t)$ – assigns to each space point a function value which is a measure of the local fitness. The evaluation function introduced in this way forms a landscape over the abstract characteristics space.

Thus, the problem of the definition and determination of a fitness function is taken into account by the fact that this is represented as a random landscape with certain statistical characteristics. As a result, it is possible to establish links to the physics of subordinate systems (Anderson 1983). The various model approaches used therefore also depend on the nature of the search space. While in a continuous search space (status or characteristics space of the competences) a metric is always defined and, with it both a distance between points and a step width of the search or mutation process, different neighbourhoods can be defined in discrete search spaces.[11]

### 6.1.1 Competence and problem spaces

In accordance with the two different applications of this model concept to competence modelling, the basic parameters introduced also have a different meaning.

| **Mathematical Symbol** | **Abstract Meaning** | **Model 1 Significance for competence development** | **Model 2 Significance for problem-solving** |
|---|---|---|---|
| $\vec{q} = \{q_1, q_2, ..., q_d\}$ | Vector of characteristics | Set of basic competences (or scaled components in accordance with KODE®) | Characteristics of the problem |
| $x(\vec{q}, t)$ | Density function; Frequency of the occupation of certain characteristics | Number or frequency of individuals with a certain competence profile | Number of individuals with a certain problem view, problem definition or problem perception |
| $w(\vec{q}, t)$ | Fitness or evaluation function | Evaluation, value, norm: expresses how well a certain competence profile suits the resolution of a task | Evaluation, value, norm: expresses how good the problem-solving behaviour of an individual is or how good the problem view and capacity for the resolution of a problem are (in a certain way the "correct" problem view is equated with resolution) |

The evaluation function $w(\vec{q}, t)$ initially depends only on the characteristics and the time. In this form, it is not dependent on the occupation of the characteristics space. This is not, however, identical to an evaluation function defined outside the system. Instead, the values and norms can also be determined internally by the group members. The existence of the evaluation function expresses that in addition to the current colonization of characteristics in the space, there is another criterion that accompanies the search process continuously in the sense of a boundary condition. In the case of competence development, the evaluation function expresses values and norms, whose existence can be assumed prior to the search, even if their location in the space is not yet known. Similar things apply for the search in the problem space. Problem solutions can certainly be negotiated socially and there will possibly be several equivalent solutions. However, the validity of solutions also depends on external objective criteria.

[11] On the connection between neighbourhood structure and fitness landscape, cf. Rosé, H. (1998). Evolutionäre Strategien und Multitome Optimierung. Berlin, Humboldt-Universität.

In an extension of the model, it is also possible to consider so-called Lotka-Volterra systems whereby the evaluation function depends explicitly on the occupation $w(\vec{q};x(\vec{q},t))$, and changes its form, depending which area of the characteristics space is occupied. This kind of representation is of special relevance for modelling processes in social systems in which the formation of values or norms is an integral component of the system dynamic.

### 6.2 Evolutionary dynamic in the phenotype space

In the continuous description used hitherto, each point in the characteristics space of the competences ***Q*** (i.e. each vector $\vec{q} = \{q_1, q_2, ..., q_d\}$ of characteristics variables $q_k$) is assigned a number (or frequency) which characterizes the implementation of certain parameter combinations. Thus, a density function $x(\vec{q},t)$ over the characteristics space of the problems is defined. The dynamic is associated with the reproduction characteristics of certain characteristics combinations. The following approach based on a general evolution dynamic of the Darwin-type (replication approach) is fundamental: (Feistel and Ebeling 1982; Ebeling, Engel et al. 1984; Feistel and Ebeling 1989)

$$\partial_t x(\vec{q},t) = w(\vec{q},\{x\})x(\vec{q},t) + Mx(\vec{q},t) \qquad (1)$$

whereby the parameter $w(\vec{q};x(\vec{q},t))$ functions as a generalized fitness function. It describes the competition and selection aspect of the evolutionary search dynamic. If $w$ only depends on the characteristics $\vec{q}$, what is involved is a stationary landscape. In the next case, $w$ depends on the characteristics and (through a social average) on time, and the following approach is applicable:

$$w(\vec{q};x(\vec{q},t)) = E(\vec{q}) - \langle E \rangle \qquad (2)$$

The angular brackets <*E*> indicate the social average (conveyed through the population):

$$\langle E \rangle = \frac{\int E(\vec{q}')\, x(\vec{q}',t)\, d\vec{q}'}{\int x(\vec{q}',t)\, d\vec{q}'}\,.$$

In the (more complicated) case of a fitness function, the evaluation of a location $\vec{q}$ also depends on the occupation function $x(\vec{q},t)$ in the entire characteristics space of the competences. A change in this occupation also leads to a change in the evaluation and there is a co-evolution between fitness function and population density. An example of such a coupling is represented by the following Lotka-Volterra approach:

$$w(\vec{q};x(\vec{q},t)) = a(\vec{q}) + \int b(\vec{q},\vec{q}')x(\vec{q}',t)d\vec{q}' \qquad (3)$$

In this case, the evaluation of a characteristics combination consists of two parts. The first term in equation (3) $a(\vec{q})$ represents an evaluation of the reproductive aspects of the characteristics. The second term describes the interaction between the characteristics, i.e. the influence of other populated places which are integrated throughout the space and weighted by the coefficients $b(\vec{q},\vec{q}')$. In this case, one refers to an adaptive landscape. (Conrad 1978; Conrad and Ebeling 1992)

In the special case of a delta function kernel it is possible to get equation (2) back from equation (3). The evolutionary dynamics is understood as a search processes in a constantly changing adaptive fitness landscape. It has been shown that this kind of model description is of particular relevance for socio-technological systems. (Ebeling, Karmeshu et al. 2001) In these, the system dynamics is largely determined by non-linear feedback between the action of the actors on the individual level and coordination and evaluation processes at macroscopic level.

The parameter *M* in equation (1) stands generally for a mutation operator. The elementary process of innovation on the micro level lies in the occupation of problem formulations with characteristics that have not been hitherto implemented.

This process of occupation (or colonization) of previously unpopulated areas in the characteristics space of the problems can be understood as an innovative search process (research and development) on the micro level. Whether this *elementary act of innovation* also lead to a global system change[12] can be decided from the outset. In Model 1, each characteristics change represents an altered application of competences. The occupation of new characteristics spaces stands for competence development. In Model 2, each of the problem-solving steps leads to a new definition of problems, i.e. to a occupation of other places in the characteristics space of the problems. The growth and selection processes which are part of a search process in a group lead to a change in the form of the occupation function. If one describes the elementary search processes within and on the periphery of the populations concentrated around certain centres as diffusion-like in an initial approach,

$$Mx(\vec{q},t) = D\Delta x(\vec{q},t) \tag{4}$$

two features become clear. Firstly, each extension of the populated part of the space initially generates a growth within the population in the competence spectrum (Model 1) or the problem perception (Model 2). In both models, this is associated with a diversification within the group. The individual variability increases. Different competence types or different problem views co-exist. Secondly, this process forms the basis for re-concentration processes in the new areas of the characteristics space. The exploration of new competitive areas (competences and norms or problem solutions) emanates from the peripheries of the existing populations through an undifferentiated search process. With continuous description, in the simplest case, the system dynamic defines (equation (1) with (2) and (4)) a temporal variation of the globally defined occupation function $x(\vec{q},t)$ through the partial differential equation (Darwin dynamics, Darwin strategy)

$$\frac{\partial x(\vec{q},t)}{\partial t} = x(\vec{q},t)\left[E(\vec{q}) - \langle E \rangle\right] + D\Delta x(\vec{q},t) \tag{5}$$

The fitness function is generally an a priori **unknown** landscape which can be estimated from empirical data and sometimes modelled as a random function with certain statistical characteristics. (Ebeling and Feistel 1990) The centres of the distribution function $x(\vec{q},t)$ that form follow in part the maxima of the evaluation function. The dynamics is very dependent on the assumptions about the structure of the fitness function. Once formed, the (localization) centres of the population clouds can be treated as analogues to the discrete types (as order parameters of the system).

---

[12] For example, in the sense of the change of localization centres between maxima of the evaluation function.

In addition, however, the description of the merging of populations, the differentiation in different populations and concentration or expansion processes requires no taxonomic extension. Instead, all of these processes are equally aspects of the system dynamics. Local maxima correspond to intermediary steps of the development. The advantages of this description are associated with growing complexity. The description chosen depends on the focus of the analysis of the concrete system and the mathematical "processability".

The dynamics defined by equation (5) is problem-solving in the sense that the density function $x(\vec{q},t)$ seeks the maxima of the evaluation function $E(\vec{q})$ in the course of time. Given that the dynamic largely rests on replication mechanisms, we also refer here to a Darwin-type dynamics or a Darwin strategy. An alternative dynamics with similar characteristics is based on the related equation:

$$\frac{\partial x(\vec{q},t)}{\partial t} = \nabla D\left[\nabla x(\vec{q},t) - \frac{1}{\Theta} x(\vec{q},t)\nabla E(\vec{q})\right] \tag{6}$$

This diffusion-type equation describes a dynamics which is referred to as a Boltzmann dynamics or Boltzmann strategy in the location space. (Asselmeyer, Ebeling et al. 1996) The parameter describes a kind of effective temperature $\epsilon = k_B T$, which is a measure of mobility (mean square displacement). In terms of a social interpretation, it expresses the willingness of individuals to embrace change. If the effective temperature is high, locality changes are frequent and far-reaching.

### 6.2.1 Competence development and problem-solving

In the previous section, we introduced various mechanisms of the evolutionary search. Of these, selection and mutation are central. Selection is described with the help of the evaluation function. Depending on their form, locations with lower evaluations are abandoned and locations with higher evaluations sought. In the case of biological evolution, we refer to the varying reproduction rates of characteristics. In the case of social search processes, decision-making processes of individuals for certain characteristics arise here. The group size remains relatively constant, at least in cases of competence development. In contrast to technological evolution or scientific evolution, growth processes play a minor role here. The groups in which competence development is measured are constant in size. Transition processes become all the more important. The occupation changes in that the individuals seek other positions in the characteristics space. The following table describes an interpretation of the different evolutionary mechanisms for the two different models.

| **Mathematical Symbol** | **Abstract Meaning** | **Model 1 Significance for competence development** | **Model 2 Significance for problem-solving** |
|---|---|---|---|
| $w(\vec{q};x(\vec{q},t)) = E(\vec{q}) - \langle E \rangle$ | Evaluation (selection) takes place as the comparison of the current position with the group average | An individual's own competence profile is compared with the group norm in relation to competences and, where applicable, changed | During the comparison of the validity of different problem solutions, technical-methodical competences and socio-communicative competences are adopted as priority (comparison with group average) |
| $w(\vec{q};x(\vec{q},t)) = a(\vec{q}) + \int b(\vec{q},\vec{q}\,')\,x(\vec{q}\,',t)\,d\vec{q}\,'$<br><br>$a(\vec{q})$<br><br>$b(\vec{q},\vec{q}\,')$ | Evaluation process with two parts<br><br>describes the evaluation and decision for a certain characteristics (selection)<br><br>describes decision-making processes for characteristics in which includes the occupation of characteristics in a particular place or different places (imitation) | Decision in favour of a certain competence profile, decision to implement competences differently<br><br>due to the value of a competence profile<br><br>because other group members support certain competence profiles | Decision in favour of a certain problem solution<br><br>the benefit of the problem solution determines the decision (selection) – technical-methodical competences are relevant for this<br><br>the views of others on a certain problem solution (imitation) determines the decision –<br>socio-communicative competences are central for this |
| $Mx(\vec{q},t) = D\Delta x(\vec{q},t)$ | Mutation, describes random search movements in the characteristics space | Different competence profiles are tested, competences are expanded playfully | Different problem definitions are tested, new problem definitions are explored, creative search – personal competences are the key to creativity |

Competences are treated differently in the two models. In the competence development model (Model 1), the competences are the variables (objects) of the evolutionary search.

Competences as variables

*Mathematical Model*

$$\partial_t x(\vec{q},t) = \underbrace{\left[a(\vec{q}) + \int b(\vec{q},\vec{q}\,')x(\vec{q}\,',t)d\vec{q}\,'\right]x(\vec{q},t)}_{\text{Selection process}} + \underbrace{D\Delta x(\vec{q},t)}_{\text{Mutation}}$$

q = {q1,q2,q3,q4} Competence profile {K, S, A, P}

x = Number of individuals with a certain competence profil

**Figure 16: Summary of the mathematical representation of the evolutionary search in the competence space**

As opposed to this, in the second model approach involving the search in a problem space, the competences are assigned to the different evolutionary search mechanisms. Here, competencies are instruments of the evolutionary search.

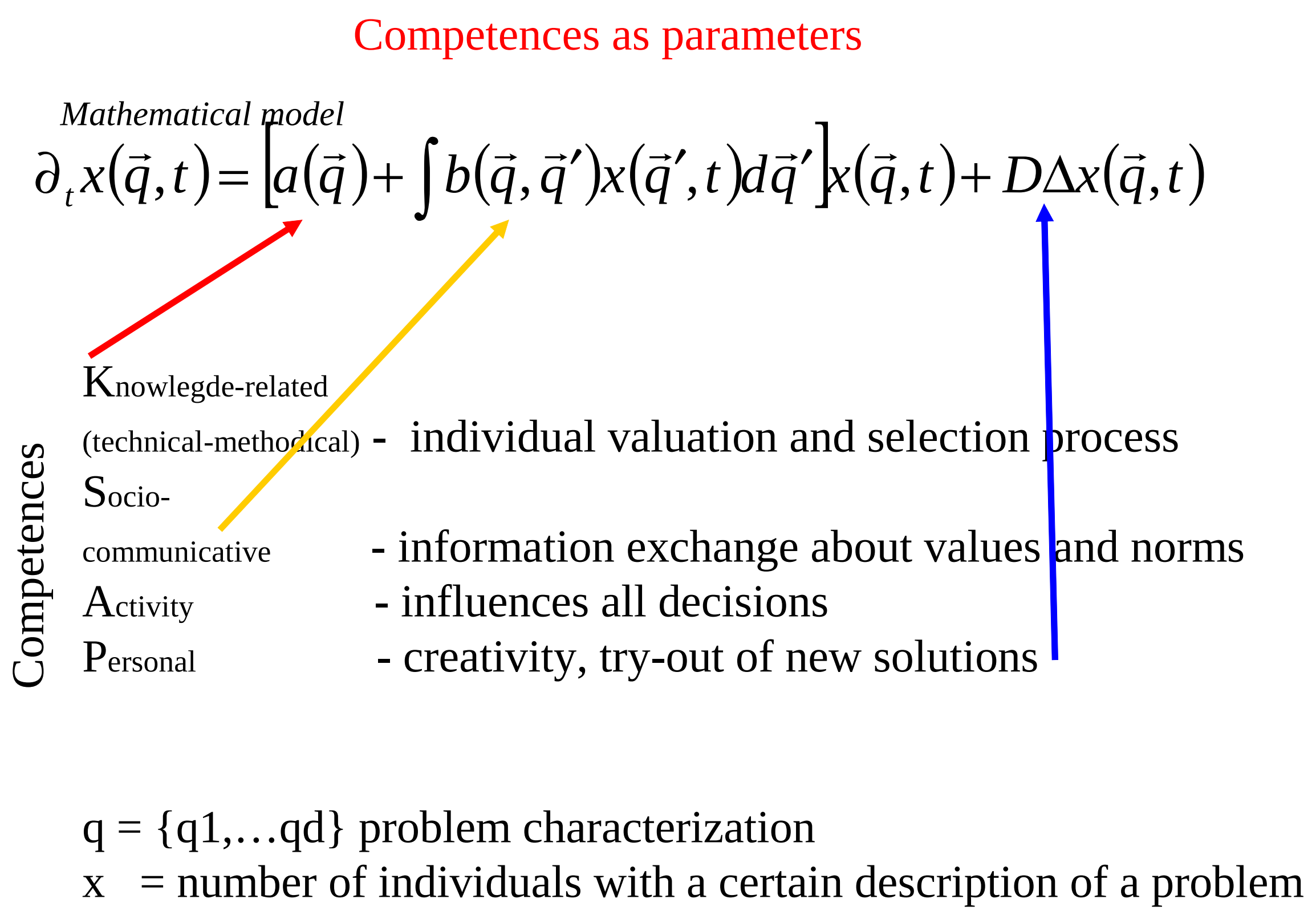

**Figure 17: Summary of the mathematical representation of the evolutionary search in the problem space**

### *6.3 Evolino – evolutionary search on a playing field – competence-controlled problem-solving*

The equations introduced in Chapter 6 describe the temporal change of the occupation. However, these equations can only be analytically resolved in very limited special cases. Computer simulations provide one way of solving the equations. The programmes required for this are highly specialized and are often written in programming languages like FORTRAN or C. However, the possibility to design the simulations interactively also exists. We initially developed a simulation which simulates a stochastic Fisher-Eigen equation as introduced above. The simulation, which we have called *Evolino*, is based on certain discretizations of the original continuous model. Thus, we describe search agents, the characteristics space is a playing field (grid) and the valuation landscape is also represented in the form of bars of different heights over the playing field. (For a website on *Evolino* see http://www.niwi.knaw.nl/web/nerdi/Evolino/english/introduction.html )

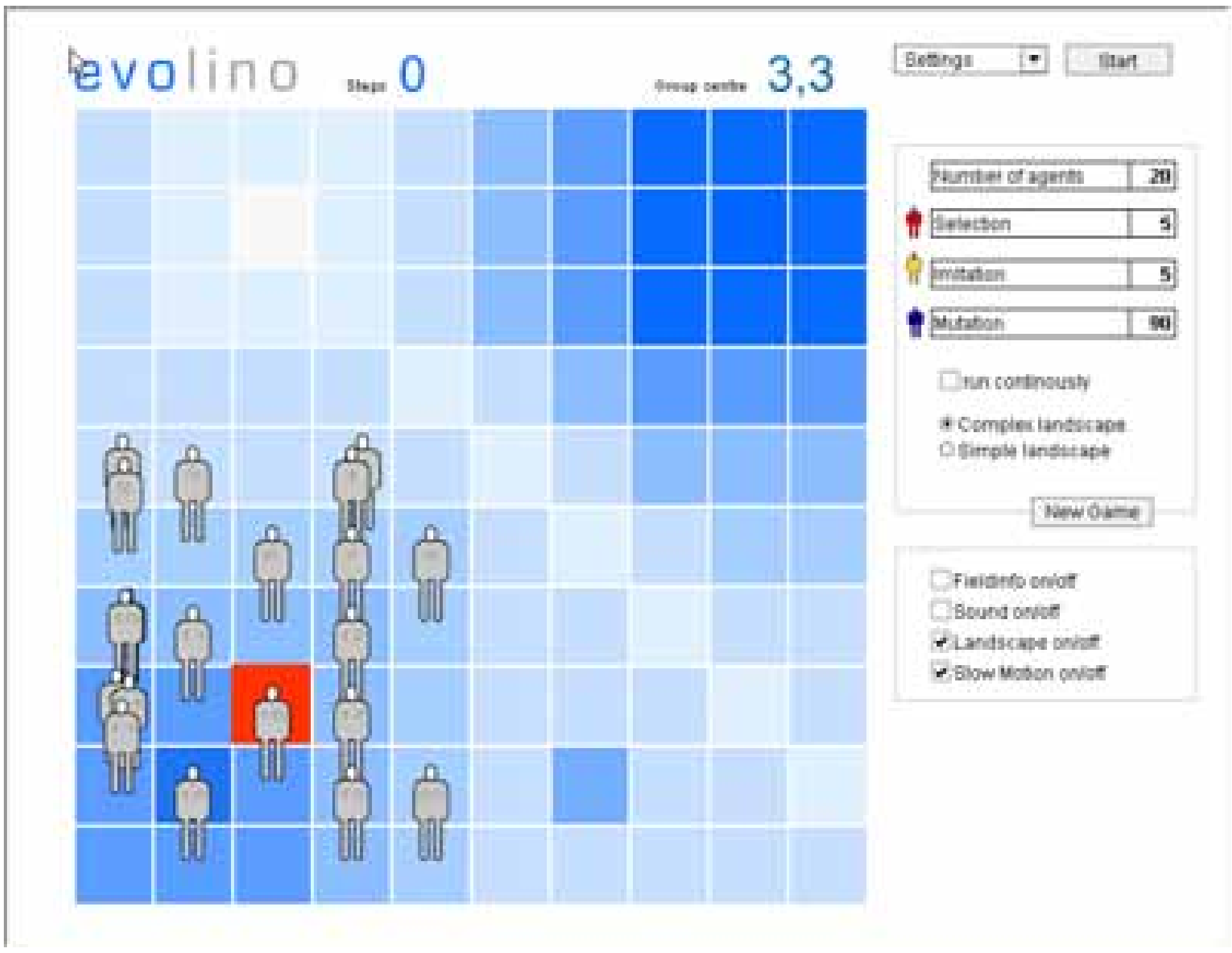

**Figure 18: Starting picture of *Evolino*. 20 players (agents) are randomly distributed in the lower left half of the playing field. The parameters for selection, imitation and mutation can be entered. The central point of the group is indicated by an orange square, the coordinates of which are provided on top in the middle. A "steps" display counts the game steps.**

In its general form, *Evolino* is an evolution game involving agents or individuals in a game space (playing field = characteristics space). The playing field which consists of 10x10 boxes is a discretization of the formerly continuous characteristics space. A landscape is defined over this playing field which is indicated by the different colours of the boxes. The depth of the blue corresponds to the height of the landscape. These values can also be displayed using a switch (field info on/off). For the most part, we worked with a landscape consisting of two peaks and an intermediate maximum. The aim of the individuals is to find the new maxima in the top right quadrant. However, the individuals only know the landscape in the places in which they are situated. The "landscape on/off" switch can be used to show the part of the landscape the search agents can see.

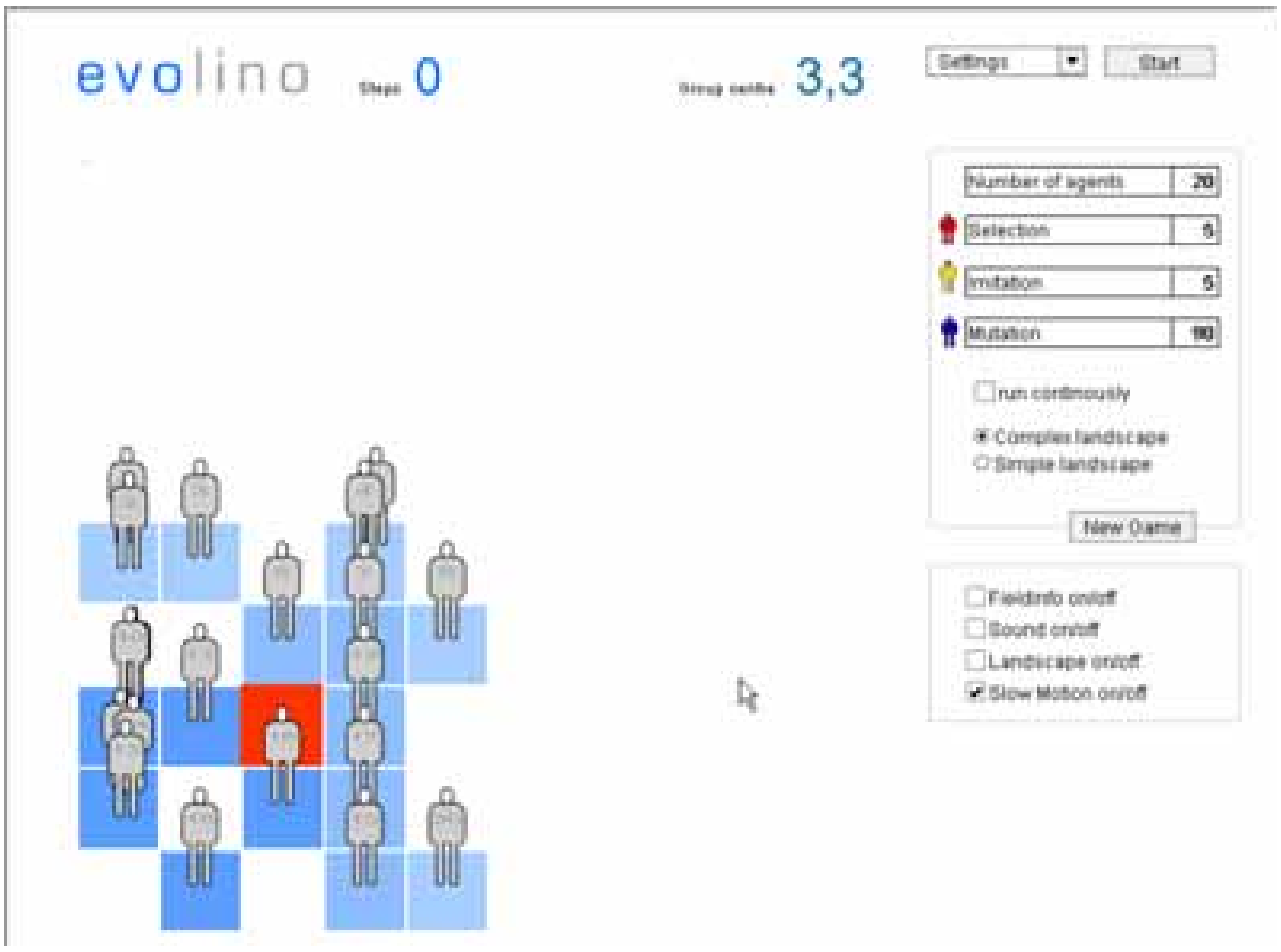

**Figure 19: Picture of the landscape in *Evolino* as seen by the players. They embark on a real search for the unknown new peak.**

As introduced in Chapter 4, the search follows three evolutionary mechanisms. These mechanisms are defined as behavioural rules for the players. The frequency with which different evolutionary mechanisms are used is defined using the parameter input. All of the parameters together total 100. A parameter combination of "selection, imitation, mutation = 10, 10, 80" means that out of 100 evolution steps, 10 are selection steps, 10 are imitation steps and 80 are mutation steps. In terms of programming technology, mutation is the basic step which is interrupted by selection and imitation.

SELECTION means that two players are randomly selected. The heights of the landscape (benefit, value, norm) in their positions are then compared and the player with the worse position goes over to the better position. If the two players have equally good positions, there is no action. The players can consciously improve themselves through this process, i.e. populate higher parts of the landscape.

IMITATION means that two players are selected. The extents to which their positions are populated are then. The player with the less populated position goes over to the player with the more populated position. If both positions are equally occupied, there is no action. With this rule, social imitation is positively evaluated. Many players in a field exercise a power of attraction.

MUTATION means that one player is randomly selected and a field adjacent to his position (above, below, left or right) is selected to which he moves. With this rule, the exploration of the landscape is modelled as a local process. Mutation is the only rule with which new areas in the landscape can be explored (innovation).

As a boundary condition with the limited playing field, so-called reflecting walls were assumed, i.e. if movements would lead out of the playing field, the players are simply returned and they effectively remain standing.

Based on the above-described rules, a game is created that displays the same characteristics for the evolutionary search as the continuous model. What is involved is a kind of cellular automat, which differs from standard cellular automats in that it also contains non-local processes or interactions (selection and imitation) and a landscape is defined as an additional element over the space.

The *Evolino* parameter space is large enough to investigate different game processes. In general, it may be said that relatively high selection leads to the group repeatedly concentrating on the peaks in the landscape that have already been reached. "Mavericks" who dare to embark on the path to the next highest peak through the valley are recaptured relatively.

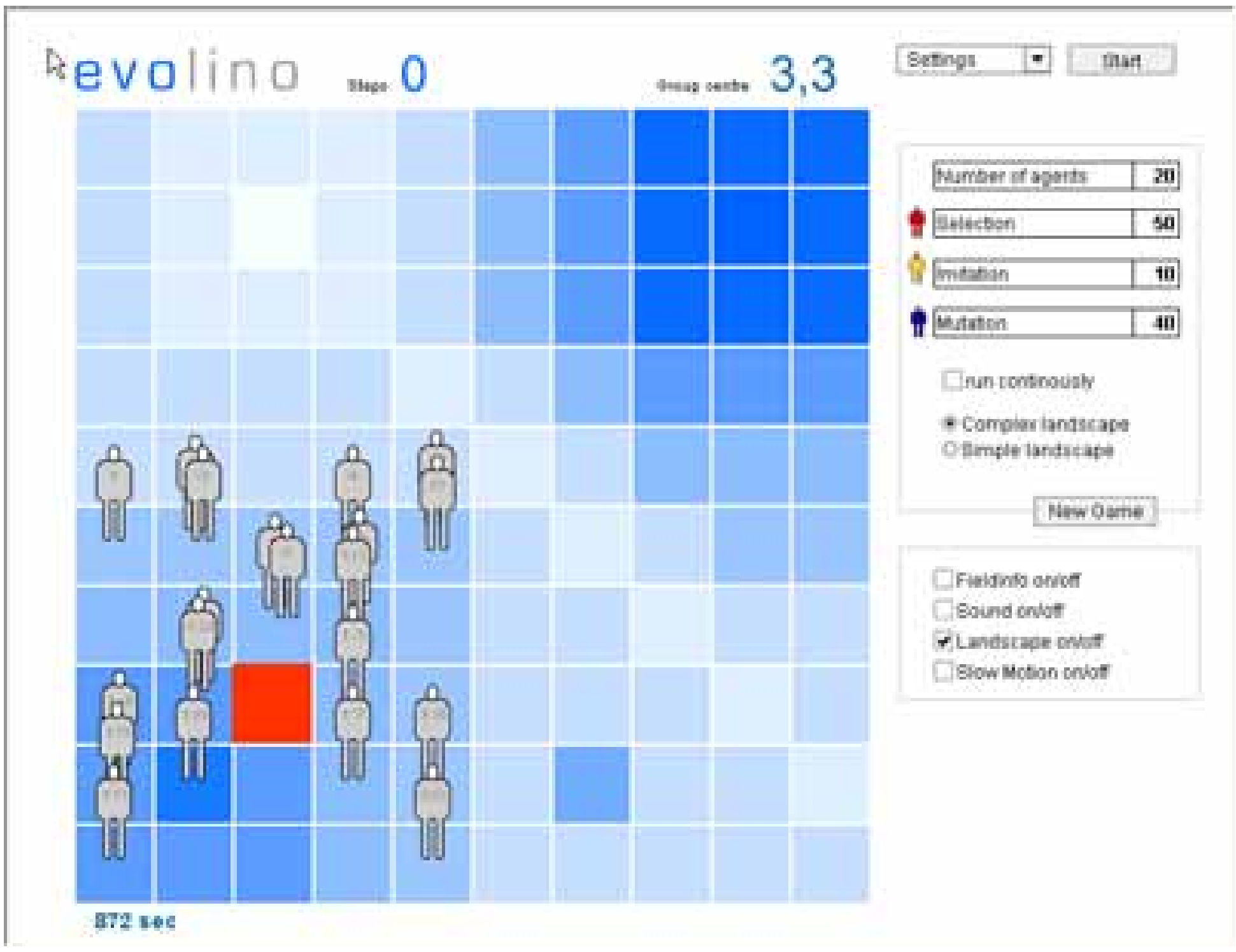

**Figure 20: Starting picture of a game with a relatively high selection rate (50).**

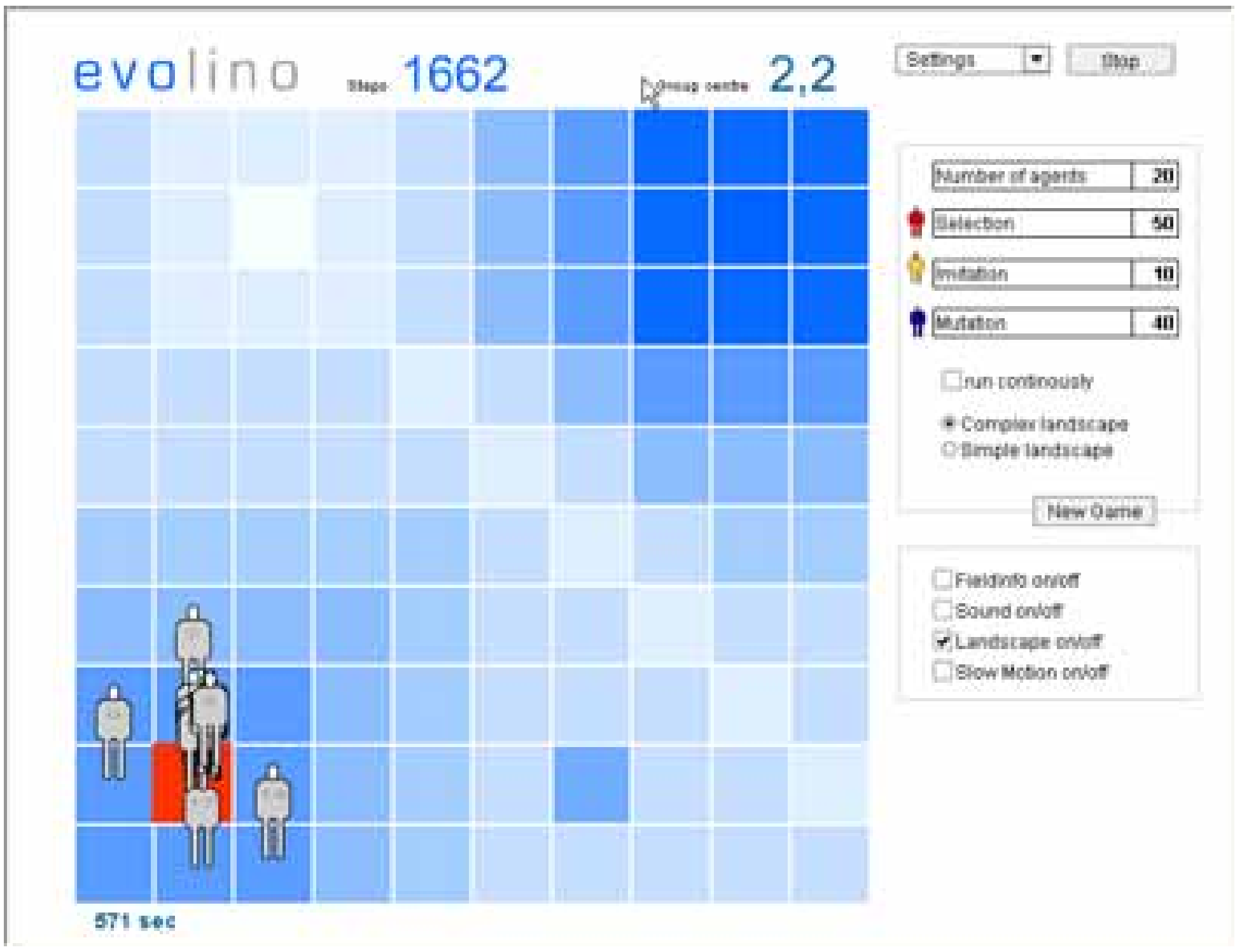

**Figure 21: Status of a game after ca. 1700 steps: the group is strongly centred around the old peak and has not found the new peak.**

In the case of a high imitation rate, random movements can easily increase, i.e. the group will stay close together but not necessarily in places with a high evaluation. The reference to the group is crucial to the individual decision-making processes, not the value landscape.

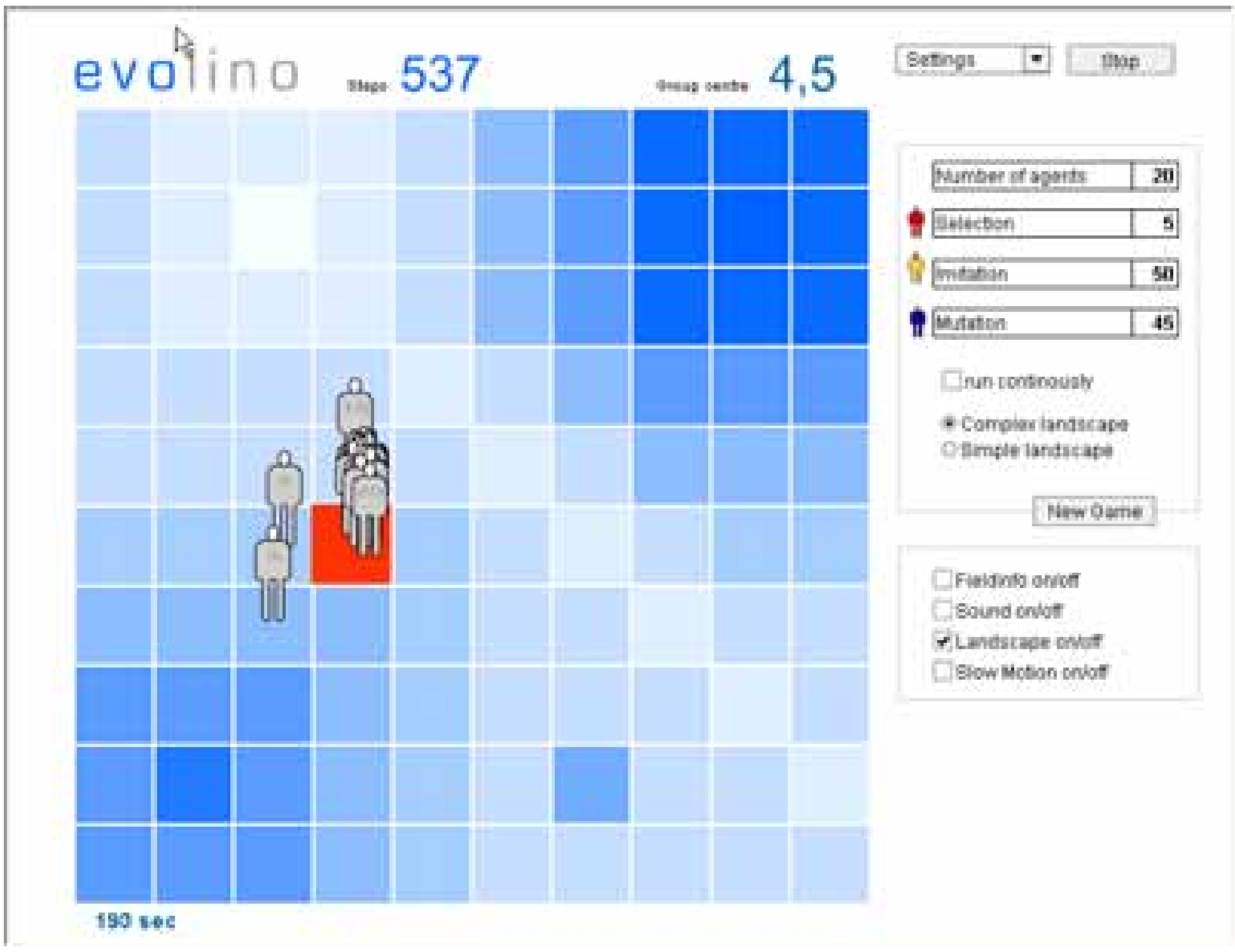

**Figure 22: Relatively high imitation rate (50). Status of the game after ca. 600 steps. The group is centred but not around a peak.**

As opposed to this, high mutation rates result in all players being distributed far and wide on the playing field and the group formation is strongly restricted.

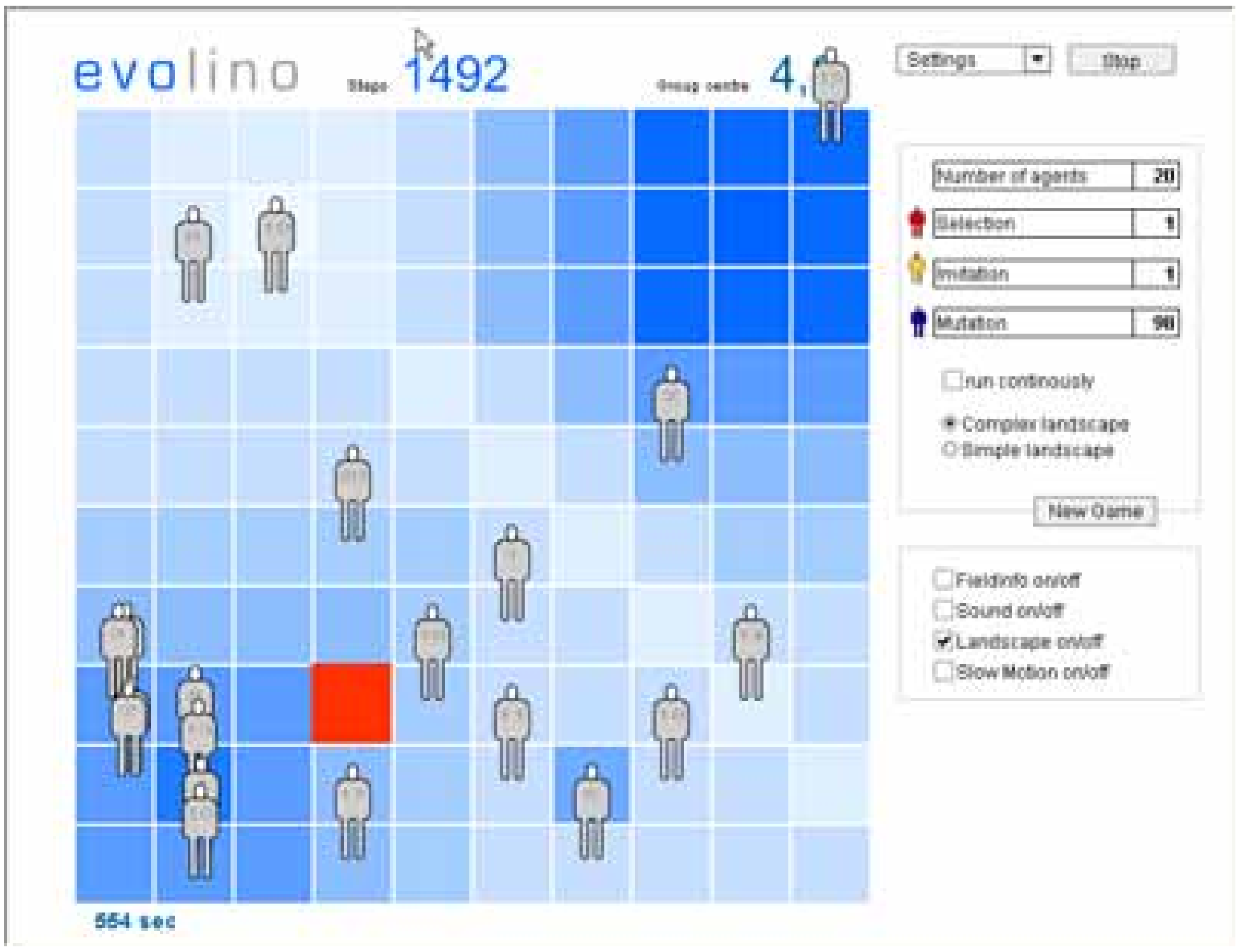

**Figure 23: Very high mutation rate (98). The players explore the entire playing field. Some also reach the new peak. However, there is no further group formation and the centre of the group remains in the valley in the middle of the playing field.**

What needs to be done is to find an optimum between all three processes. This is not always easy. The simulation should also demonstrate the influence that random processes have on the search. Courses can differ completely even with the same sets of parameters. The simulation can be carried out quickly and slowly. In slow mode (slow motion), individual decision-making steps can be followed and it becomes clearer how many individual and communicated decision-making steps are necessary to change the average position of the group (marked by the centre). In a real problem-solving process, communication protocols could be produced and these courses could be interpreted as trajectories of an evolutionary search. (Tschacher, Schiepek et al. 1992; Tschacher and Dauwalder 1999)

In the course of the game, the group can temporarily gather on the intermediate maximum before the final global maximum is reached. Another effect that was observed is that competing sub-groups only arise before the members reassemble relatively evenly around the group centre.

The *Evolino* game provides the starting point for three different competence games: *EvoKom*, *SynKom* and *SynKom_Berg*. All three game variants use Model 2, i.e. the competences are interpreted as different mechanisms of the search.

### 6.3.1 EvoKom – competence-controlled problem search

The *EvoKom* game uses the same set of rules as *Evolino*.

The following rules are used in the game:

*1 Mutation*

*2 Imitation*

*3 Selection*

We apply hereby the following interpretations of evolution steps as competence-controlled decision-making processes during problem-solving:

1. Creativity is the centre of *personal competence*. In the evolutionary picture, it is interpreted as notional mutation. The more creative a person is, the more often such mutations arise.
2. Social action obeys rules, norms and values. Adaptation to this is interpreted as imitation, the capacity for it as *social competence*. Excessive imitation/adaptation leads to frictionless behaviour which is, however, hostile to innovation.
3. The evaluation of the knowledge of one's own position and that of others requires *technical-methodical competences*. The appropriation of knowledge and, thus, the alteration of one's own position, is a selective decision-making process whereby group values and norms are interiorized individually.

The red square represents the relevant problem resolution by the group. Thus, the movement of the red square describes the finding of new problem solutions in the group process.

Three limit value games can be defined: 1. The technical-methodical competence dominates, only direct solution improvements are accepted. The group finds the next best solution, but no innovative new solutions. 2. There is only social imitation, all players will want the same thing and the problem solution also remains unchanged. 3. There is only mutation in the sense of creativity; the consequence will be a clear deferral of the problem solution. The combination of these action-propelling competences give rise to new game courses.

The different game courses can be stored in their results (end coordinates of the group centre, number of steps and parameter) and demonstrated. This makes it possible to compare the search behaviour of groups with different competence strategies.

### 6.3.2 SynKom – group order parameter is formed synergetically and internally

In the *EvoKom* game, the players search in an external landscape. With favourable parameter combinations, the group is re-ordered around the higher innovative maximum. However, the internal process of group formation is often of importance in terms of social processes. For this reason, we also designed a game without an external landscape.

The following rules are used in the game:

*1 Mutation*

*2 Imitation*

*3 Centrifugal repulsion*

In this game, we interpret the rules as follows:

1. Creativity is the centre of *personal competence*. In the evolutionary picture, it is interpreted as notional mutation. The more creative a person is, the more often such mutations arise.
2. Social action obeys rules, norms and values. Adaptation to this is interpreted as imitation, the capacity for it as *social competence*. Excessive imitation/adaptation leads to frictionless behaviour which is, however, hostile to innovation.
3. To turn away from the familiar and penetrate into new directions can be interpreted as activity and the capacity to do it as *activity-related competence*.

Creativity, imitation and activity are not interpreted as physical but intellectual problem solving efforts. The red square represents the relevant problem resolution by the group.

The movement of the red square then describes the finding of new problem solutions in the group process. The coordinates of the group order centre are calculated from the distribution of the group members on the playing field.

Given that no selection (goal predefinition of the problem solution) is prescribed, the mobility rate of the red square becomes the yardstick of innovation.

Three limit value games can be defined: 1. There is only the (repulsion) activity; the problem solution hardly changes because the players want something fundamentally conflictive. 2. There is only social imitation; all players will want the same thing and the problem solution also remains unchanged. 3. There is only mutation in the sense of creativity; the consequence will be a clear deferral of the problem solution. The combinations of these action-propelling competences give rise to new game courses.

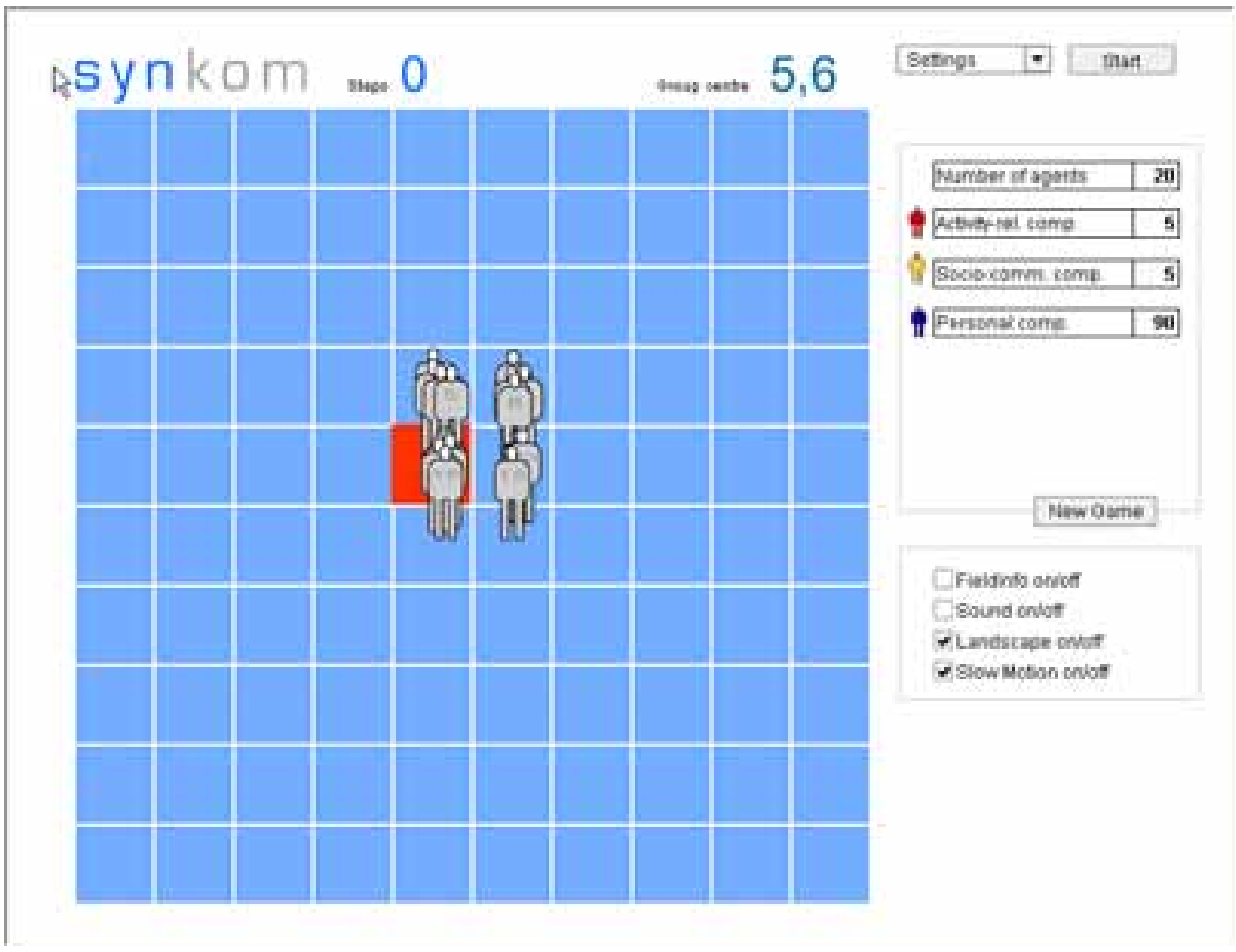

**Figure 24: *Synkom* game. The group starts from the middle of the playing field. There is no landscape. The orange box represents the group centre.**

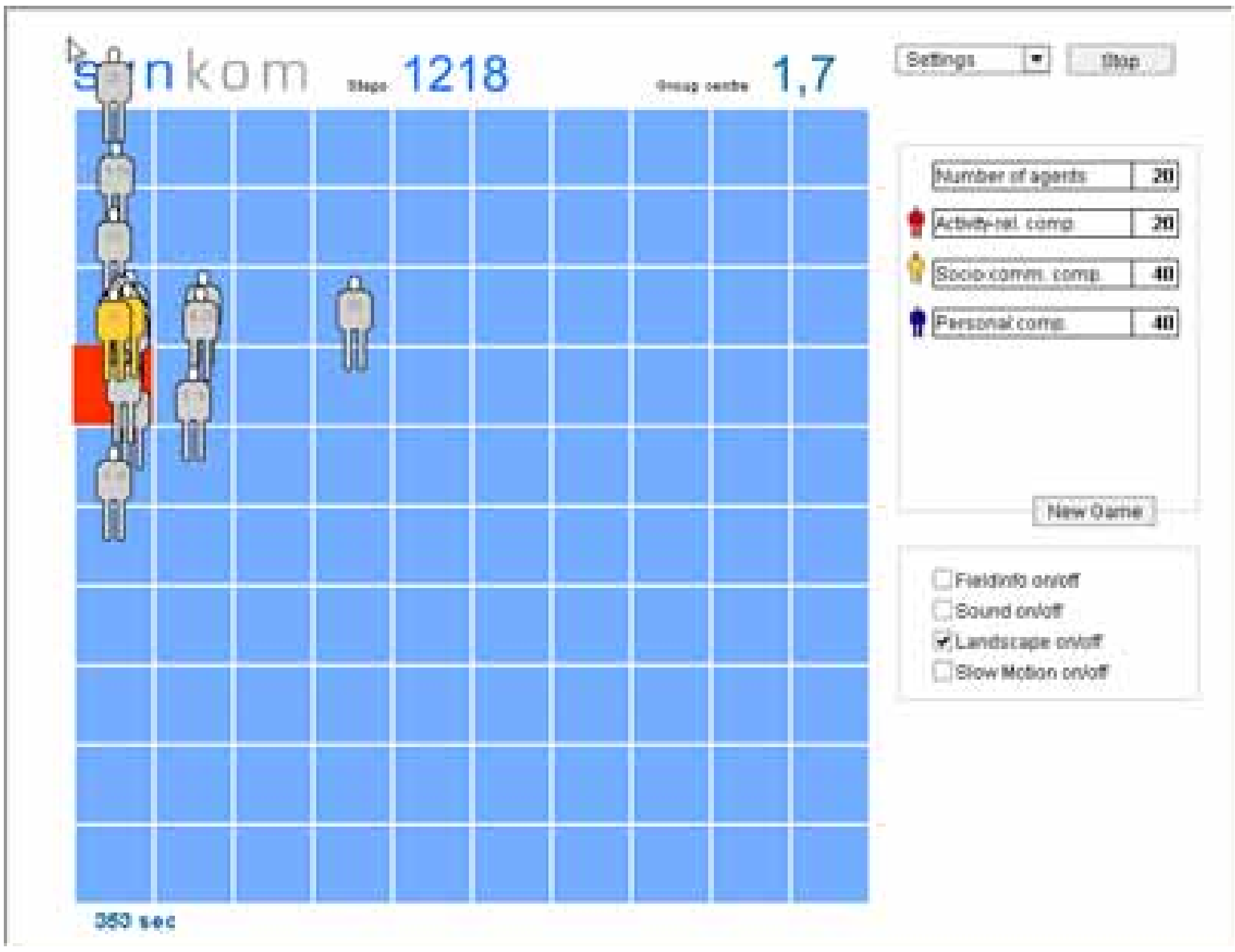

**Figure 25: The group order parameter has wandered up to the upper left corner of the playing field.**

### 6.3.3 SynKom_Berg – group reformation and gradient strategy

For the game variant *SynKom*, the innovation yardstick is the speed of movement of the group centre, i.e. how quickly a group centre can emerge in new locations. However, if the level of creativity within the group is high, the simulation leads to an unsatisfactory result. The group centre stabilizes itself in the middle of the playing field. This is due to the relatively small size of the playing field. Thus, the idea arose to create a combination of *EvoKom* and *SynKom*. We re-introduce an external landscape, but this time a very simple one that rises from the lower left to the upper right. The task is no longer to find a high peak among different peaks, but to climb the only mountain as quickly as possible. The selection element of the *Evolino* game always corresponds to a gradient strategy. In the case of a single mountain, this strategy must always lead to success. The game criterion is now the period of time it takes to reach the mountain.

The following rules are used in *SynKom_Berg*:

*1 Mutation*

*2 Imitation*

*3 Selection*

We interpret these rules in this game as follows:

1. Creativity is the centre of *personal competence*.
   In the evolutionary picture, it is interpreted as notional mutation.
   The more creative a person is, the more often such mutations arise.

2. Social action obeys rules, norms and values.
   Adaptation to this is interpreted as imitation, the capacity for it as *social competence*.
   Excessive imitation/adaptation leads to frictionless behaviour which is, however, hostile to innovation.

3. The active selection and targeting of goals is interpreted here as selection and the capacity to do this as *technical-methodical competence*.
   This competence is less an external one and more related to substantial and deliberate goals.

Creativity, imitation and goal-setting are not interpreted as physical but intellectual problem-solving efforts.

The red square represents the relevant problem resolution by the group. Thus, the movement of the red square describes the finding of new problem solutions in the group process; the distance from the starting point measures the strength of the direction criterion (selection yardstick).

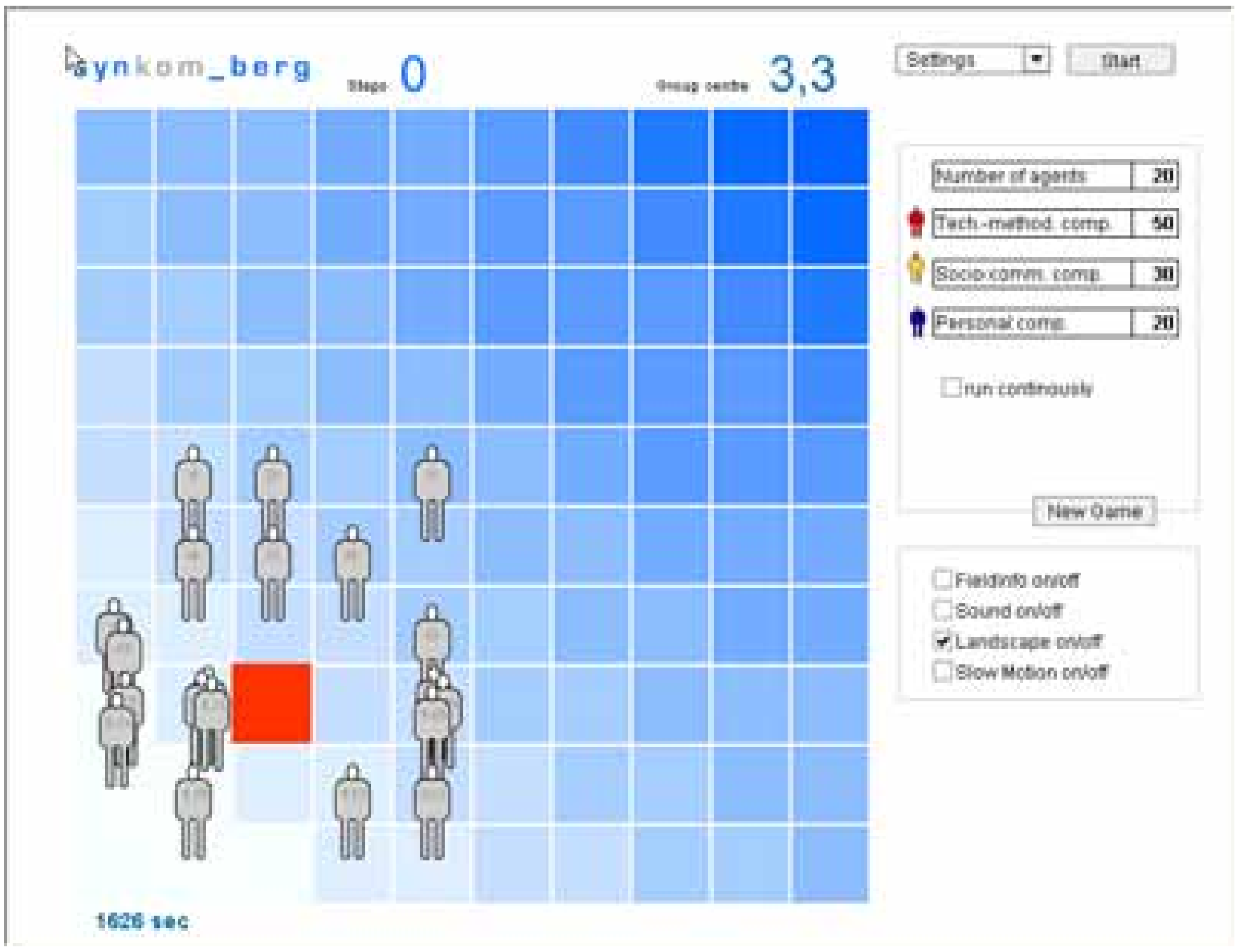

**Figure 26: *SynKom_Berg* game. Starting position in the lower half of the playing field. The landscape rises continuously to the top right.**

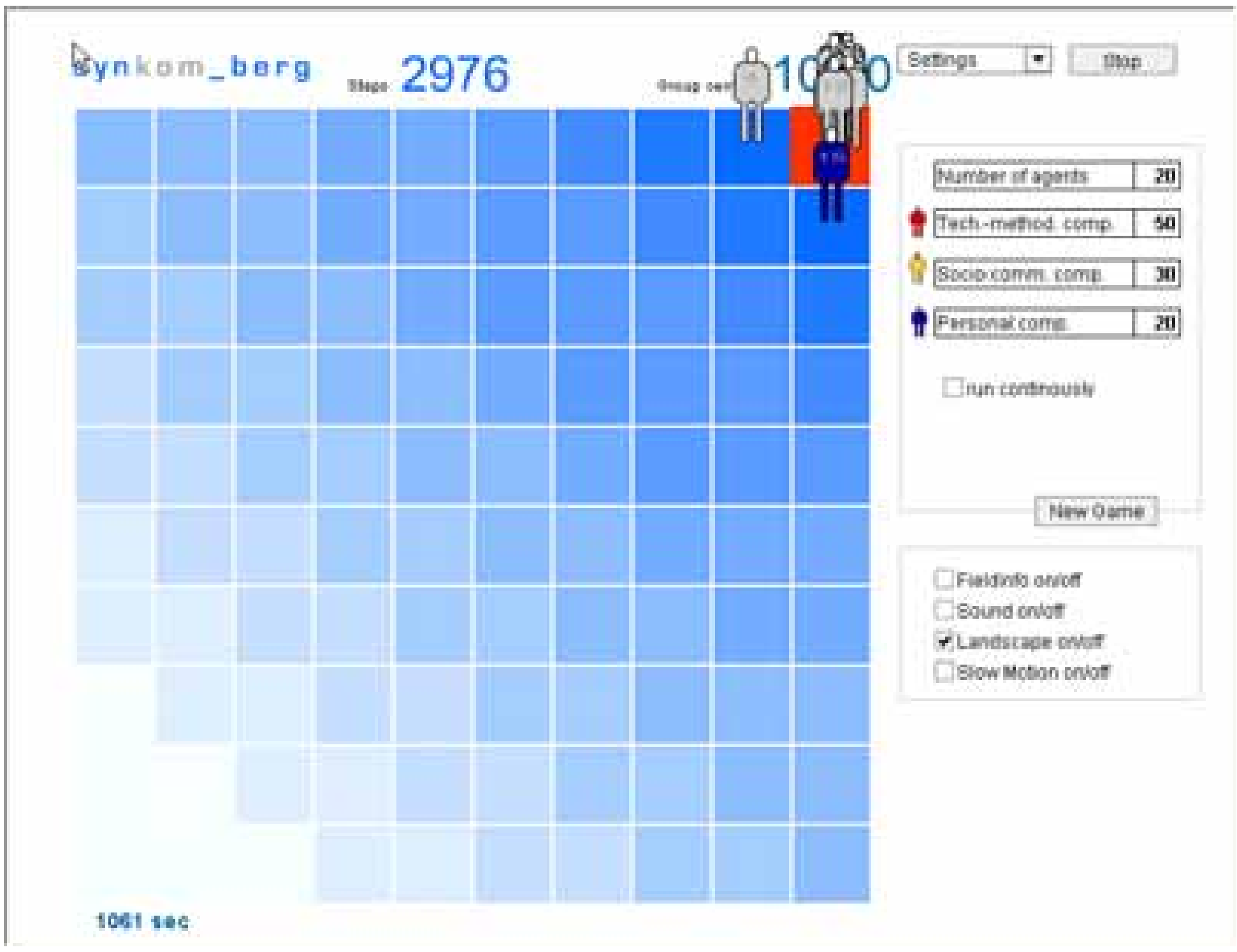

**Figure 27: In this case, after about 3000 game steps all players have assembled in the upper left quadrant of the playing field.**

# 7 The Modelling of Competence Development in Groups with the Help of Active Brownian Agents in an Evolution Landscape

## 7.1 Description of active Brownian Agents in the Characteristics Space

We will now develop a new model which differs from the previous one in that, in addition to the characteristics coordinates, we introduce the velocities of the change as a new variable.

$$v_1 = \frac{d}{dt} q_1 = \dot{q}_1, v_2 = \dot{q}_2, ..., v_d = \dot{q}_d$$

This is similar to the description of particles in physics by coordinates and velocities. Alternatively, one could introduce the impulses $p_1, p_2, ..., p_d$ as in Newtonian mechanics.

In the simplest case of Cartesian coordinates, the following simple connections apply:

$$v_1 = \dot{q}_1 = \frac{d}{dt} \mathrm{q}_1 \;, v_2 = \dot{q}_2, ..., v_d = \dot{q}_d$$
$$p_1 = m\dot{q}_1 = mv_1 \;, p_2 = m\dot{q}_2, ..., p_d = m\dot{q}_d$$

In the interest of clarity and since masses are not well defined in our picture we use the velocities here instead of the impulses. We assume that the dynamics should depend not only on the coordinates in the characteristics space, but also from the velocities (impulses). We consider the dependence on the velocities as a model for the role of flexibility. Competence and flexibility are mutually requisite. The parameter *m* indicates a kind of inertia *vis à vis* changes.

The new model relates analogously to approaches from physics in which velocities/impulses have proven a significant variable in dynamics and statistical physics. This is in particular connected with the work of Ludwig Boltzmann. For reasons discussed above, we will use the language of velocity consistently in the following account.

In our view, a transition to the Boltzmann concept of particles with coordinates and velocities is also relevant for social systems. An additional class of variables is introduced as a result. As in natural systems described in mechanics and statistical physics, the speed of change, i.e. flexibility, is also crucial in social systems. For example, an employee who reacts faster to changes in conditions and new requirements, i.e. who is "flexible", has better chances on the labour market. Flexibility is an important component of evaluations. The acquisition of flexibility is undoubtedly associated with the (life-long) process of learning. Thus, flexibility can also be described as the disposition for self-organized learning.

In accordance with this insight, we define the status in this new model as the totality

$$\vec{q} = [q_1, q_2, ..., q_d]; \vec{v} = [v_1, v_2, ..., v_d]$$

and refer to the phase space ***Q,V***. On some occasions we do not operate with generalized coordinates and impulses $q_i, p_i$, but with simple Cartesian coordinates $x_i$, then $v_i = \dot{x}_i$ applies for velocity. In the general case, the link between coordinates and velocities can be more complicated.

Instead of the density in the characteristics space $x(\vec{q}, t)$, which was used in the earlier model, we now define a density $f(\vec{q}, \vec{v}, t)$ over the phase space ***Q,V***, just as Ludwig Boltzmann did in his time in the physical coordinate-velocity space (phase space) of molecules. This density $f(\vec{q}, \vec{v}, t)$ is the decisive basic parameter of the theory in the new picture.

We will now concentrate on the development of the dynamics in the continuous phase space. The advantage of this approach lies in the ability to describe individual variability explicitly. However, this is associated with a higher mathematical complexity which makes it difficult to make analytical statements.

### 7.1.1 Competence development and meta-competence

If we apply the extension of the mathematical model to our interpretation of the search in the competence space, the new velocity variables represent the speed in the change of the use of certain competences. As already shown in Chapter 5.1, a change of location in the competence space corresponds to the selection of a different competence profile. A person can decide, for example, to refer more to socio-communicative competences in a certain learning situation, and, just a few moments later, he can refer again to his technical-methodical competences and make use of personal competences at another stage in the problem-solving process. This change between certain competence patterns can occur quickly or slowly. The speed of change says something about the extent to which an individual can avail of his own competences. In our view, this characteristic as a generalized self-organization disposition can be viewed as a self-organization competence which includes the ability to integrate oneself into self-organization processes in the first place. As a meta-competence, this characteristic is inextricably linked with the other basic competences. In our specific model, we can furthermore distinguish between different parts of meta-competence (depending on the basic competence to which the velocity of change belongs).

If we think of competence as ability which already is located on a meta-level (first order phenomena) meta-competence as flexibility is a second order phenomena. In this view, the interaction of different personal and social features of a person determines his different competences in a self-organized process. Meta-competence described as flexibility is than an additional phenomena based not only on the existence of the different competences in action but also how this action proceeds (fast or slow). So, it is linked to an additional temporal dimension in the process of using competences.

Competences are characteristics of individuals. In the model they appear as characteristics of a process (parameters) of individuals and groups.

Change of location in the problem space
= search for a problem solution while using different competences

competences = characteristics = *static picture*

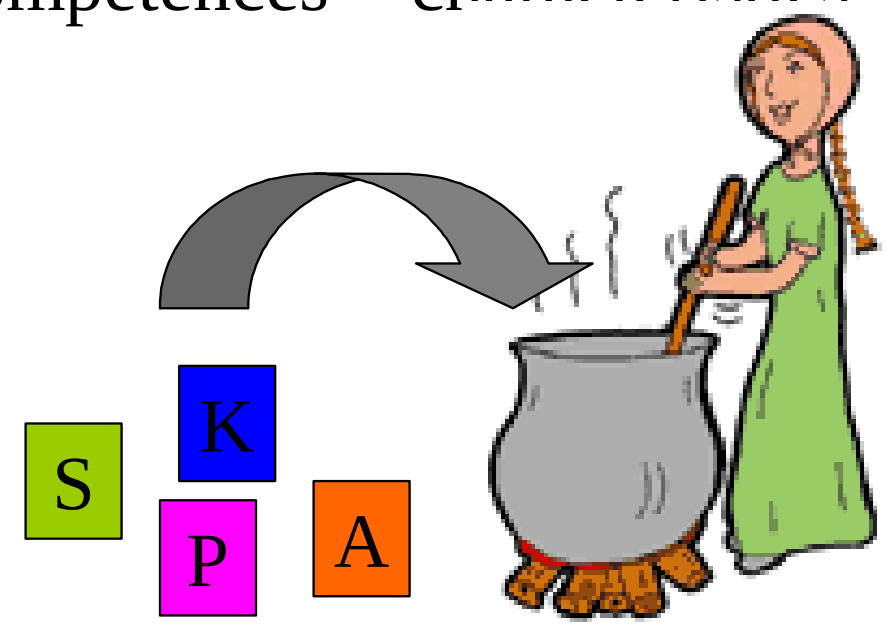

Meta-Competence = ability to use different competences
context dependent = *dynamical picture*

Velocity of change = flexibility to develop different solutions in a certain period of time (meta-competence)

**Figure 28: Describing meta-competence through the introduction of the new velocity variables**

### 7.1.2 Problem-solving and activity

If we apply the extension of the mathematical model to our second interpretation of the problem-solving search, the new variables describe how quickly certain problem views can be changed. In this case, they describe how quickly an individual is in a position to change between different problem views. In this picture, the velocity can be linked with the activity-related competence. Compared with the other three basic competences, the activity-related competence plays a particular role. It specifies the activity level on which the individual is acting (cf. also Figure 2). Thus, this competence complements the previous evolution mechanisms. We shall see below that an additional group of parameters is associated with the introduction of velocities.

In the language of physics, the previously described evolution models correspond to a so-called over-damped (i.e. strongly braked through friction) motion. In the new evolution model developed here, we add the property of inertia which is important not only in the context of physics, but also in sociological processes.

In physics, inertia refers to all effects which are triggered by mass and are associated with Newton's law. A body preserves in its state of rest or motion as long as there are no forces to change it. A body tends to maintain a state of motion, due to inertia. In a certain sense, this also applies, of course, to societal actors. Everyone tends to continue with a certain mobility status associated, for example, with professional activity or recreational activities. Our new model should be used to model this aspect of inertia (in the metaphorical sense). Another concept which will be used is "active friction or active forces". The model of so-called "active friction" is used in physics to model self-induced vibrations in mechanics and acoustics in particular basic processes of music (organ pipes, violin strings etc.) and also in electronics (electron pipes, vibration generators on semi-conductor basis etc.). Here the term active friction expresses the fact, that these effects act not as "damping" by usual passive friction but instead in the opposite way as self-acceleration. Active friction provides energy input and allows self-organization. "Active forces" are specific forces which accelerate motions in a given direction, they are of internal origin) based on internal driving factors or motor-like organs in organisms. Active forces play an important role in the motion of all animals.

Active processes which generate movement also exist in sociological processes; we speak of motivation, drive and the urge to act. We know from our observations that social actors are strongly influenced by "internal forces". Motivation is just one side of this dynamics. In order to model this aspect, at least, in the simplest possible form, we provide our agents with internal forces to improve their ability to self-organization. Thus, the agents are driven by a kind of internal engines. Once in motion, the change is accelerated. This model property should reflect that competence is largely connected with motivation and the existence of internal impulses. We link this kind of internal impulse with the picture of activity-related competence. This kind of inner drive prompts us to refer to active agents.

It is possible, for example, to imagine a creative writing process. While at the beginning it is difficult to put pen to paper, with time it gets easier when one has surrendered oneself completely to the world of writing. Another example is a *brainstorming* process in a group. While the participants find it difficult to find associations at the beginning, after a while the associations really flow. Intellection processes can accelerate as though a chain reaction has been triggered. In terms of our interpretation of velocity in the sense of activity, this means that the activity itself intensifies in the sense of positive feedback.

Briefly, the following difference exists between passive and active agents: the passive agent mainly follows external influences and forces; he follows them passively. In other words, he follows any changes without inertia of reaction. The active agent has an inner drive; he may accelerate motion even in the case that there are no external forces, he moves himself apparently aimlessly, he cannot stand still but buzzes around like an insect in a swarm. If external forces exist, he has an advantage that results from his mobility. In addition, he has the characteristic of a certain inertia; he needs time to adapt. It is our belief that such characteristics which can be more or less pronounced in an individual, are of interest for the modelling of competence development.

### 7.2 Dynamics in the general characteristics space – general model frame

In the continuous description used in the previous chapter, a number (or frequency) is assigned to each point in the characteristics space $Q$ (i.e. each vector $\vec{q} = \{q_1, q_2, ..., q_d\}$ of characteristics variables $q_k$) which indicates the implementation of certain parameter combinations. Thus, a density function $x(\vec{q},t)$ was applied over the characteristics space of the problems or characteristics. The dynamic was associated in an earlier model with the reproduction characteristics of certain characteristics combinations. In order to incorporate the role of the velocities, we require a further dynamic in the phase space. To formulate this extended dynamics, we create a kind of generalized Fokker-Planck equation.

$$\frac{\partial f(\vec{q},\vec{p},t)}{\partial t} = f(\vec{q},\vec{p},t)w(\vec{q},\vec{v};\{x\}) - \vec{v}\frac{\partial f}{\partial \vec{q}} - \frac{\partial U(\vec{q})}{\partial \vec{q}}\frac{\partial f}{m\partial \vec{v}} + \frac{\partial}{\partial \vec{v}}\left[m\gamma(\vec{q},\vec{v})\vec{v}f + D_v\frac{\partial f}{\partial \vec{v}}\right] \tag{7}$$

Let us discuss the meaning of the different terms in equation (7). The change of the distribution (term of the l.h.s.) is determined by four terms on the right hand side of the equation. In the comparison with the first model (which is determined by the evaluation function $w = E - \langle E \rangle$ and the diffusion *D*), two new functions arise here, the "potential" $U(\vec{q})$ and the "active friction" $\gamma(\vec{q},\vec{v})$. In the comparison to the standard Fokker-Planck equation, which is used in many physical applications (Ebeling and Sokolov 2005) the first term *fw* on the r.h.s. is new. This way our basic equation (7) generalizes the Fisher-Eigen equation as well as the Fokker-Planck equation. For $w = 0$ follows the standard Fokker-Planck equation. Assuming, as in our earlier model, $w = E - \langle E \rangle$ we see that the evaluation by *E* is supplemented now by an evaluation through a potential function *U*. While the first term in equation (7) favours distributions around the maxima of $E(\vec{q})$, the other terms favour distributions around the maxima of the negative potential ($-U(\vec{q})$). Both tendencies supplement each other and will go in the same direction only by assuming the special relation $U(\vec{q}) = const(-E(\vec{q}))$. The expression $m\gamma(\vec{q},\vec{v})$, which we find in the brackets on the r.h.s. plays the role of the (negative) active force $F_a = -m\gamma(\vec{q},\vec{v})$. If the friction is active, it has some negative contributions to friction and this means that the force has (for some velocities) the same direction of the velocity, this motion is accelerated.

Thus, the dynamics is far richer than it was in the first case involving a purely location-dependent dynamic. It can be shown that the purely space-dependent dynamics follows as a limit case of the phase space dynamics. In order to outline this limit case, we integrate equation (7) over the velocities. Under the precondition that $w$ given by equation (2) is purely space-dependent, the continuity equation follows

$$\frac{\partial}{\partial t}x(\vec{q},t) = (E(q) - \langle E \rangle)x(q,t) - \nabla\vec{J}(\vec{q},t) \tag{8}$$

whereby density and flux are defined using the following equations

$$x(\vec{q},t)=\int f(\vec{q},\vec{v},t)d\vec{v} \tag{9}$$

$$\vec{J}(\vec{q},t)=\int \vec{v}f(\vec{q},\vec{v},t)d\vec{v} \tag{10}$$

In order to obtain a closed equation for $x(\vec{q},t)$, the flux $\vec{J}(\vec{q},t)$ must be eliminated. This is possible in the case of constant and large friction $\gamma(\vec{q},\vec{v})=\gamma_0=const, \gamma_0 - \propto$. Following the general formula of the kinetic theory, we multiply the Fokker-Planck equation (7) by $\vec{v}$ and integrate over the velocities. In the limit case $\gamma_0 - \propto$, then

$$\gamma_0\vec{J}(\vec{q},t)=-\Theta\nabla x(\vec{q},t)-x(\vec{q},t)\nabla U(\vec{q},t) \tag{11}$$

results approximately, whereby represents an effective temperature $\epsilon = k_B T$ (dispersal of the Gauss distribution of the velocities). We have replaced the potential with the negative value function $U(q) = - E(q)$. Using the so-called Einstein relation

$$D=\frac{\epsilon}{m\gamma_0}=\frac{D_v}{m\gamma_0^2} \tag{12}$$

we come back from equation (7) to the evolution equations (5) and (6). Thus, we have proven that the dynamics in the phase space actually represents a generalization of the dynamic in the location space. However, the new dynamic is significantly more complicated and, according to our theory, in terms of its approach more suited to the description of sociological processes.

The extended mathematical model – including the evaluation term *fw* as well as the Fokker-Planck terms - has not been considered in this way in the physics literature. Our quest for an operationalization of meta-competence led us to a model in which active agents with characteristics and velocities move interactively and search in a landscape as a group. Previous models considered either the search in a landscape, but by simple diffusion agents (Darwin strategy), or the search with passive agents in a potential landscape (Boltzmann strategy). A first study on search processes with active agents in 2D potential profiles has been published recently. (Ebeling et al. 2005) We discuss here the main results of our model. First, we discuss the evolution dynamics of Fisher-Eigen type in the coordinate space and then the Fokker-Planck dynamics in the phase space.

## *7.3 Analysis of the evolution dynamics – consideration of simple cases*

### 7.3.1 <u>Evolution dynamics in the coordinate space without inertial effects</u>

The new model is very rich in relation to the dynamics, we shall now consider in more detail certain special cases where explicit analysis is possible. As a first special case, we shall consider again the borderline case of so-called damped dynamic. Extensive results and examples already exist for this. (Ebeling, Karmeshu et al. 2001)

To explicitly implement the transition from the new to the old model, we set in accordance with equation (2) $w = w(\vec{q}) = \eta[E(\vec{q}) - \langle E \rangle]$, whereby $\eta > 0$ is a free parameter with which we can make the share of the Fisher-Eigen dynamics stronger or weaker. Furthermore, as in the previous section, we assume that $\gamma = \gamma_0 = const$ and go to the boundary $\gamma_0 \to \infty$, i.e. to stronger damping. Then, for the space-dependent density $x(\vec{q},t) = \int f(\vec{q},\vec{p},t) d\vec{p}$ , as described above, the following equation follows

$$\frac{\partial x(\vec{q},t)}{\partial t} = \eta x(\vec{q},t)[E(q) - \langle E \rangle] + D\nabla\left[\nabla x(\vec{q},t) - \frac{1}{\Theta} x(\vec{q},t)\nabla U\right] \qquad (13)$$

This corresponds to a generalization of the dynamics of the earlier models in the characteristics space according to equation (5) or (6). In other words, the Fisher-Eigen dynamics is extended by a flow term proportional to the gradient of $U(\vec{q})$. Consequently, the new model of the characteristics-space dynamics is already a real generalization of the old model. In the context of optimization problems, a dynamics based on equation (13) is called a combined Darwin-Boltzmann strategy in the characteristics space.

(Boseniuk, Ebeling et al. 1987; Boseniuk and Ebeling 1991; Asselmeyer, Ebeling et al. 1996; Asselmeyer and Ebeling 1997)

This search strategy can be described as follows. Agents search a landscape, they communicate and compare their positions, they move to better positions looking for higher maxima of the valuation function $E(\vec{q})$ (Darwin component). At the same time they move also under the influence of the external potential $U(\vec{q})$ looking for positions located at the minima of the potential. The temperature parameter defines here the range of the stochastic (random) changes by diffusion. It could also be said, it defines with what level of individual variability and spread of the mutations the agents move. If the temperature is decreased over time, the agent cloud is concentrated around the maxima of $E(\vec{q})$ or the minima of $U(\vec{q})$.

It makes sense to limit the variety of the model further and to assume a simple permanent relation between the value function $E(\vec{q})$ and the potential $U(\vec{q})$. In the simplest case, $U(\vec{q}) = U_0 - E(\vec{q})$, where $U_0$ is some constant parameter.

The minus sign ensures that both partial processes go in the same direction, i.e. towards the maxima of the value function. For the residual potential, we can use a constant, e.g. zero, or a simple function, e.g. a paraboloid, with which a “confinement” can be achieved in a certain space area. It is also possible to model attracting or repulsing interactions with this residual potential. We shall return to this aspect later.

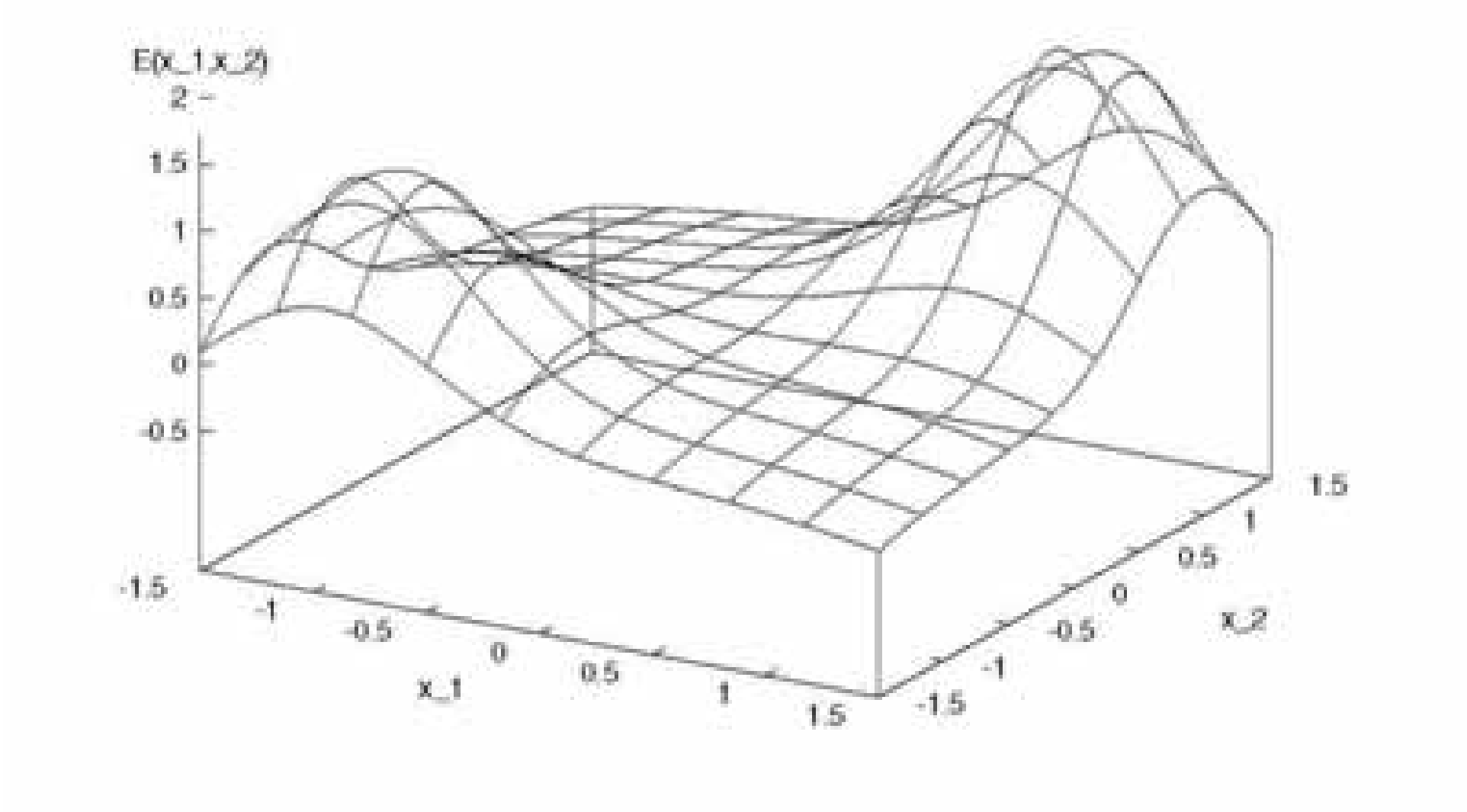

**Figure 29: Example of an evolution landscape with two peaks**

Figure 29 shows a special two-dimensional landscape $\vec{q} = (x, y)$ with two peaks. The right peak is somewhat higher than the left one. The analytical representation of the landscape in the particular example is:

$$E(x, y) = 1.5\exp\left(-2(x+1)^2 - 2(y+1)^2\right) + 2\exp\left(-2(x-1)^2 - 2(y-1)^2\right) - 0.1x^2 - 0.1y^2 \quad (14)$$

The height of the left peak at (-1,-1) has the value 1.5 and the height of the right peak at (+1,+1) is 2. Between the two peaks, in the coordinate origin (0,0), there is a saddle. A transition from left to right would thus correspond to an improvement of the evaluation of a characteristics combination. Such transition processes require a special dynamics. The above-specified Fokker-Planck equation defines such a dynamics.

The derived equation system including elements self-reproduction of the Fisher-Eigen type as well as Boltzmann-type diffusional flow has become known as a model for so-called mixed evolutionary strategies. (Schweitzer, Ebeling et al. 1996; Asselmeyer and Ebeling 1997) Mixed strategies, which are included as a special case in our new model, have proven a very effective instrument for the resolution of complex problems. (Asselmeyer, Ebeling et al. 1996; Schweitzer, Ebeling et al. 1997; Schweitzer 2002)

### 7.3.2 The active Brownian dynamic model

Another special case of much recent interest, which is still not very well explored is active Brownian motion in a landscape. This model which follows from equation (13) for the case that there is no evaluation, $w = 0$, or $= 0$. A real Fokker-Planck equation results which dynamical basis can also be formulated as a Langevin equation. The theory of active Brownian motion has also undergone intensive development and been subject to numerous applications. (Steuernagel, Ebeling et al. 1994; Ebeling, Schweitzer et al. 1999; Schweitzer, Ebeling et al. 2001; Schweitzer 2002)

The previous applications of the active Brownian dynamics were concentrated on biological problems, e.g. the dynamics of swarms of biological objects (Ebeling and Schweitzer 2001) and so-called Brownian agents. (Schweitzer 2002)

As our model contains both the mixed evolutionary dynamics and the dynamics of active Brownian systems as a special case, the possibility of combining the results of both directions arises.

In the simplest case, which contains the two borderline cases, we have the dynamics:

$$\frac{\partial f(\vec{q},\vec{v},t)}{\partial t} = f\eta\left[E(\vec{q}) - \langle E(\vec{q})\rangle\right] - \vec{v}\frac{\partial f}{\partial \vec{q}} + \frac{\partial E(\vec{q})}{\partial \vec{q}}\frac{\partial f}{m\partial \vec{v}} + \frac{\partial}{\partial \vec{v}}\left[m\gamma(\vec{q},\vec{v})f + D_v\frac{\partial f}{\partial \vec{v}}\right] \qquad (15)$$

These rather very complex partial differential equations are very difficult to analyse. However, some special solutions are known. (Asselmeyer, Ebeling et al. 1996; Erdmann, Ebeling et al. 2000)

The simulation of populations, which can be described by equation (15), can be implemented in the special case of $= 0$ , i.e. there is no self-reproduction, with a relatively simple algorithm which is developed and applied below. By applying the general relation between the Fokker-Planck equations and the so-called Langevin-equations (Anishchenko, Astakhov et al. 2002), we obtain the motion equations for the individuals of the population.

### 7.4 From occupation landscapes to swarms of agents – stochastic Langevin dynamics of the individuals – the “Brownian agent” simulation

The above-described partial differential equations (15) can only be solved analytically in the simplest cases and numerically with a high level of complexity. It is easier and clearer to simulate the dynamics of individuals. We are specializing here on the case of the dynamic of individuals with Cartesian coordinates in a low-dimensional space (in our simulations generally two-dimensional) and represent the coordinates and velocities of the i-th individual as vectors $r_i$, $v_i$ in this space. The Langevin equation which corresponds to equation (15) with $\eta = 0$ (no reproduction) reads:

$$m\frac{dv_i}{dt} = K_i + F_i + \sqrt{2D}\xi_i(t) \tag{16}$$

This is a Newton motion equation extended by a stochastic Langevin source. The terms on the right side of the motion equation represent the forces that affect the individuals, symbolically represented as “particles”. The term $K_i$ models the external force that drives the agents on the mountains in the landscape of the function $E$. We assume that this force has “potential character” as well as the conservative forces of mechanics.

$$K_i = \frac{\partial E(r_1,\ldots,r_N)}{\partial r_i} \tag{17}$$

The second force term $F_i$ models "dissipative forces", which we specify as follows (Schweitzer, Ebeling et al. 1998; Ebeling, Schweitzer et al. 1999, Erdmann et al. 2000):

$$F_i = - m\gamma_0 v_i + de_i v_i \tag{18}$$

Here, $_0$ designates a coefficient of passive friction. This coefficient has the dimension of a frequency. The other term ($de_iv_i$) models an acceleration of agents in the direction of the velocity $v_i$ (a forward thrust) which is based on the conversion of a kind of “energy” from a reservoir $e_i$. As in a biologically motivated model, we assume a simple monotone function with decreasing velocity (this is sometimes called SET model). (Schweitzer, Ebeling and Tilch, 1998; Ebeling, Schweitzer and Tilch 1999).

$$e_i = m\frac{q}{c + dv_i^2} \tag{19}$$

Thus, we obtain the dissipative force

$$F_i = m\left[\frac{dq}{c + dv_i^2} - \gamma_0\right]v_i \tag{20}$$

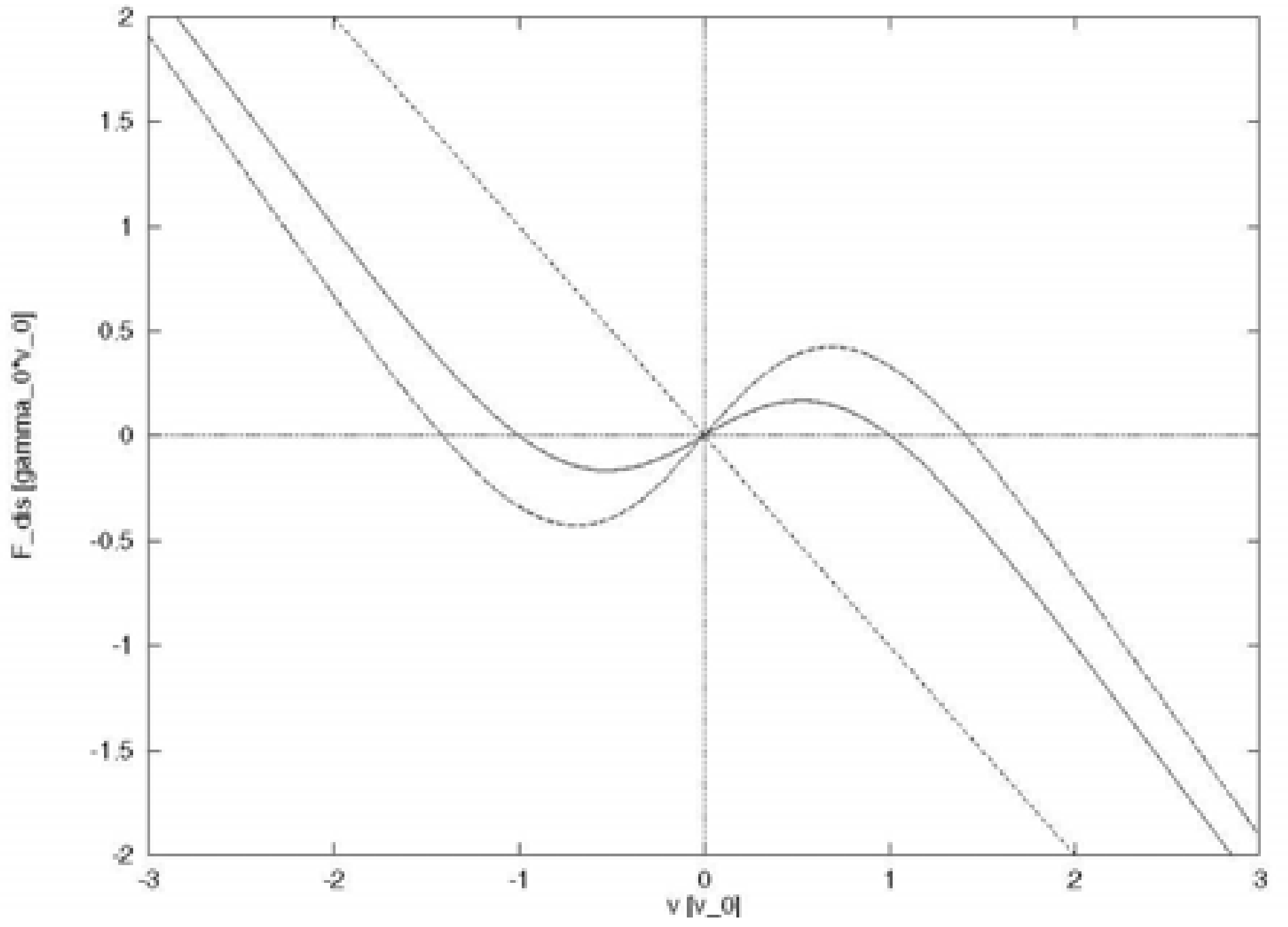

**Figure 30: Dissipative forces: two examples of a dissipative force with accelerating effect for v < $v_0$ as compared with a purely passive frictional force (straight line) which is constantly opposed to the velocity. The idea behind formula (20) is that slow "particles" can be accelerated while very fast "particles" are damped.**

The last force term on the right side of equation (16) refers to the stochastic force which affects the Brownian particle *i* with the strength *D*. The random force $\xi_i(t)$ is assumed as Gaussian white noise. As standard in statistical physics, we presuppose that the passive friction is connected with the noise level through an Einstein relation (see above):

$$D = m\gamma_0 kT \tag{21}$$

A FORTRAN program which in the basic version was written by Alexander Neiman was used to simulate Langevin equation. In particular we used the program – extended in different directions – to simulate the accelerated (active) agents in a two-dimensional landscape. We call the entities of our simulation *Brownian Agents*. The aim here was not to create an interactive interface, but to carry out systematic evaluations.

In the simulations, the population is represented as a point cloud. In this way, we obtain an impression of the development of the probability density. Figure 31 to Figure 33 show, by way of example, how the point cloud develops in the case of two hills which are located in the lower left and top right quadrants of the plane.

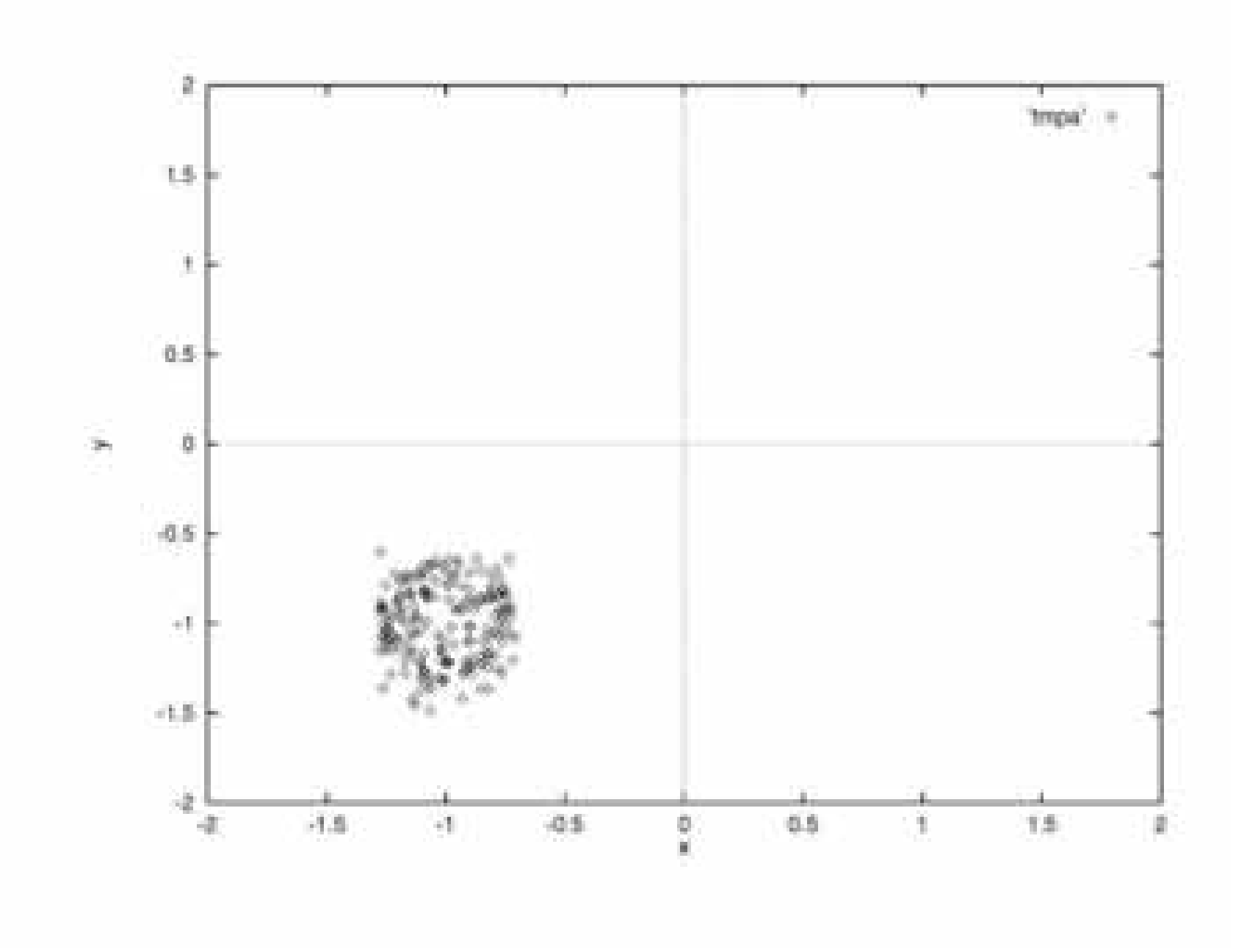

**Figure 31: The population occupies a relatively high peak in the evolution landscape. In the sample landscape shown in Figure 29, this corresponds to the occupation of the left peak.**

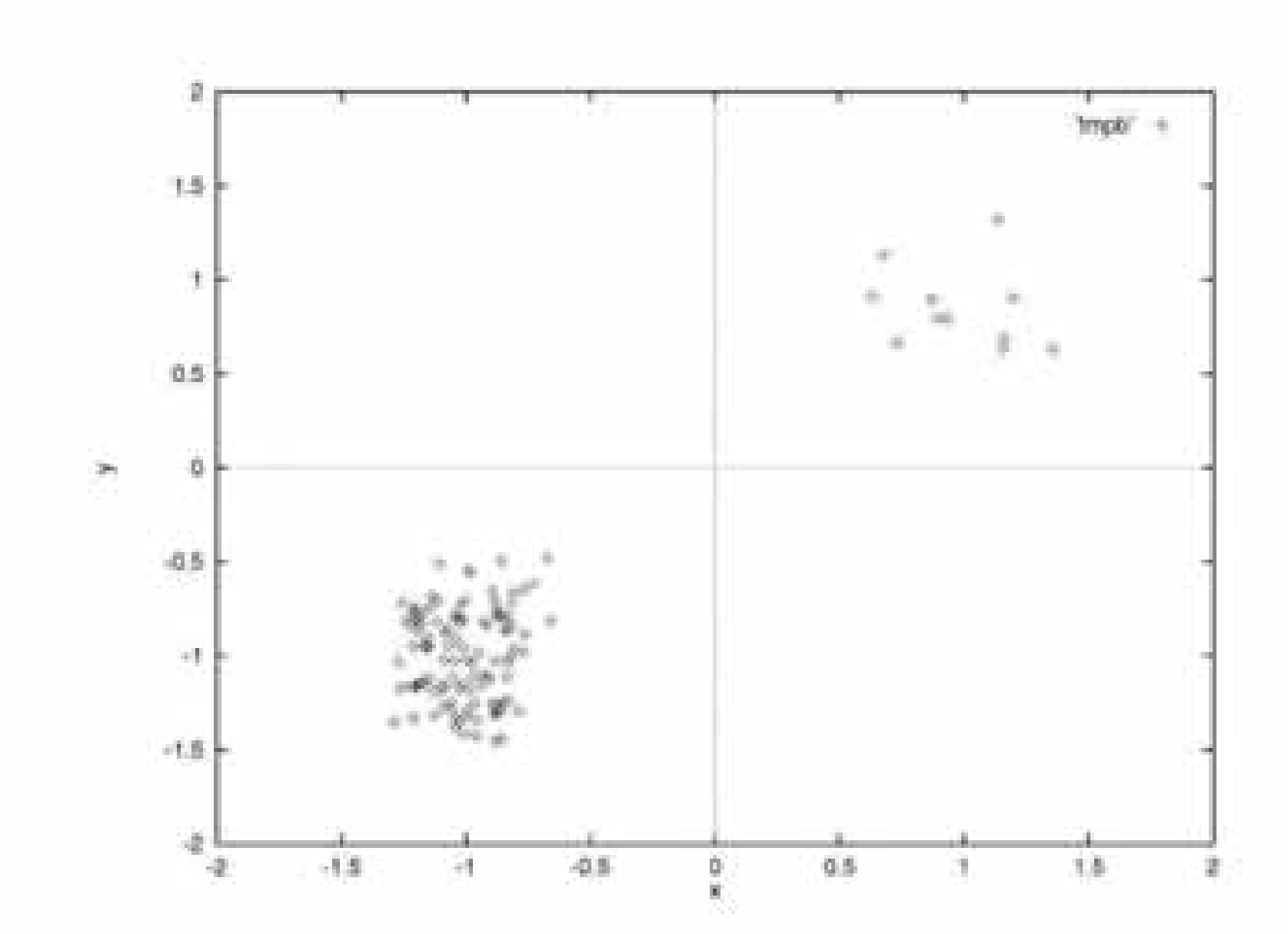

**Figure 32: The population completes the transition to a new (higher) peak in the evolution landscape. In the sample landscape shown in Figure 29, this corresponds to a hike across the valley in the centre of the landscape.**

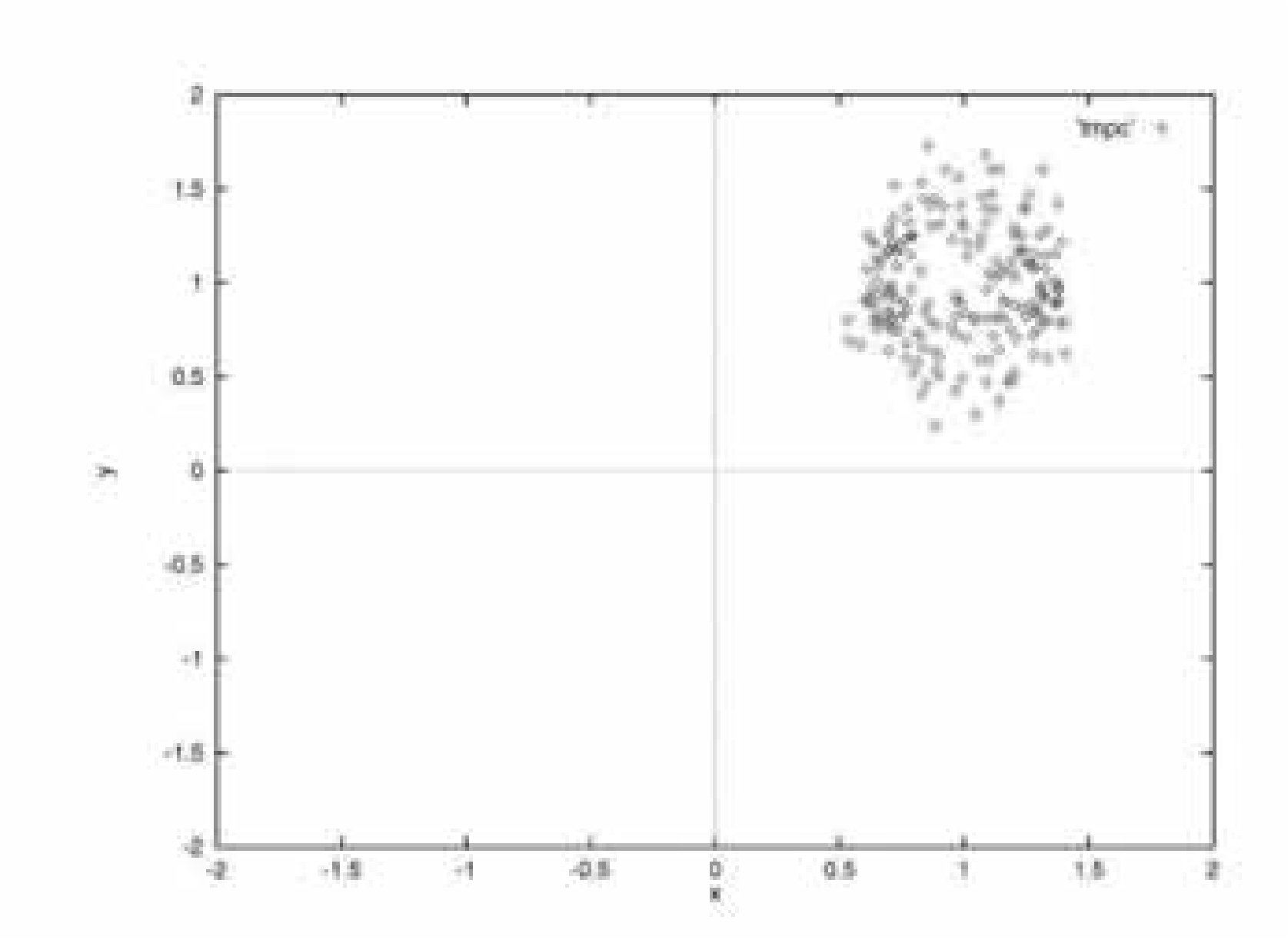

**Figure 33: Final stage in the transition to a new (higher) mountain in the evolution landscape. In the example, this correspond to the population of the right peak in Figure 29**

### 7.4.1 The average time for transitions of active independent agents between the wells in bistable potentials

We start again with the model of the active Brownian agents whose location in the characteristics space and velocity is described by the following set of coupled differential equations:

$$\begin{aligned} &\dot{x}_1 \quad v_1 \\ &\dot{x}_2 \quad v_2 \\ &\dot{v}_1 \quad \frac{1}{m}\frac{U}{x_1} \qquad v\ v_1 \quad \sqrt{2D}\ {}_1\ t \\ &\dot{v}_2 \quad \frac{1}{m}\frac{U}{x_2} \qquad v\ v_2 \quad \sqrt{2D}\ {}_2\ t \end{aligned} \qquad (22)$$

Here, $U(x_1,x_2)$ represents the potential which, as introduced in Chapter 7.3.1, is linked with the valuation landscape through a negative algebraic sign, i.e. where the valuation landscape has its maxima, the potential has a minimum. The second term on the left side describes the conservative forces that drive the agents on to the peaks in the valuation landscape or into the valleys of the potential. The first term on the right side describes the so-called dissipative force; this can be a normal passive friction, in which case $= {}_0$=const, or active friction, i.e. an additional impulse. For the case of active friction, according to equation (20) the following formula can be developed through the introduction of new variables:

$$\gamma(v) = \gamma_0\left(1 - \frac{\delta}{1+\left(v_1^2+v_2^2\right)/v_d^2}\right) \qquad (23)$$

In the following, we consider a landscape which displays two maxima of equal height. Thus, the corresponding bistable potential takes the form of:

$$U(x,y) = a\left(\frac{1}{4}z^4 - \frac{1}{2}z^2 - cz\right) + \frac{1}{2}\omega^2\left(x-y\right)^2 \qquad (24)$$

whereby *z=(x+y)/2*, i.e. the potential pots lie on the diagonals. The parameter c determines the asymmetry in the depth of the potential pots. If *c = 0*, the two valleys are equally deep, i.e. both peaks are the same height. The height of the barrier *U* (depth of the valley) between the two minima (maxima) results as *a/4*.

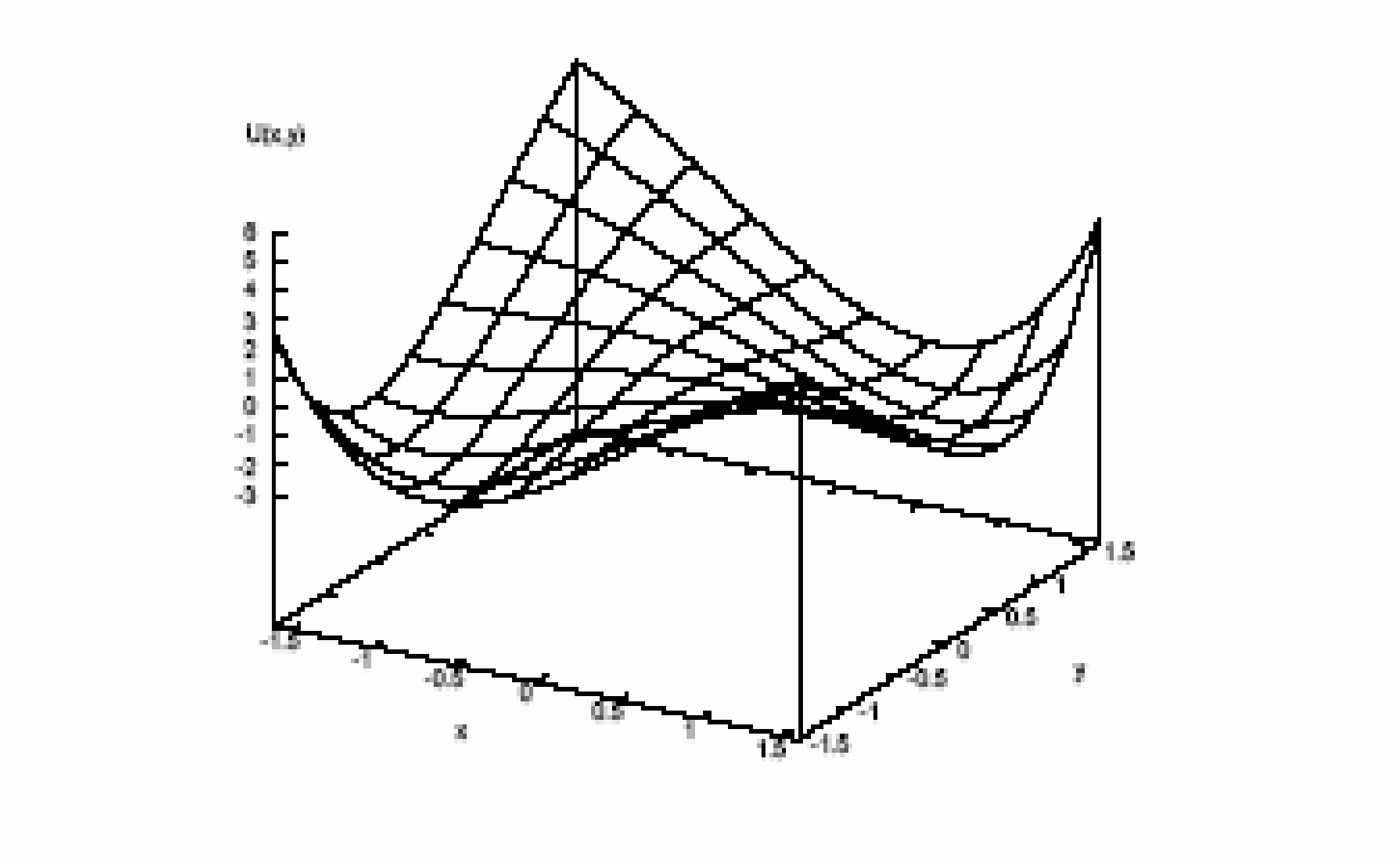

**Figure 34: Representation of the potential U with potential minima lying on the diagonal.**

The following parameters arise for the simulations from equations 22, 23 and 24:

- D – Strength of the random noise or mutation strength
- U – Height of the barrier between the minima – measure of the difficulty of the transition, difficulty of the problem solution
- c – Measure of the value difference between the different problem solutions. If c=0, the two problem solutions are equal in value and the transition only describes the process for finding something new but alternatively equivalent. If c is not equal to 0, there is not just another solution available but a better one.
- Measure of the inner impulse, activity strength.

Given that the representation of the group of agents as a point cloud and their transition always gives rise to similar pictures, we looked for another parameter with which to describe the transition process. As with the *Evolino* simulations, we firstly formed the group order parameter, i.e. the group centre calculated as the focus. This initially remains in the potential minima in which the agent's motion starts. The more agents that swarm out and reach the new objective, the more the group order parameter will shift in this direction. The wandering motion of the group centre coordinates were measured over time. This makes it possible to determine the transition times from one optimum to the next highest. If one understands the change between the different maxima in the competence or problem space as a crucial transformation for the group and its members (value and norm transformation), the transition period could also be described as a "time of change" in the metaphorical sense.

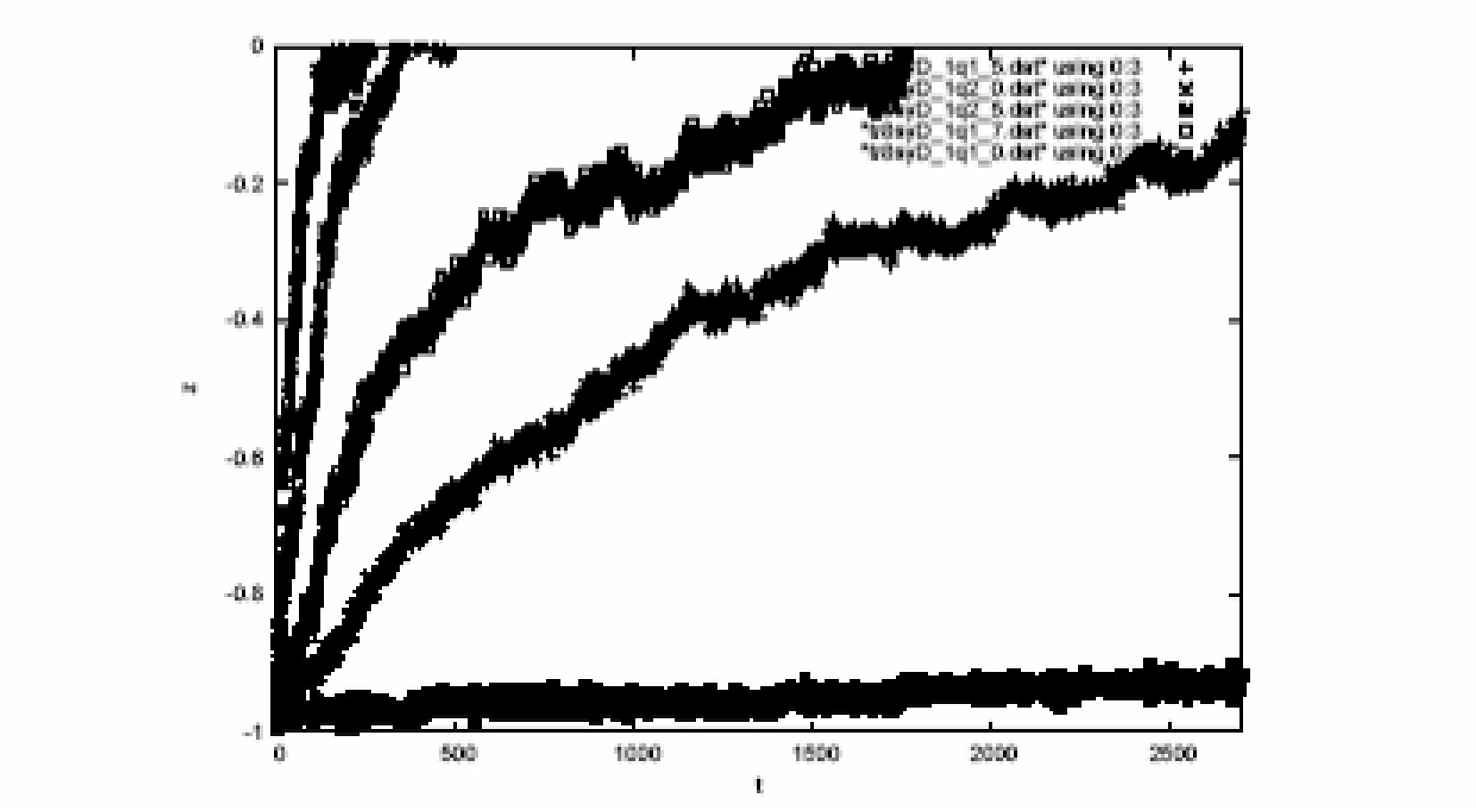

**Figure 35: Transition of the group order parameter from the starting point (coordinates z=-1) to the next peak with a symmetrical potential and average impulse. Parameters: D=0.1,   U=1, c=0 and curves from left to right:   =2.5,   =2.0,   =1.7,   =1.5,   =1.0.**

In this simulation, there is no interaction between the agents, nor do they compare their position among themselves. Instead, they all make the transition independently of each other. That there is a collective effect despite this lies in the fact that all agents act under the same boundary conditions. This is comparable with a situation whereby employees in a company are given the same task which they must resolve independently, the only criterion being that the task is resolved. In such a case, the individual solution paths can be very different. Figure 35 demonstrates clearly that the agents may not find the new solution without inner drive or activity. The stronger the inner drive, the faster the transition to the new optimum will be.

Figure 36 shows the same process under the condition that the new goal now has a higher value, i.e. not only must the valley between two peaks be overcome, the second peak is higher and has therefore a higher attractiveness.

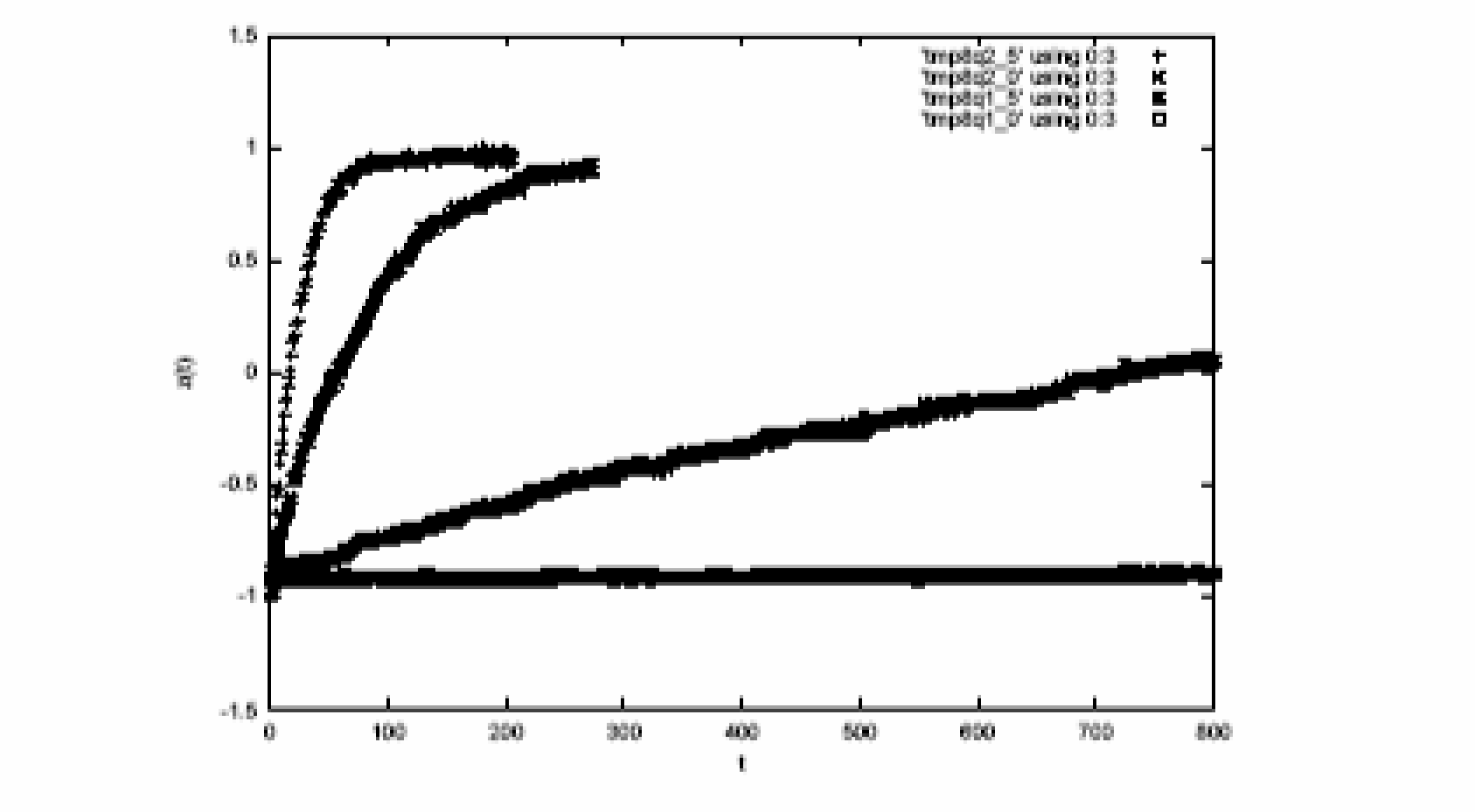

**Figure 36: Transition of the group order parameter from the starting point (coordinates z=-1) to the next peak (coordinates with z=1) with an unsymmetrical potential and strong and average drive. Parameters: D=0.05, U=1, c=0.1 and curves from left to right: =2.5, =2.0, =1.7, =1.5, =1.0.**

The comparison of the two figures shows that if the impulse level is low the transition takes place smoothly. If the inner drive increases, there is a sudden transition. Thus, the inner drive represents a kind of critical parameter which can also change the transition behaviour qualitatively. Moreover, the transition takes place generally faster than in the case of the symmetrical potential. The higher peak exercises a greater attraction and accelerates the finding of the solution. In other words, the clearer the evaluation differences, the faster the acceptance of the better solution.

### 7.4.2 The influence of group interaction on the transition times – communicating active agents

In this section we introduce attracting or repulsing interactions between the agents which are modelled as conservative interactions in the physical sense. In the case of a social dynamic, interactions describe how individual decision-making processes are influenced by other individual or by the group. Communication is a key mechanism of such influencing.

In the simplest case, we assume that the individuals (agents) orient themselves in accordance with the group order parameter (common focus in the sense of an average, mass centre or geometrical centre in the characteristic space). An attracting interaction then describes the mechanism whereby the individual would like to be as similar as possible to the group order parameter (average) and feels attracted by the group norm. In this case, it may be assumed that the individual does not want to stand out, either through a special competence profile or through original problem solutions.

In order to model such interaction, we revert to the aforementioned residual potential (Chapter 7.3.1). We initially assumed that we can use a constant, e.g. zero, or a simple function, e.g. a paraboloid, for the residual potential with the "*confinement*" which is limited to a certain space area. In equation (14) we applied

$$U_0(x,y) = a\left(x^2 + y^2\right)$$

where a=0.1. This gave rise to the individuals with their dynamics being localized weakly in a certain space area around the zero point.

Furthermore, it was already mentioned that one can also model interactions between the individuals with this residual potential. A relatively simple and manageable case arises when parabolic interaction potentials (linear forces between the partners) are applied. This corresponds to the approach:

$$U_0 = a\sum_i \left(x_i^2 + y_i^2\right) + \sum_{ij} b_{ij}\left(\left(x_i - x_j\right)^2 + \left(y_i - y_j\right)^2\right)$$

This leads to the force function

$$K_i = -\frac{\partial U_0(r_1,...,r_N)}{\partial r_i}$$

with the x-y components

$$K_{ix} = -ax_i - b\left(x_i - R_x\right)$$
$$K_{iy} = -ay_i - b\left(y_i - R_y\right)$$

Here, R is the vector of the centre of gravity of particles with the components

$$R_x = \frac{\sum_j m_j x_j}{\sum_j m_j}$$
$$R_y = \frac{\sum_j m_j y_j}{\sum_j m_j}$$

Positive parameters mean attraction to the centre of gravity and negative mean repulsion from the centre.

Figure 37 shows the movement of a group of agents who are only held together by interaction. This movement is comparable with the movement with which the *SynKom* simulation was generated. There is no valuation landscape on the basis of which the players can orient themselves. The individuals relate to each other or the group with their decision-making processes. The value or norm is generated internally. The individuals either relate positively to each other (imitation) or try to separate themselves, swarm out (activity).

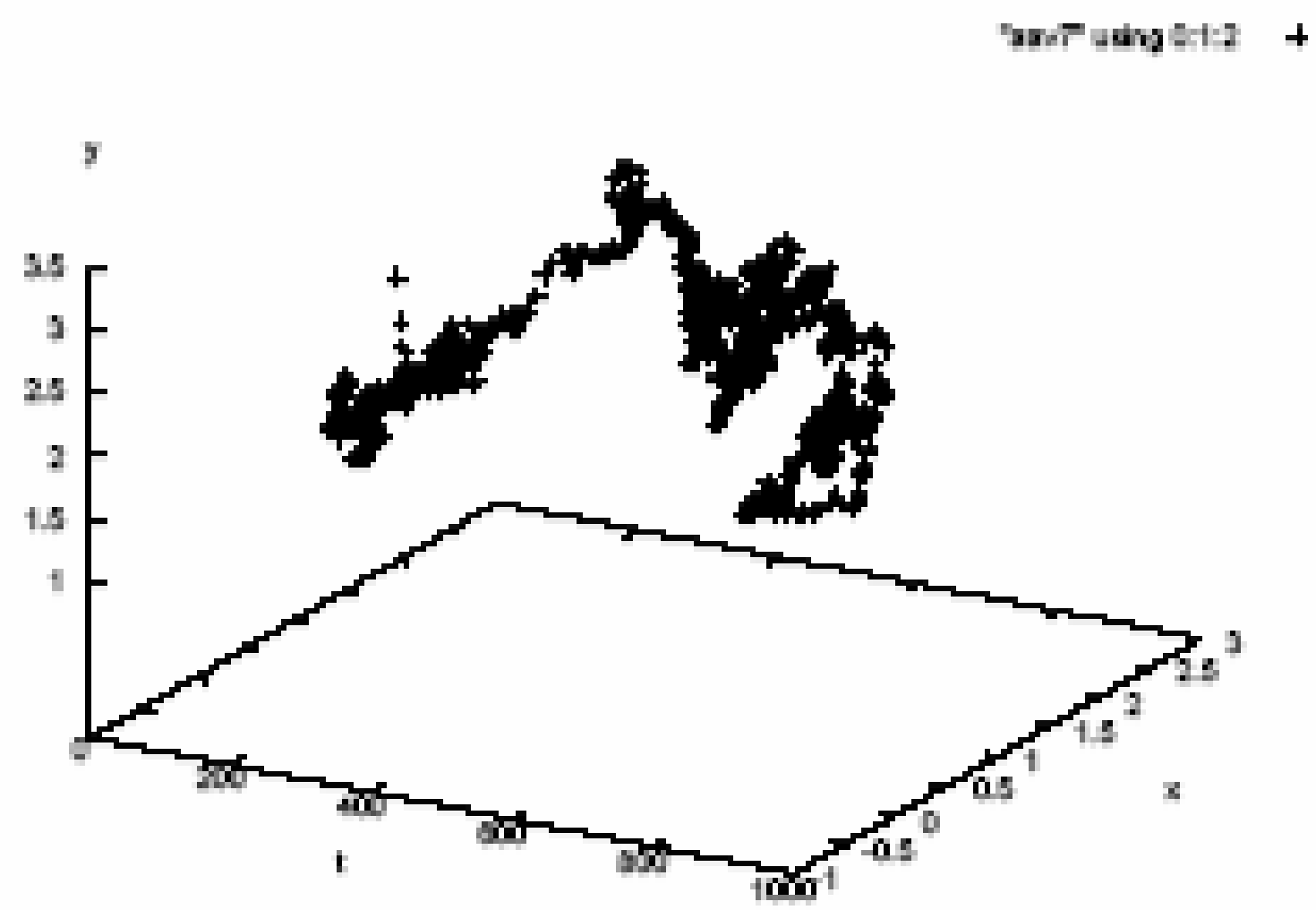

**Figure 37: Free motion of group of agents who are held together by interaction (modelled as linear interaction). Parameters:   =3, a=1, D=0.5, N=300.**

The influence of conservative collective forces on a transition in a bistable valuation landscape can be described as follows: the individuals in the left minimum hold together and delay the transition which then occurs very suddenly. Figure 38 shows that no common transition is achieved for strong collective interaction a=1.2. For somewhat smaller values a= 1.0, after around 250 temporal steps a sudden common transition is observed and for the half value a=0.5, things go far faster after around 20 temporal steps.

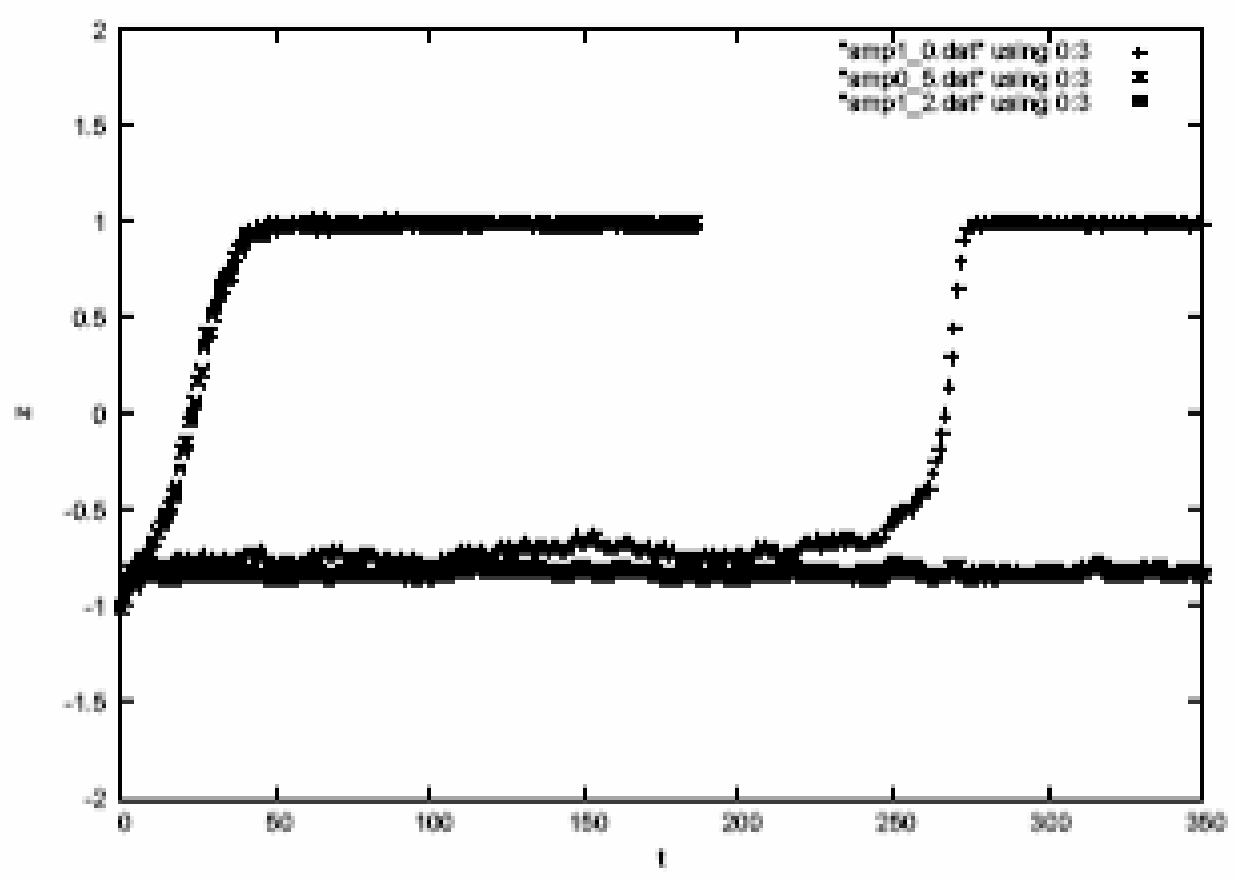

**Figure 38: Transition of the group order parameter to a higher peak in an unsymmetrical potential. The individuals interact with each other. The parameter "a" specifies the strength of an interaction. Other parameters: D=0.1, c=0.1,   =2.0 for the following interaction strengths a=0.5,1.0,1.2 (from left to right)**

In Figure 39 the transition for small collective attraction a=0.5 is compared with the case whereby there is no reaction a=0 and the case whereby there is collective repulsion a =-0.2. In the case that the group centre is dominant, the group remains centred around the old value for longer, than it would without interaction. On the other hand, if the majority of the agents is initially in the new maximum, the re-formation of the group centre then occurs more quickly at the new position. Thus, attracting group interactions decelerate the first transition and then accelerate it later. As opposed to this, a repulsion, i.e. the tendency of individuals to distinguish themselves from the group, to break away, the higher individual activity does not appear to have a positive effect in any case. The leaving of the old group value takes place as quickly as without interaction, i.e. faster than with a dominant group centre. On the other hand, the group finds it more difficult to agree again on a new value order parameter.

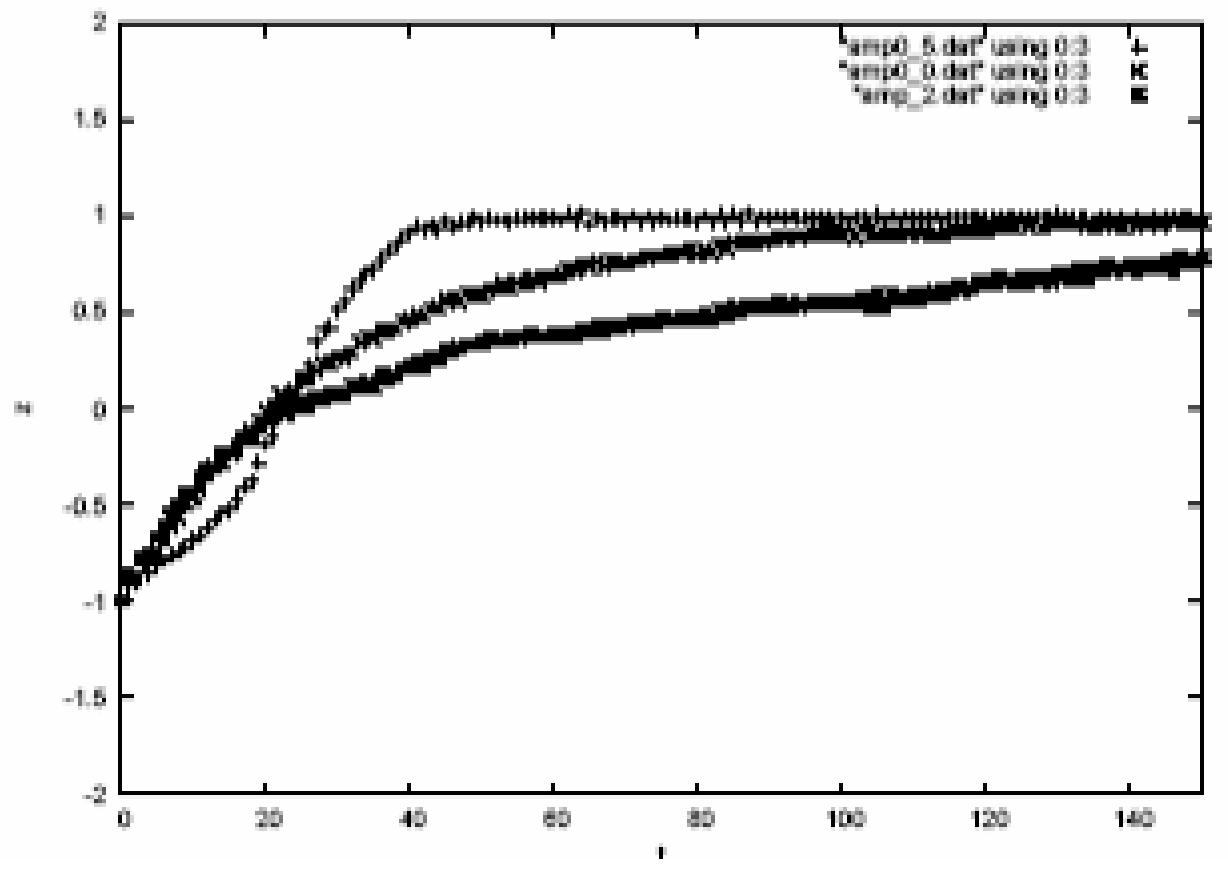

**Figure 39: Transition between peaks of different heights in the case of attracting and repulsing interaction. Parameters: D=0.1, c=0.1, d=2.0 and strength of the interaction a=0.5,0.0,-0.2**

Another way of including interactions involves interactions which link the velocities with each other. Thus, situations can be modelled in which individuals try to change their characteristics in the same direction as the group does. This leads to a targeted group dynamics.

# 8 Summary and Outlook

We understand evolution as the outcome of self-organization processes whereby the appearance of new characteristics plays a central role. In terms of the universal properties of non-linear phenomena, the scientific theories that describe them have adopted a bridging function between the understanding of nature and society. (Mainzer 1997) We adopted the position that it is legitimate to examine the extent to which the instruments developed in modern scientific theories can be used to describe development processes in the area of educational processes and, in particular, with regard to the formation of competence.

Since the development of the classical theory of self-organization, concepts and methods from the natural sciences have been transferred to social-scientific fields of application (cf. Weidlich and Haag 1983; Haag, Mueller et al. 1992; Laszlo 1992 and earlier works, e.g. Haken 1973; Haken 1977; Prigogine and Sanglier 1985). Together with the extension of the methodical instruments available, this also led to the formation of new problem fields in the disciplines in question. In terms of an example, one need only think of the area of evolutionary economics (cf. Witt 1993; Hodgson 1998) and the research area "artificial societies" (Conte and Gilbert 1994; Epstein and Axtell 1996; Hegselmann 1996; Gilbert 1997). In this context, this study can be classified as belonging to the transfer of concepts and models.

Out of the wide variety of mathematical models, those associated with the concept of landscapes in high-dimensional spaces are dealt with in particular. Evolution is described as the collective search of interacting individuals and groups for better local solutions in an unknown landscape. In this *landscape picture*, evolution dynamics are examined from the perspective of population-oriented concept and model approaches. Other applications have also been implemented by the authors for additional social phenomena. (Bruckner, Ebeling et al. 1989; Bruckner, Ebeling et al. 1990; Ebeling, Karmeshu et al. 2001; Scharnhorst 2001; Scharnhorst 2003)

In the call for project proposals, attention was drawn to two points which are often neglected in today's model literature. "Firstly, there are very few studies on the dynamic of self-organizing systems from clearly differing dissimilar 'particles'. However, this is the normal case in every socially interacting group. The actors contribute very different and often opposing competences. Their interaction in the self-organizing action of the group can accelerate or hinder successes. In particular, if they change significantly during the thought and action processes, this can lead to a completely unpredicted system dynamic" and "Secondly, a basic question concerning such modelling is the extent to which a generalized self-organizations disposition, a self-organizations competence exists, in addition to the usual basic competences (i.e. the personal, activity-related, technical-methodical and socio-communicative), which includes the ability to integrate oneself into self-organization processes. It would have the character of a meta-competence and would be a precondition of all other basic competences".

This study answers the first question by proposing a specific model formation. We do not describe individual processes of competence development in the sense of a personality model, but group processes of a larger number of individuals. However, given that these individuals are described in a competence space, an individuality of the agents is automatically given. It results from the fact that the group is distributed in the competence space. Variability and variance around the group order parameter is an expression of individual idiosyncrasy. Each place in the competence space is unique and corresponds to a unique competence profile. In that this location can be travelled through, competence development can be modelled. The individual trajectories in the competence space correspond to the individual biographical experiences. The extension of the model through the observation of velocities in this individual space also makes it possible to conceptualize meta-competence. Simulations like *EvoKom*, *SynKom*, and *SynKom_Berg* make it possible to follow the individual development steps. At the same time, the spotlight is on group processes. The parameters that appear in the models specify the boundary conditions for all group members. While the competence space model treats individual competences as variables, the problem space model returns to the group view of competences. In the second model, competences are treated as group parameters. In this case, the introduction of velocity makes it possible to represent activity-related competence as a special basic competence.

In all of the models presented in this study, the objective is the resolution of a basic evolution problem: How does evolution leave optima that have once been reached and find new solutions? How does innovation arise through the instabilization of existing self-organized patterns?

In the case of the competence centre, what is involved is the transition to new norms and values in a group. The question that arises here is how the existence of different competence profiles in a group (variability achieved through mutation), the mutual reference and dominance of the group value and the group communication of the value of certain competence profiles contribute the fact that the group reaches an agreement on another value in the competence space. The simulations show that what is involved here is an optimal combination of variability and standardization. In the case of the problem space, it is the competences that assume the role of evolution mechanisms like selection, mutation and imitation. In the corresponding simulation games, the “optimal” combination of such competences can be exercised.

In terms of possible extensions of the approach developed in this study, the following points can be raised:

1. The linking of search strategies with different landscape forms. In the *Evolino*-based simulations, two prototypical landscapes are assumed in each case. It would be conceivable here to design games that more different landscape types and thus test the strategies against landscapes. This has already been done in the *Brownian Agents* simulation, however without the possibility to play the simulations on-line.
2. Development of the simulations on the training of the motivation of certain competences in problem-solving processes: the practical reference of different landscape forms (external and internal) must be further developed here. Which learning situation corresponds to which landscape type?
3. Introduction of individual agents in the *Evolino*-based simulations: what happens when different agents with different competence strategies not only resolve a task one after the other but simultaneously and still interact with each other while doing it? Such an extension would of course dramatically increase the number of parameters in the model. If this is interactively designed, the players would have the option of designing “their” agents. This means that the parameter diversity would be limited in a natural way. Furthermore, such a game could be used for the observation of selected competence strategies.

**Acknowledgement**: This project was funded by the Federal Ministry for Education and Research, Germany and the European Social Fund. Special thanks goes to John Erpenbeck for his fruitful comments on the operationalization of competences in the light of self-organization theories. The authors are grateful to Anne Beaulieu in her function co-leader of the project on "Competence and innovation in research networks - modeling self-organized learning of heterogeneous agents" this report is part of. In particular, we thank Anne for her comments of the design of the simulations. We thank Thomas Hüsing, who programmed most of the simulations and also contributed substantially to our debates. Further, we would like to thank Alexander Neiman to make his program available to us. We thank all colleagues of the group Networked Research and Digital Information (Nerdi) at NIWI-KNAW which commented on the interactive simulations.